\documentclass{jfm}

\usepackage{graphicx}
\usepackage{newtxtext}
\usepackage{newtxmath}
\usepackage{natbib}
\usepackage{hyperref}
\hypersetup{
    colorlinks = true,
    urlcolor   = blue,
    citecolor  = black,
}
\usepackage{subcaption}

\newcommand{\RomanNumeralCaps}[1]
\linenumbers

\title{Developing a Non-Newtonian Fluid Model for Dust, for Application to Astrophysical Flows}

\author{Elliot M. Lynch \aff{1}
  \corresp{\email{elliot.lynch@ens-lyon.fr}},
 \and Guillaume Laibe,\aff{1}}

\affiliation{\aff{1}Univ Lyon, Univ Lyon1, Ens de Lyon, CNRS, Centre de Recherche Astrophysique de Lyon UMR5574, F-69230, Saint-Genis,-Laval, France}

\begin{document}
\maketitle

\begin{abstract}
In the astrophysics community it is common practice to model collisionless dust, entrained in a gas flow, as a pressureless fluid. However a pressureless fluid is fundamentally different from a collisionless fluid - the latter of which generically possess a non-zero anisotropic pressure or stress tensor. In this paper we derive a fluid model for collisionless dust, entrained in a turbulent gas, starting from the equations describing the motion of individual dust grains. We adopt a covariant formulation of our model to allow for the geometry and coordinate systems prevalent in astrophysics, and provide a closure valid for the accretion disc context. We show that the continuum mechanics properties of a dust fluid corresponds to a higher-dimensional anisotropic Maxwell fluid, after the extra dimensions are averaged out, with a dynamically important rheological stress tensor. This higher-dimensional treatment has the advantage of keeping the dust velocity and velocity of the fluid seen, and their respective moments, on the same footing. This results in a simplification of the constitutive relation describing the evolution of the dust Rheological stress.

\end{abstract}


{\bf MSC Codes }  {\it(Optional)} Please enter your MSC Codes here

\section{Introduction}



The dynamics of dust in turbulent flows is important to a wide array of astrophysical, geophysical and engineering applications. In the case of astrophysical applications, dusty astrophysical fluids often combine a high Mach number with subsonic turbulence which feeds off of a Rayleigh stable shear flow. The dust number density is typically much lower than that of the gas, such that dust-dust collisions are infrequent. However, dust particles are typically too numerous to be kept track of individually. As such there is a need to be able to model the dynamics of weakly collisional/collisionless dust in turbulent gasses effectively.


The most physically accurate method of evolving dust grains in fluids is an N-body approach where each solid particle is evolved independently \citep[although this approach can still exhibit spurious trapping behaviour,][]{Commercon23}. However, this approach is typically prohibitively expensive for practical computations in the astrophysics setting, due to the large range of lengthscales and number of dust particles involved, except on the smallest of scales. Two common methods are used to make modelling dust dynamics computationally tractable. One is to significantly reduce the number of dust grains compared with reality, or to treat N-body particles as a dust aggregate; for instance the dust module in Athena, \citep{Bai10,Zhu14}, and PLUTO, \citep{Mignone19}, and superparticle implementations by \citet{Youdin07b,Balsara09} and \citet{Yang16b}. This is commonly used when there is no back reaction or interaction between dust grains as the number of particles required to achieved convergence will be much lower. In accretion disc simulations, making use of such such methods, it is common to employ of order 10 particles per cell \citep{Laibe12a}, which is not sufficient to adequately sample the particle velocity distribution \citep{Peirano06}. On smaller scales, in particular for the small/incompressible shearing box \citep{Latter17}, adequate particle resolution may be possible with current computational resources and would provide an excellent check on models capable of simulating the global disc scale. The second method is to treat the dust as a continuous fluid \citep{Barriere05,Laibe12a,Laibe12b,Laibe14,Lin17,Lin19,Bi21}. In this paper we shall derive such a fluid model, starting from a stochastic differential equation for the motion of individual grains entrained in a turbulent gas flow.

The most common model of a dust fluid (in the astrophysics community) is to model it as a pressureless fluid coupled to the gas via the drag terms \citep[as has been done in][]{Barriere05,Laibe12a,Laibe12b,Laibe14,Lin17,Lin19,Bi21}. The justification for treating the dust as a pressureless fluid is that when the dust number density is much lower than that of the gas, dust-dust collisions are unimportant to the dust dynamics (although could be important for fragmentation/coagulation) which is dominated by gravity and the dust-gas interaction. As dust collisions are unimportant the dust, according to the literature, can be treated as pressureless. Unfortunately this argument for pressureless dust is flawed due to a misunderstanding about the micro-physical origin of pressure in a fluid.

The issue with this argument is that it conflates fluid pressure with collisionality. However fluid pressure is not a measure of fluid collisionality, but instead is a measure of the mean squared (density weighted) velocity dispersion of the particles. Crucially a collisionless fluid can have a non-zero velocity dispersion, and will thus have a non-zero pressure tensor. In fact weakly collisional/collisionless fluids often have large anisotropic pressure tensors  and the hydrodynamical description of the fluid breaks down, not because fluid properties such as pressure and density are not defined, but because of the difficulty in truncating the moment expansion, used to derive hydrodynamics from kinetic theory, at finite order \citep{Chapman90,Grad48,Grad49,Bobylev82,Bobylev17}. Collisions in a fluid are not the source of pressure - instead the effect of collisions is to ensure that the moment expansion truncates by damping higher order moments, along with isotropising the fluid pressure tensor \citep[e.g.][see also Boltzman's H-theorem]{Levermore96}. In conclusion, while there is a strong argument that dust in astrophysical fluids (and many geophysical fluids) can be approximated as being collisionless, we cannot conclude, \textit{a priori}, that the dust pressure is negligible. In addition to this pressure from the particle motion, in turbulent gas-dust mixture there is an additional dust Reynolds stress from the turbulent motion.



Stochastic differential equations (SDEs) have been used to model turbulent motion in fluids \citep[e.g.][]{Thomson87, Pope87,Sawford91,Minier04}. Various authors have extended such stochastic models of turbulent fluids to describe the motion of dust grains entrained in the flow \citep[e.g.][]{Dubrulle95,Minier01,Carballido06,Youdin07,Minier14,Minier15,Ormel18,Laibe20,Booth21}. \citet{Dubrulle95}, \citet{Carballido06}, \citet{Fromang06}, \citet{Ormel18}, \citet{Laibe20} and \citet{Booth21} used their models to calculate the steady state vertical structure of a dust layer in an astrophysical disc. \citet{Youdin07} calculated the dust velocity correlations in a rotating shear flow and, importantly for our work, calculated a dust fluid model by preforming a moment expansion of the Fokker-Planck equation associated with the stochastic dust motion.

In this paper we develop a dust fluid model starting from a system of stochastic differential equations (SDEs) describing the motion of a single dust grain in a turbulent gas. To do this we shall preform a moment expansion of the Fokker-Planck equation associated with the SDEs, similar to that preformed \citet{Youdin07} but without the restrictive assumption that the correlation time is the shortest timescale in the problem, and adopt a closure valid for the accretion disc context. This approach differs from the more commonly adopted method of Reynolds averaging the pressureless two-fluid model and including a closure relation motivated by the interaction of dust grains with individual turbulent eddies \citep[e.g. adopted by][]{Binkert23}. Our approach makes use of a novel 6-dimensional formulation, which keeps the dust velocity and velocity of the fluid seen, along with their moments, on the same footing. In this formulation the dust Kinetic tensor, Reynolds stress for the fluid seen and dust-gas cross correlation tensor combine into a single 6-dimensional stress tensor, which is advected by the flow. We adopt a covariant formulation of the dust fluid equation so that the model can be adapted to non-Cartesian coordinates often adopted in astrophysics problems. This will also allow for the adoption of orbital coordinates systems \citep[e.g.][]{Ogilvie13a,Ogilvie14}, which will facilitate the study of distorted (elliptical or warped) dust discs. Finally, we explore the physical properties of our dust fluid model and consider the behaviour of the dust stress tensor in a rotating shear flow. Studying the behaviour of the dust fluid in rotating shear flows allows us to connect our model to problems in astrophysical and experimental fluid dynamics (accretion discs and dusty Taylor-Couette flows respectively). This may provide a basis to experimentally test the model in the lab.


In Section \ref{single grain stocastic} we consider a SDE for motion of a single dust grain in a turbulent gas disc. In Section \ref{fluid model derivation} we derive the dust fluid equations by performing a moment expansion of the Fokker-Planck equation associated with the SDE introduced in Section \ref{single grain stocastic} and discuss our closure scheme.  Section \ref{dust fluid properties}-\ref{Steady state section} describes the physics of the model. Section \ref{dust fluid properties} discusses the dust fluid physics and highlights key properties of the model. Section \ref{inerial hyperbolic} considers the hyperbolic structure, and wave modes, of the dust fluid equations. Sections \ref{Steady state section} looks at the behaviour of the dust rheological\footnote{whenever we speak of the dust rheology or rheological stress we are referring to the rheology of the dust fluid and not the, entirely separate, rheology of the individual solid dust grains.} stress tensor in a rotating shear flows. In Section \ref{discussion} we suggest possible refinements that could be made to the model. We present our conclusions in Section \ref{conclusion} and further mathematical derivations are given in the Appendices.


\section{Overview of Astrophysical Flows}

In this section we shall briefly outline the key properties of the astrophysical fluids, which are the primary motivation for developing this model, for the benefit of non-astrophysicists. The primary flow of interest are protoplanetary discs and other dusty accretion discs, with an additional interest in dusty quasi-spherical flows present in star formation and dusty planetary atmospheres. Focusing on accretion discs - these are disc like structures of gas and solid matter in approximately Keplerian rotation about 1 (or more) central object which dominate the gravitational field. The gas in such a system has the following properties

\begin{itemize}
 \item The flow in the inertial frame, stationary with respect to the centre of mass of the system, is highly hypersonic. However, in the fluid frame it principally behaves like a subsonic shear flow in a rapidly rotating frame. 
 \item The geometry of the flow naturally lends itself to using cylindrical or spherical coordinates, both for simplifying analytical treatment and for improved angular momentum conservation, diffusivity and speed of numerical schemes.
 \item The discs are Rayleigh stable, however they can exhibit subsonic hydrodynamical or magnetohydrodynamical turbulence. Magnetohydrodynamical turbulence in discs \citep[due to the magneto-rotational instability][]{Balbus91,Hawley91,Hawley95} is much stronger than hydrodynamical turbulence (e.g. Vertical Shear Instability \citet{Nelson13,Lin15,Flock17,Svanberg22} or parametric instability \citet{Papaloizou05b,Papaloizou05a,Ogilvie13b,Barker14}). However, discs that are cool enough for the presence of dust are typically too cool to be well ionised, which tends to suppress the action of the magnetic fields. Thus turbulence in such discs is expected to be hydrodynamical and very subsonic.
 \item The disc is stratified with a pressure scale height $H \sim R/M$, where $R$ is the cylindrical distance from the central object and $M$ is the Mach number. This vertical confinement gives the disc a shallow-water like character and is also important for setting the maximum size of turbulent eddies. The rapid rotation means the eddies (inertial waves) are predominantly vertical with vertical extend $\sim$ the scale height.
 \item Characteristic timescales are the orbital period of $\sim 1$ day-$10^3$ years (depending on the position in the disc). Characteristic lengthscale are the scale height $H \lesssim 0.1 R$ and cylindrical radius $R \sim 0.1-100 \, \mathrm{Au}$ (astronomical units) $\sim 10^7-10^{10} \, \mathrm{km}$.
 \item Molecular viscosity is typically sufficient low that it can be neglected (Although the Kolmogorov scale is $\sim 10 \mathrm{m}$ \citet{Armitage20}).
\end{itemize}
The typical properties of dust in protoplanetary discs and prestellar cores are

\begin{itemize}
 \item The dust is polydispersed with size $\sim \mu \mathrm{m} - 10 \, \mathrm{cm}$ and forms a near continuous distribution in size space, however we will only consider the monodispersed case in this paper. For computational reasons most simulations of dusty accretion discs are monodispersed at present. The monodispersed case is also of observational interest as observations tend to be sensitive to a narrow range in size space dependant on the observational wavelength. 
 \item Dust to gas mass ratio is typically $\gtrsim 0.01$, with the vast majority of the mass in the largest grains \citep{Testi14}.
 \item Total number of grains $\gtrsim 1 \, \mathrm{mm}$ is $\sim 10^{32}$. The dust number density is $n \sim 10^{-9} \mathrm{cm}^{-3}$ this corresponds to $\sim 10^{27}$ particles per cubic scale height \citep{Testi14,Lesur22}.
 \item The mean free path for dust-dust collision is $\sim 10^{5} \, \mathrm{km}$, with the collision timescale being typically much longer than the stopping time.
\end{itemize}

\section{Stochastic differential equation for dust particle motion in a dust disc.} \label{single grain stocastic}




Consider a dust grain entrained in a gas flow, in the Epstein regime,  where the gas velocity at the dust grain position is denoted $\mathbf{v}^{\rm g}$. The position $\mathbf{x}$ and velocity $\mathbf{v}$ for a dust particle, subject to force per unit mass, $\mathbf{f}$, and gas drag, are given by the following set of differential equations:

\begin{align}
d x_i &= v_i d t , \\
d v_i &= f_i d t - \frac{1}{t_s} (v_i - v_i^{\rm g}) d t  ,
\label{dust particle arb turb}
\end{align}
where $t_s$ is the stopping time for the dust particle under consideration. Typically we take the force per unit mass to be due to gravity with $f_i = - \nabla_i \phi$, where $\phi$ is the gravitational potential. Here $x_i$, $v_i$, $v_i^{\rm g}$ and $f_i$ are the covariant components of the vectors $\mathbf{x}$, $\mathbf{v}$, $\mathbf{v}^{\rm g}$ and $\mathbf{f}$ respectively. These are related to the contravariant components, $x^i$, $v^i$, $v^i_{\rm g}$ and $f^i$ via the metric tensor $\gamma_{i j}$, where $x_i = \gamma_{i j} x^{j}$, $v_i = \gamma_{i j} v^{j}$, $v_i^{\rm g} = \gamma_{i j} v_{\rm g}^{j}$ and $f_i = \gamma_{i j} f^{j}$ and we have adopted the Einstein summation convention such that pairs of matching covariant, contravariant indices are implicitly summed over \citep[see e.g.][for details]{Hobson06}.

The stopping time, in the Epstein regime, for a spherical dust grain of size $s$ and grain density $\rho_{\rm grain}$ in a gas of density $\rho_{\rm g}$ is

\begin{equation}
 t_{s} = \frac{\rho_{\rm grain} s}{\rho_{\rm g} c_s} \sqrt{\frac{\pi \gamma}{8}} ,
\end{equation}
where $\gamma$ is the adiabatic index of the gas and $c_s$ is the gas sound speed \citep{Epstein24,Baines65,Whipple72}. The relative importance of gas drag is dictated by a comparison between the stopping time and some characteristic timescale of the fluid flow, $t_f$. This is encapsulated by the Stokes number $\mathrm{St} = t_s/t_f$ which is a dimensionless number which controls how strongly the gas and dust are coupled. In rotating shear flows, with angular velocity $\Omega$, it is typical to take $t_f = \Omega^{-1}$ (although in some applications it can be useful to instead set $t_f$ to be the timescale associated with the fluid shear).


A commonly used model for the stochastic gas velocity, subject to homogeneous turbulence, is to model it as a Ornstein-Uhlenbeck process,

\begin{equation}
d v_i^{\rm g} = -\frac{1}{t_c} v_i^{\rm g} d t +  \sqrt{\frac{2 \alpha}{t_c}} c_s  d W_{i} ,
\label{O-U proc}
\end{equation}
where $t_c$ is the correlation time (or ``eddy turnover'' time) of the turbulence, $c_s$ is the gas sound speed, $\alpha$ is a dimensionless measure of the strength of the fluid turbulence and $W_i$ is a Wiener process. This model of turbulence regards the turbulent flow as a member of a statistical ensemble of similar flows \citep{Thomson87}, with each ``draw'' following a fluid element in a single realisation of the flow.


As with the stopping time it is useful to introduce a dimensionless correlation time $\tau_c = t_c/t_f$. Some authors define the Stokes number to be $\mathrm{St} = t_s/t_c$, however this only really makes sense in homogeneous turbulence applications where $t_c$ is the only fluid timescale.

For more complex fluid flows, in the infinite Reynolds number limit, we can model turbulence as undergoing an Ornstein-Uhlenbeck walk about the mean flow. In this model the gas velocity evolves according to

\begin{equation}
d v_i^{\rm g} = f_i^{\rm g} d t - \frac{1}{t_c} (v_i^{\rm g} - u_i^{\rm g}) d t + \sqrt{\frac{2 \alpha}{t_c}} c_s d W_i , \label{gas velocity seen}
\end{equation}
where $f_i^{\rm g}$ is the force per unit mass on the gas and $u_i^{\rm g} = \mathbb{E}_g (v_i^{\rm g})$ is the mean gas velocity at the dust location. This mean gas velocity needs to be solved for separately, for which we use the Equations \ref{continuity total form}-\ref{constituative total form} in Appendix \ref{gas model appendix} In the absence of back reaction the force per unit mass on the gas is due to gravity and pressure gradients with $f_i^{\rm g} = -\nabla_i \phi - \rho_g^{-1} \nabla_i p_g$ where $p_g$ is the gas pressure and $\rho_g$ is the gas density. With this choice of $f_i^{\rm g}$, Equation \ref{gas velocity seen} amounts to modelling the pressure fluctuation and dissipation terms as being responsible for the Ornstein-Uhlenbeck terms present above \citep{Pope00}. $f_i^g$, $\alpha$, $t_c$, $c_s$ and $\mathbf{u}^g$ are all functions of space and, in general, time. For instance, in accretion discs, $t_c$ is typically proportional to the orbital period and is thus an increasing function of cylindrical radius. Likewise, the soundspeed and $\alpha$ vary (typically slowly) throughout the disc, although $\alpha$ is often assumed to be constant. All these quantities must be evaluated at the dust particle position. In principle one may be able to incorporate the effects of back reaction into $f_i^g$, we shall give a brief discussion of this possibility in Section \ref{discussion}.







Combining the model for the gas and dust, we arrive at a system of stochastic differential equations describing the motion of a dust grain in a turbulent gas,

\begin{align}
d x_i &= v_i d t , \\
\label{dust particle O-U x}
d v_i &= f_i d t - \frac{1}{t_s} (v_i - v_i^{\rm g}) d t  , \\
D_d v_i^{\rm g} &= f_i^{\rm g} d t - \frac{1}{t_c} (v_i^{\rm g} - u_i^{\rm g}) d t + \sqrt{\frac{2 \alpha}{t_c}} c_s d W_i  .
\label{dust particle O-U}
\end{align}
Now one can regards each `draw' as selecting, and following, a single dust grain entrained with the turbulent flow. The gas fluid elements do not, in general, follow the dust grains, so we must correct for the fact we are taking a sample of the gas along the trajectory of the dust. Following \citet{Minier04,Minier14} we take the operator $D_d$ to be

\begin{equation}
D_d v_i^{\rm g} = d v_i^{\rm g} - (u^k - u^{k}_{g}) \nabla_{k} u_{i}^{\rm g} d t .
\end{equation}
This can be thought of as a separate `advection' step which, on average, corrects for the difference in the gas and dust trajectories. One can more compactly write these equations in terms of the dynamics of a particle in 6-dimensions, subject to drag, stochastic forcing and force per unit mass $F_{\alpha}$ (which contains contributions from the force on the dust and gas $f_i$, $f_i^{\rm g}$, along with the shift correction, $(u^k - u^{k}_{g}) \nabla_{k} u_{i}^{\rm g}$),





\begin{align}
\begin{split}
d X_{\alpha} &= V_{\alpha} d t , \\
d V_{\alpha} &= F_{\alpha} d t - C_{\alpha \beta} \left(V^{\beta} - U_g^{\beta} \right) d t + \sigma_{\alpha \beta} d W^{\beta} ,
\end{split}
\label{6-d O-U proc}
\end{align}
where we have adopted the convention that Greek indices are over the 6-dimensional space and Latin indices are taken over the 3-dimensional space. These 6-dimensional indices are raised and lowered with a 6-dimensional metric tensor $g_{\alpha \beta}$, constructed from $\gamma_{i j}$, which will be properly defined in the next section. We have introduced $U_g^{\beta}$, the mean gas velocity ``seen'' by the dust; the $6 \times 6$ drag tensor $C_{\alpha \beta}$, which incorporate both the gas-dust drag on the stopping time along with the return of the stochastic gas velocity towards the mean on the turbulent correlation time which in the 6D picture acts like a ``drag'' between the gas components of the velocity and the mean gas flow. We have also introduced $\sigma_{\alpha \beta}$ which controls the strength of the stochastic forcing in each component of the momentum equation - i.e. it's the 6D form of the last term in equation \ref{dust particle O-U}. In addition to simplifying the subsequent derivations, Equation \ref{6-d O-U proc} allows us to derive the fluid model for more general drag and turbulence models without increasing the complexity. For instance the subsequent derivations works equally well for anisotropic stochastic driving. 

One can also include anisotropic correlation times as seen in some two-phase turbulence models \citep[e.g.][]{Minier04,Minier14}, based on the analysis of \citet{Csanady63}, which attempts to incorporate the effects of spatial correlation on the fluid seen by the dust particles. We have chosen not to include this correction as the proposed form of the correction in the literature \citep[as described in][]{Minier04,Minier14} predicts that rapidly drifting particle in rotating shear flows experience the same turbulence as that in homogeneous-isotropic turbulence. This likely arises due to the Csanady correction neglecting the anisotropy in the correlation length induced by the shear. It is possible that the two-step stochastic model \citep[as discussed in][]{Minier23} will better account for the effects of spatial correlations and improve the modelling of dusty anisotropic turbulence in the future.

\subsection{Geometry of the 6-dimensional space}


The 3 additional dimensions in the 6-D system are a set of dummy gas degrees of freedom corresponding to the gas displacement. These should not be thought of as the gas position vector as the gas is coincident with the dust. These additional dimensions are, in a sense, non-physical and, in order that the 6D system agree with the 3D system, the 6D system must posses translational invariance along these dummy directions. The coordinate basis of the gas displacement are independent of the basis of the dust position vector. However it is useful to choose the basis of the gas displacement dimensions such that it reflects the underlying (physical) 3D coordinate system.



To construct this coordinate system we first consider the coordinates of the underlying 3D system with metric tensor $\gamma_{i j}$ and associated Christoffel symbols $\mathcal{T}_{i j}^k$. Introducing basis vectors for the 6-D system, $\{ \hat{e}_{\alpha} \}$, and the notation $\alpha_d  \in \{ 1, 2, 3\}$ and $\alpha_{\rm g} \in \{4, 5, 6\}$ such that $\hat{e}_{\alpha_d}$ give the basis vectors of the dust position vector and $\hat{e}_{\alpha_g}$ gives the basis vectors of the gas displacement vector. Additionally it is useful to introduce the bijection $\cdot^{*} : \{1...6\} \rightarrow \{1...6\}$, which interchanges the `dummy gas' and position indices with $1,2,3 \mapsto 4,5,6$ and $4,5,6 \mapsto 1,2,3$. 

Throughout this work we shall make use of symmetrising/anti-symmetrising operations on the tensor indices with $E_{(\alpha_1 \cdots \alpha_n)}$ and $E_{[\alpha_1 \cdots \alpha_n]}$, for some tensor $\mathbf{E}$, being symmetrisation and anti-symmetrisation of the indices in brackets, where $E_{(\alpha \beta)} = \frac{1}{2} (E_{\alpha \beta} +  E_{\beta \alpha})$ and $E_{[\alpha \beta]} = \frac{1}{2} (E_{\alpha \beta} -  E_{\beta \alpha})$. The operation $*$ does not commute with symmetrisation/anti-symmetrisation operations, but instead follows the obvious order of operations such that

\begin{align}
 E_{(\alpha, \beta^{*})} &= \frac{1}{2} ( E_{\alpha \beta^{*}} +  E_{\beta^{*} \alpha}) , \\
 E_{(\alpha, \beta)^{*}} &= \frac{1}{2} ( E_{\alpha \beta^{*}} +  E_{\beta \alpha^{*}}) ,
\end{align}
with equivalent expressions for antisymmetrisation.





The physical solutions must be independent of the gas displacement, we can therefore integrate out the dummy gas dimensions. Introducing an integral over the dummy gas directions,

\begin{equation}
\overline{\cdot} := \int \cdot J_{g} d^3 x_g ,
\end{equation}
where $J_g$ is the Jacobian determinant of the dummy gas coordinates. Thus for $\mathbf{E}$, an arbitrary tensoral quantity, we have

\begin{equation}
\overline{\nabla_{\alpha_g} \mathbf{E}} = 0 .
\label{turbulent aved deriv}
\end{equation}

For Cartesian gas displacement coordinates this integrating out of the non-physical space is straight-forward. Unfortunately if the coordinate system describing the dust position is non-Cartesian then we need to rotate the ``dummy'' components of vectors so that they reflect the underlying 3D coordinate system (e.g. when calculating the gas drag). It is instead useful to setup the geometry of our 6D space so that the rotation happens automatically. To do this we first introduce the metric tensor of the 6D coordinate system

\begin{equation}
g_{\alpha \beta} = \begin{cases}
 \gamma_{\alpha \beta} , & \alpha, \beta \in \{1,2,3\} , \\
\gamma_{\alpha^{*} \beta^{*}} , & \alpha, \beta \in \{4,5,6\} , \\
0 , & \mathrm{otherwise} . \\
\end{cases}
\label{6D metric}
\end{equation}
We also introduce a metric connection $\overline{\nabla}_{\alpha}$ that is responsible for rotating the dummy gas coordinate system. We require that this connection satisfy the following properties:


\,
\begin{enumerate}
 \item $\overline{\nabla}_{\alpha}$ is a metric connection, so that $\overline{\nabla}_{\alpha} g_{\beta \gamma} = \overline{\nabla}_{\alpha} g^{\beta \gamma} = 0$.
 \item Translational invariance with respect to the gas displacement such that $\overline{\nabla}_{\alpha_g} \mathbf{E} (\mathbf{x}_d) = 0$ for tensoral quantity $\mathbf{E}$.
 \item Alignment of the dummy gas coordinates with the position coordinates. For vectors $\mathbf{A}$, $\mathbf{B}$ and $\tilde{\mathbf{B}}$ with $B^{\alpha_d} = 0$ and $\tilde{B}^{\alpha} = B^{\alpha^{*}}$ then we require $(\mathbf{A} \cdot \overline{\nabla} \mathbf{B})^{\beta} = (\mathbf{A} \cdot \overline{\nabla} \tilde{\mathbf{B}})^{\beta^*} $ .
\end{enumerate}
\,

Property (ii) ensures that $\overline{\nabla}_{\alpha_g} \mathbf{E} (\mathbf{x}_d) = \overline{\nabla_{\alpha_g} \mathbf{E}}$, where $\nabla_i$ is the covariant derivative, and allows us to carry out the integral over the dummy gas directions by replacing covariant derivatives with $\overline{\nabla}_i$. Property (iii) is required to ensure that the geometric terms in Lagrangian time derivatives act the same on the dust and gas components of the 6D vectors. This can be seen considering $\mathbf{A} = \mathbf{U}$ and considering the action of the Lagrangian time derivative, $D = \partial_{t} + \mathbf{U} \cdot \overline{\nabla}$, on the vectors $\mathbf{B}$ and $\tilde{\mathbf{B}}$. As $B^{\alpha} = \tilde{B}^{\alpha^{*}}$ one requires $(D \mathbf{B})^{\alpha} = (D \tilde{\mathbf{B}})^{\alpha^{*}}$ this requires condition (iii) as $\mathbf{U}$ is arbitrary and $(\partial_t \mathbf{B})^{\alpha} = (\partial_t \tilde{\mathbf{B}})^{\alpha^{*}}$.


The connection which satisfies these properties, given the metric tensor \eqref{6D metric}, acts on the basis vectors $\{ \hat{e}_{\alpha} \}$ as follows,

\begin{equation}
 \overline{\nabla}_{\alpha_d} \hat{e}_{\beta_d} = \mathcal{T}_{\alpha_d \beta_d}^{\gamma_d} \hat{e}_{\gamma_d} ,  \quad \overline{\nabla}_{\alpha_d} \hat{e}_{\beta_g} = \mathcal{T}_{\alpha_d \beta_g^{*}}^{\gamma_g^{*}} \hat{e}_{\gamma_g} ,
\end{equation}
with $\overline{\nabla}_{\alpha_g} \hat{e}_{\beta} = 0$. As $\mathcal{T}_{i j}^{k}$ are the Christoffel symbols components for the 3D coordinate system associated with the metric $\gamma_{i j}$ it is straight-forward to show that this connection satisfies property 1. Property 2 follows from $\overline{\nabla}_{\alpha_g} \hat{e}_{\beta} = 0$. Finally for property 3,


\begin{align}
 (\mathbf{A} \cdot \overline{\nabla} \mathbf{B})^{\beta} = A^{\alpha} \overline{\nabla}_{\alpha} B^{\beta} &= A^{\alpha} \partial_{\alpha} B^{\beta}  + A^{\alpha} \mathcal{T}_{\alpha \gamma}^{\beta} B^{\gamma} ,\\
\begin{split}
 (\mathbf{A} \cdot \overline{\nabla} \tilde{\mathbf{B}})^{\beta} = A^{\alpha} \overline{\nabla}_{\alpha} \tilde{B}^{\beta} &= A^{\alpha} \partial_{\alpha} \tilde{B}^{\beta}  + A^{\alpha} \mathcal{T}_{\alpha \gamma^{*}}^{\beta^{*}} \tilde{B}^{\gamma} \\
&= A^{\alpha} \partial_{\alpha} B^{\beta^{*}}  + A^{\alpha} \mathcal{T}_{\alpha \gamma^{*}}^{\beta^{*}} B^{\gamma^{*}} =  
 (\mathbf{A} \cdot \overline{\nabla} \mathbf{B})^{\beta*} .
\end{split}
\end{align}
While this connection has the advantage of keeping the gas coordinate system aligned and avoids the necessity of including rotation matrices in the equation of motion it does have one major drawback in that it is not torsion free (since it isn't the Levi-Civita connection). This torsion arises when $\bar{\nabla}_{\alpha_d} \hat{e}_{\beta_g} \ne 0$ as $\overline{\nabla}_{\alpha_t} \hat{e}_{\beta_d} = 0$, by construction, and is associated with the rotation of the dummy gas coordinate system. The torsion tensor, $S_{\alpha \beta}^{\gamma}$, is given by

\begin{equation}
 S_{\alpha \beta}^{\gamma} \hat{e}_{\gamma} = \overline{\nabla}_{\alpha} \hat{e}_{\beta} - \overline{\nabla}_{\beta} \hat{e}_{\alpha} ,
\end{equation}
making use of the properties of the connection the torsion tensor components are 

\begin{equation}
 S_{\alpha_d \beta_g}^{\gamma_g}  = - S_{\beta_g \alpha_d}^{\gamma_g} = \mathcal{T}_{\alpha_d \beta_g^{*}}^{\gamma_g^{*}} ,
\end{equation}
with all other components zero.

Finally, after specialising to the oriented 6-D geometry one can write $U_g^{\alpha}$ in terms of the mean gas velocity in the gas frame, $u_g^{i}$,

\begin{equation}
 U_g^{\alpha} = \begin{cases}
u_g^{\alpha} \, , \quad \alpha \in \{1, 2, 3 \} , \\
u_g^{\alpha^{*}} \, , \quad \alpha \in \{4, 5, 6 \}  ,
\end{cases} 
\end{equation}
while the drag and diffusion tensors can be written in terms of the metric tensor. The 6-D force per unit mass is

\begin{equation}
 F_{\alpha} = \begin{cases}
f_{\alpha} \, , \quad \beta \in \{1, 2, 3 \} , \\
f^g_{\alpha^{*}} + (U^{\beta} - U^{\beta}_{g}) \nabla_{\beta} U_{\alpha}^{\rm g} \, , \quad \alpha \in \{4, 5, 6 \}  ,
\end{cases} 
\end{equation}
while the drag tensor is

\begin{equation}
C_{\alpha \beta} = \begin{cases}
\frac{1}{t_s} g_{\alpha \beta} , & \alpha, \beta \in \{1,2,3\} ,  \\
-\frac{1}{t_s} g_{\alpha \beta^{*}} , & \alpha \in \{1,2,3\} , \quad  \beta \in \{4,5,6 \} , \\
0 , & \alpha \in \{4,5,6\} , \quad \beta \in \{1,2,3 \} , \\
\frac{1}{t_c} g_{\alpha \beta} , & \alpha, \beta \in \{4,5,6\} , 
\end{cases}
\label{C OU turb}
\end{equation}
while the diffusion tensor is 

\begin{equation}
D_{\alpha \beta} = \begin{cases}
\frac{\alpha c_s^2}{t_c} g_{\alpha \beta} , & \alpha, \beta \in \{4,5,6\} , \\
 0 , & \mathrm{otherwise} . \\
\end{cases} 
\label{D OU turb}
\end{equation}
This diffusion tensor is applicable to isotropic diffusivity. More generally one can include an anisotropic diffusivity by introducing an $\alpha$-tensor $a_{\alpha \beta}$, in which case the diffusion tensor will be

\begin{equation}
D_{\alpha \beta} = \begin{cases}
 \frac{ c_s^2}{t_c} a_{\alpha \beta} , & \alpha, \beta \in \{4,5,6\} , \\
 0 , & \mathrm{otherwise} . \\
\end{cases} .
\end{equation}
If one were to instead use the more usual Levi-Civita connection the above expressions would be considerably more complex as they would need to include the rotation of the dummy gas directions.

\section{Derivation of the dust-fluid model.} \label{fluid model derivation}

\subsection{Derivation of the Fokker-Planck equation}

In order to derive the dust fluid model we must first obtain the Fokker-Planck equation associate with Equation \ref{6-d O-U proc}, and then perform a moment expansion to derive the fluid model. To do this consider an arbitrary ($C^2$) function of the dust particle position, velocity and stochastic gas displacement,  $A = A(\mathbf{X},\mathbf{V})$. By use of Ito's lemma this evolves according to

\begin{equation}
d A = \frac{\partial A}{\partial X_\alpha} d X_\alpha + \frac{\partial A}{\partial V_{\alpha}} d V_{\alpha} + \frac{1}{2} \frac{\partial^2 A}{\partial V_\alpha \partial V_{\beta} } \langle d V_{\alpha}, d V_{\beta} \rangle  ,
 \label{d A inital eq}
\end{equation}
where the angle bracket $\langle \cdot, \cdot \rangle$ denotes the covariance. Where the covariance of a Wiener process $d W_{\alpha}$ is given by

\begin{equation}
 \langle d W^{\alpha}, d W^{\beta} \rangle = g^{\alpha \beta} \, d t .
\end{equation}
This leads to the following for the covariance of the velocity,

\begin{equation}
 \langle d V_{\alpha}, d V_{\beta} \rangle = \sigma_{\alpha \mu}  \sigma_{\beta \nu} \langle d W^{\mu} d W^{\nu} \rangle = 2 D_{\alpha \beta} d t ,
\end{equation}
where we have introduced the diffusion tensor, $D_{\alpha \beta} = \frac{1}{2} g^{\mu \nu} \sigma_{\alpha \mu} \sigma_{\beta \nu}$. Substituting Equation \ref{6-d O-U proc} into Equation \ref{d A inital eq}, we arrive at

\begin{align}
\begin{split}
d A =& \frac{\partial A}{\partial X_\alpha} V_{\alpha} d t + \frac{\partial A}{\partial V_{\alpha}} \left[F_{\alpha} - C_{\alpha \beta} \left(U^{\beta} - U_g^{\beta} \right) \right] d t \\
&+ \frac{\partial A}{\partial V_{\alpha}}  \sigma_{\alpha \beta} d W^{\beta} + D_{\alpha \beta} \frac{\partial^2 A}{\partial V_\alpha  \partial V_{\beta} } d t .
\end{split}
\label{d A}
\end{align}
The expectation of $A$ is given by,

\begin{equation}
\mathbb{E} [A] = \int p^{L} (\mathbf{X},\mathbf{V},t,\mathbf{X}_0,\mathbf{V}_0,t_0) A(\mathbf{X},\mathbf{V}) d^6 \mathbf{X} d^6 \mathbf{V} ,
\end{equation}
where $p^{L} (\mathbf{X},\mathbf{V},t,\mathbf{X}_0,\mathbf{V}_0,t_0)$ is the probability for the system to arrive at state $(\mathbf{X},\mathbf{V},t)$ from an initial state $(\mathbf{X}_0,\mathbf{V}_0,t_0)$. $\mathbb{E} [d A]$ is given by,

\begin{equation}
 \mathbb{E} [d A] =  \int d p^{L} (\mathbf{X},\mathbf{V},t,\mathbf{X}_0,\mathbf{V}_0,t_0) A(\mathbf{X},\mathbf{V}) d^6 \mathbf{X} d^6 \mathbf{V} .
\end{equation}
Substituting Equation \ref{d A} into the above and, after appropriate integration by parts (assuming appropriate regularity conditions for $p$, namely that $p$ and $\frac{\partial p}{\partial V_{\alpha}}$ vanish as $V^{\beta} \rightarrow \infty$) we arrive at


\begin{align}
\begin{split}
\int A \Biggl\{d p^{L} &+ \frac{\partial}{\partial X_{\alpha}} (p^{L} V_{\alpha}) d t + \frac{\partial}{\partial V_{\alpha}} \left[ \left( F_{\alpha} - C_{\alpha \beta} \left(U^{\beta} - U_g^{\beta} \right) \right) p^{L}\right] d t \\
& - D_{\alpha \beta}\frac{\partial^2 p^{L}}{\partial V_{\alpha} V_{\beta} } d t  \Biggr\} d^6 \mathbf{X} d^6 \mathbf{V} ,
\end{split}
\end{align}
provided that $\int_{\partial} p^{L} A \mathbf{V} \cdot d \mathbf{S} d^6 \mathbf{V} = 0$, where $\int_{\partial} d \mathbf{S}$ denotes an integral over the spatial boundaries. i.e. the expected net flux of $A$ through the domain boundaries is zero. 

As $A$ is arbitrary (baring being $C^{2}$, and the boundary conditions) we arrive at the Fokker-Planck equation for $p$,

\begin{equation}
\frac{\partial p^{L}}{\partial t} + \frac{\partial}{\partial X^{\alpha}} (p^{L} V^{\alpha}) + \frac{\partial}{\partial V_{\alpha}} \left[ \left(F_{\alpha} -  C_{\alpha \beta} \left(V^{\beta} - U_g^{\beta} \right)\right) p^{L} \right]  = D_{\alpha \beta}\frac{\partial^2 p^{L}}{\partial V_{\alpha} V_{\beta} } .
\label{Fokker-Planck Equation pl}
\end{equation}
This equation gives an evolutionary equation for the Lagrangian transition PDF, describing the probability of finding a particle at $\mathbf{X}$, $\mathbf{V}$ at time $t$ conditional on it being located at $\mathbf{X}_0, \mathbf{V}_0$ at time $t_0$. The fluid model will consist of a set of Eulerian fields located at a given position in Space and must be obtained from the Eulerian Mass Density Function (MDF), $p (\mathbf{X}, \mathbf{V}, t)$, which is the expected mass density of particles at $\mathbf{X}$, $\mathbf{V}$ at time $t$ \citep{Pope85,Pope00,Minier01b}. This will contain contributions from particle s with different initial condition $(\mathbf{X}_0, \mathbf{V}_0)$, arriving from differing trajectories. This can be obtained from the Eulerian mass density function at $t_0$, $p (\mathbf{X}_0, \mathbf{V}_0, t_0)$, by using the transition PDF and integrating over the initial positions and velocities \citep{Pope85,Pope00,Minier01b},

\begin{equation}
 p (\mathbf{X}, \mathbf{V}, t) = \int p^{L} (\mathbf{X},\mathbf{V},t,\mathbf{X}_0,\mathbf{V}_0,t_0) p (\mathbf{X}_0, \mathbf{V}_0, t_0) \, d^6 \mathbf{X}_0 d^6 \mathbf{V} .
\end{equation}
We can obtain the Fokker-Planck equation for $p$ by multiplying Equation \ref{Fokker-Planck Equation pl} by $p (\mathbf{X}_0, \mathbf{V}_0, t_0)$ and integrating over the initial position and velocities. This leaves the form of the Fokker-Planck Equation unchanged and we obtain

\begin{equation}
\frac{\partial p}{\partial t} + \frac{\partial}{\partial X^{\alpha}} (p V^{\alpha}) + \frac{\partial}{\partial V_{\alpha}} \left[ \left(F_{\alpha} -  C_{\alpha \beta} \left(V^{\beta} - U_g^{\beta} \right)\right) p \right]  = D_{\alpha \beta}\frac{\partial^2 p}{\partial V_{\alpha} V_{\beta} } .
\label{Fokker-Planck Equation}
\end{equation}

\subsection{Moment expansion of the Fokker-Planck equation}

Fluid dynamical models can be derived from the Fokker-Planck equation via a moment expansion, in a similar manor to that done in kinetic theory. In performing this moment expansion we wish to arrive at a a set of PDEs in space and time from the initial PDE in $(t,\mathbf{X},\mathbf{V})$. This means we need to compute a moment expansion in $\mathbf{V}$.  A similar procedure was carried out by \citet{Youdin07}. Defining the velocity moments of $p$ as follows,

\begin{align}
\rho_6 &:= \int p d^6 \mathbf{V}  , \\
\rho_6 U_{\alpha} &:= \int p V_{\alpha} d^6 \mathbf{V} , \\
\Pi_{\alpha \beta} &:= \int p (V_{\alpha} - U_{\alpha}) (V_{\beta} - U_{\beta}) d^6 \mathbf{V}  , \\
\Pi_{\alpha_1 \cdots \alpha_k} &:= \int p (V_{\alpha} - U_{\alpha}) \cdots (V_{\alpha_k} - U_{\alpha_k}) d^6 \mathbf{V} .
\end{align}
Note that this moment expansion is in the 6-dimensional space, so that $\rho_6$ is the 6-dimensional mass density and $\Pi_{\alpha \beta}$ is the 6-dimensional rheological stress tensor\footnote{We have chosen to call the second velocity moment the rheological stress tensor rather than the dust pressure tensor as it contains contributions from both the dust pressure (particle velocity dispersion) and the dust Reynolds stress. These two stresses are indistinguishable due to the way we have formulated the averaging. This can run into issues when dust-dust collisions are included as the dust collisional velocity is principally sensitive to the particle (rather than turbulent) velocity dispersion \citep{Fox14,Capecelatro16b}.} and we have chosen a normalisation such that

\begin{equation}
 \int p d^6 \mathbf{V} d^3 \mathbf{x}_{\rm g} = \rho_d ,
\end{equation}
where $\rho_d$ is the dust density (i.e. the density of the dust phase, this is equal to the grain density, $\rho_{\rm grain}$, multiplied by the dust volume fraction). We have opted to normalise with respect to the dust mass density rather than the dust number density so that $\int \Pi_{\alpha \beta} d^3 \mathbf{x}_{\rm g}$ has the same units as the gas pressure.


Taking the zeroth velocity moment of Equation \ref{Fokker-Planck Equation} we arrive at the (6-D) dust continuity equation,

\begin{equation}
\dot{\rho}_6 + \nabla_{\alpha} \left[ \rho_6 U^{\alpha} \right] = 0 .
\end{equation}
The first $\mathbf{V}$ moment of  Equation \ref{Fokker-Planck Equation} leads to the (6-D) dust momentum equation,

\begin{equation}
\frac{\partial}{\partial t} [\rho_6 U_{\alpha}] + \nabla^{\beta} [\Pi_{\alpha \beta} + \rho_6 U_{\alpha} U_{\beta}] - \rho_6 F_{\alpha} + \rho_6 C_{\alpha \beta} (U^{\beta} - U_g^{\beta}) = 0 .
\end{equation}
Taking the second $\mathbf{V}$  moment yields a constitutive relation for the (6-D) dust stress tensor,
\begin{align}
 \begin{split}
 \frac{\partial}{\partial t} & [\Pi_{\alpha \beta} + \rho_6 U_{\alpha} U_{\beta}] + \nabla^{\gamma} \left[\Pi_{\alpha \beta \gamma} + 3 U_{(\alpha} \Pi_{\beta \gamma)} + \rho_6 U_{\alpha} U_{\beta} U_{\gamma} \right] \\
&- 2 \rho_6 U_{(\alpha} \left[F_{\beta)} - C_{\beta) \gamma} \left(U^{\gamma} - U_g^{\gamma} \right) \right] + 2 \Pi^{\gamma}_{\;\; (\alpha} C^{\;}_{\beta) \gamma}= 2 \rho_6 D_{(\alpha \beta)} .
 \end{split}
\end{align}
Here we have made use of the notation for the symmetrisation of the tensor indices. As we shall make extensive use of this notation we give explicit expressions for the symmetrised terms in the above equation as a illustrative example, $U_{(\alpha} \Pi_{\beta \gamma)} = \frac{1}{3} (U_{\alpha} \Pi_{\beta \gamma} + U_{\beta} \Pi_{\alpha \gamma} + U_{\gamma} \Pi_{\alpha \beta} )$ and $2 U_{(\alpha} \left[F_{\beta)} - C_{\beta) \gamma} \left(U^{\gamma} - U_g^{\gamma} \right) \right] = U_{\alpha} \left[F_{\beta} - C_{\beta \gamma} \left(U^{\gamma} - U_g^{\gamma} \right) \right] + U_{\beta} \left[F_{\alpha} - C_{\alpha \gamma} \left(U^{\gamma} - U_g^{\gamma} \right) \right]$.

Higher velocity moments can be computed in a similar manor. Making use of the expressions for the velocity moments of the terms of the Fokker-Planck equation given in Appendix \ref{Higher moment F-P terms}, we can take the k-th velocity moment of the Fokker-Planck equation to obtain

\begin{align}
\begin{split}
 \frac{\partial \Pi_{\alpha_1 \cdots \alpha_k}}{\partial t} &+ k \Pi^{\;}_{(\alpha_1 \cdots \alpha_{k-1}} \left[ D U^{\;}_{\alpha_k}  + \nabla^{\;}_{\alpha_k )} \phi + C_{\alpha_k)}^{\;\; \gamma} (U_{\gamma} - U_{\gamma}^{g}) \right] +  \nabla^{\sigma} \left[ \Pi_{\alpha_1 \cdots \alpha_{k} \sigma} + U_{\sigma} \Pi_{\alpha_1 \cdots \alpha_k} \right]  \\
&+ k \Pi^{\;}_{\sigma (\alpha_1 \cdots \alpha_{k-1}} \left[\nabla^{\sigma} U^{\;}_{\alpha_k)} +  C_{\alpha_k)}^{\;\; \sigma} \right] = k (k - 1) \Pi_{(\alpha_1 \cdots \alpha_{k-2}} D_{\alpha_{k-1} \alpha_{k})} .
\end{split}
\end{align}
Making use of the dust momentum equation this simplifies to

\begin{align}
\begin{split}
 (D + \nabla_{\sigma} U^{\sigma}) &\Pi_{\alpha_1 \cdots \alpha_{k}} + k \Pi_{\sigma ( \alpha_1 \cdots \alpha_{k-1}} \nabla^{\sigma}U_{\alpha_k)} +\nabla^{\sigma} \Pi_{\alpha_1 \cdots \alpha_k \sigma}  \\
&= - k \left[ \Pi^{\sigma}_{\;\; (\alpha_1 \cdots \alpha_{k-1}} C^{\;}_{\alpha_k) \sigma} - \frac{1}{\rho} \Pi_{(\alpha_1 \cdots \alpha_{k-1}} \nabla^{\sigma} \Pi_{\alpha_{k } ) \sigma} - (k - 1) \Pi_{(\alpha_1 \cdots \alpha_{k-2}} D_{\alpha_{k-1} \alpha_{k})}\right]  .
\end{split}
\end{align}
Taking $k=2$ in the above equation we recover the constitutive relation for $\Pi_{\alpha \beta}$ (to obtain this we note that $\Pi_{\alpha} = 0$ by the definition of $U^{\alpha}$).

It is useful to define various tensor advection operators $\mathcal{D}$, $\mathcal{D}_1$ and $\mathcal{D}_2$. When acting on the k-th velocity moment these are given by

\begin{align}
 \mathcal{D} \Pi_{\alpha_1 \cdots \alpha_k} &= D \Pi_{\alpha_1 \cdots \alpha_k} +k \Pi_{\gamma (\alpha_1 \cdots \alpha_{k - 1}} \nabla_{\alpha_k)} U^{\gamma} + \Pi_{\alpha_1 \cdots \alpha_k} \nabla_{\gamma} U^{\gamma} , \\
 \mathcal{D}_1 \Pi_{\alpha_1 \cdots \alpha_k} &= D \Pi_{\alpha_1 \cdots \alpha_k} + \Pi_{\alpha_1 \cdots \alpha_k} \nabla_{\gamma} U^{\gamma} , \\
 \mathcal{D}_2 \Pi_{\alpha_1 \cdots \alpha_k} &= D \Pi_{\alpha_1 \cdots \alpha_k} + k \Pi_{\gamma (\alpha_1 \cdots \alpha_{k - 1}} \nabla^{\gamma} U_{\alpha_k)} + \Pi_{\alpha_1 \cdots \alpha_k} \nabla_{\gamma} U^{\gamma} .
\end{align}
The first of these is closely related to the convective Maxwell derivative, with $\mathcal{D} \Pi_{\alpha_1 \cdots \alpha_k = 0}$ implying that the the tensoral quantity $\rho_6^{-1} \Pi_{\alpha_1 \cdots \alpha_k = 0}$ (i.e. the k-th velocity correlation) is passively advective by the flow. The other operators $\mathcal{D}_1$ and $\mathcal{D}_2$ are defined for convenience. This highlights one advantage of the 6-d formalisation as couplings between the dust kinetic tensor ($T_{\alpha_d \beta_d}$), cross correlation tensor ($T_{\alpha_d \beta_g}$) and fluid seen Reynolds stress ($R_{\alpha_g \beta_g}$), are shown to arise from the advection of the dust Rheological stress by the 6D flow.


Rearranging the continuity, momentum and constitutive equations, and making use of the operator $\mathcal{D}_2$, we obtain

\begin{align}
D \rho_6 &= - \rho_6 \nabla_{\alpha} u^{\alpha} , \\
\rho_6 D U_{\alpha} &= \rho_6 F_{\alpha} - \nabla^{\beta} \Pi_{\alpha \beta} - \rho_6 C_{\alpha \beta} (U^{\beta} - U_g^{\beta}) , \\
\mathcal{D}_2 \Pi_{\alpha \beta} &= -\nabla^{\gamma} \Pi_{\alpha \beta \gamma}  -2 \left( \Pi^{\gamma}_{\;\; (\alpha} C^{\;}_{\beta) \gamma} - \rho_6 D_{(\alpha \beta)} \right) . \label{constituative pre integration}
\end{align}
 As the right hand side of Equation \ref{constituative pre integration} is symmetrised this ensures that $\Pi_{\alpha \beta}$ remains symmetric for symmetric initial conditions. Using a similar argument to that advanced in \citet{Ogilvie03,Lynch21}, $\Pi_{\alpha \beta}$ is positive semi-definite for positive semi-definite initial conditions (See Appendix \ref{realisability of P} for a details). The evolutionary equation for the k-th velocity moment simplifies to

\begin{equation}
 \mathcal{D}_2 \Pi_{\alpha_1 \cdots \alpha_k} = -\nabla^{\gamma} \Pi_{\gamma \alpha_1 \cdots \alpha_k} - k \left[ \Pi^{\sigma}_{\;\; (\alpha_1 \cdots \alpha_{k-1}} C^{\;}_{\alpha_k) \sigma } - \frac{1}{\rho} \Pi_{(\alpha_1 \cdots \alpha_{k-1}} \nabla^{\sigma} \Pi_{\alpha_{k } ) \sigma} - (k - 1) \Pi_{(\alpha_1 \cdots \alpha_{k-2}} D_{\alpha_{k-1} \alpha_{k})} \right] .
\end{equation}



Alternatively one can write the constitutive equation in terms of the operator $\mathcal{D}$ and obtain the following alternative form of Equation \ref{constituative pre integration},

\begin{align}
\mathcal{D} \Pi_{\alpha \beta}  = -2 \left( \Pi^{\gamma}_{\;\; (\alpha} A^{\;}_{\beta) \gamma} - \rho_6 D_{(\alpha \beta)} \right) , \label{constituative alternative}
\end{align}
where we have defined 

\begin{equation}
A_{\alpha \beta} = C_{\alpha \beta} - \omega^{\gamma} \varepsilon_{\alpha \beta \gamma},
\end{equation}
where $\omega^{\gamma}$ is the dust-fluid vorticity and 

\begin{equation}
\omega^{\gamma} \varepsilon_{\gamma \alpha \beta}  = \nabla_{\alpha} U_{\beta} - \nabla_{\beta} U_{\alpha} .
\end{equation}
The evolutionary equation for the k-th velocity moment can be similarly rewritten. In the full 6-D model, with the Levi-Civita connection, Equation \ref{constituative alternative} is the more useful form of the constitutive relation as it is independent of the Christoffel symbol components (by symmetry) and it is more connected to the underlying physics of the rheological stress tensor where the operator $\mathcal{D}$ is responsible for passively advecting the pressure tensor and the drag, vorticity and turbulent ``heating'' on the right hand side of Equation \ref{constituative alternative} act like sources/sinks for the stress tensor. Unfortunately in the presence of torsion, the constitutive equation based on Equation \ref{constituative alternative} ends up more complicated to manipulate than that based on Equation \ref{constituative pre integration} owing to the addition of terms involving the torsion tensor. As such we shall stick to Equation \ref{constituative pre integration} for the constitutive relation from this point onwards.

Finally, for the purposes of developing the closure scheme for the moment expansion, it is useful to express the evolutionary equation for the k-th velocity moment in terms of the operator $\mathcal{D}_1$, 

\begin{align}
\begin{split}
 \mathcal{D}_1 \Pi_{\alpha_1 \cdots \alpha_{k}} &= -\nabla^{\sigma} \Pi_{\alpha_1 \cdots \alpha_k \sigma}  -k \Biggl[ \Pi^{\sigma}_{\;\; (\alpha_1 \cdots \alpha_{k-1}} B^{\;}_{\alpha_k) \sigma } \\
& - \frac{1}{\rho} \Pi_{(\alpha_1 \cdots \alpha_{k-1}} \nabla^{\sigma} \Pi_{\alpha_{k } ) \sigma} - (k - 1) \Pi_{(\alpha_1 \cdots \alpha_{k-2}} D_{\alpha_{k-1} \alpha_{k})} \Biggr] , \label{D1 k-th velocity moment}
\end{split}
\end{align}
where we have introduced $B_{\alpha \beta} = C_{\alpha \beta} + \nabla_{\beta} U_{\alpha}$.






\subsection{Closure scheme.}

As is usual for a moment expansion we now have an infinite tower of velocity moments that is not useful for practical computations and must now consider a closure scheme. In this section we shall show that when the fluid is thermally stable, and the turbulent velocity small relative the fluid velocity, the third velocity moment typically decays until it is asymptotically small relative to the stress tensor and we can therefore drop the $\nabla^{\gamma} \Pi_{\alpha \beta \gamma}$ in the constitutive relation and close the moment expansion at the 2nd velocity moment.

\subsubsection{Well coupled ordering scheme.}

Previous authors have noted that when the dust is well coupled to the gas $(\mathrm{St} \ll 1)$ it can be approximated with a fluid description. We can consider such a `well coupled' ordering scheme by introducing a small parameter $\epsilon > 0$, which can be regarded as a characteristic Stokes number such that $\mathrm{St} = O(\epsilon)$. We consider units such that $U^{\alpha} = O(1)$, $\mathcal{D}_1 = O(1)$ and sufficiently weak turbulence heating such that $D_{\alpha \beta} = O(\epsilon^2)$. In our units the spatial gradients are limited such that $\nabla^{\sigma} = O(\epsilon^{-1})$ (In that the magnitude of the spatial gradients cannot significantly exceed $\epsilon^{-1}$, they can, however, be $\ll \epsilon^{-1}$).


Introducing rescaled velocity moment, $\tilde{\Pi}_{\alpha_1 \cdots \alpha_k}$, such that $\Pi_{\alpha_1 \cdots \alpha_k} = \epsilon^{\delta_k} \tilde{\Pi}_{\alpha_1 \cdots \alpha_k}$, and stretched/rescaled variable $\tilde{X} = X/\epsilon$, such that $\nabla = \epsilon^{-1} \tilde{\nabla}$, then we arrive at a rescaled equation for the k-th velocity moment

\begin{align}
 \epsilon^{\delta_k} \mathcal{D}_1 \Pi_{\alpha_1 \cdots \alpha_{k}} &= - \epsilon^{\delta_{k+1} - 1} \nabla^{\sigma} \Pi_{\alpha_1 \cdots \alpha_k \sigma}  - k \Biggl[ \epsilon^{\delta_k - 1} \Pi^{\sigma}_{\;\; (\alpha_1 \cdots \alpha_{k-1}} B^{\;}_{\alpha_k) \sigma } \nonumber \\
&- \frac{\epsilon^{\delta_{k-1} + \delta_2 -1}}{\rho} \Pi_{(\alpha_1 \cdots \alpha_{k-1}} \nabla^{\sigma} \Pi_{\alpha_{k } ) \sigma} - (k - 1) \epsilon^{\delta_{k-2} + 2} \Pi_{(\alpha_1 \cdots \alpha_{k-2}} D_{\alpha_{k-1} \alpha_{k})} \Biggr] .
\end{align}

Proposing $\delta_k = 3 \mathrm{ceil}(k/2)$, we can rearrange the above to obtain, for even $k$,

\begin{align}
\epsilon \mathcal{D}_1 \Pi_{\alpha_1 \cdots \alpha_{k}} + k &\left[ \Pi^{\sigma}_{\;\; (\alpha_1 \cdots \alpha_{k-1}} B^{\;}_{\alpha_k) \sigma }  - (k - 1) \Pi_{(\alpha_1 \cdots \alpha_{k-2}} D_{\alpha_{k-1} \alpha_{k})} \right] \nonumber \\
&= \epsilon^{3} \left[ -\nabla^{\sigma} \Pi_{\alpha_1 \cdots \alpha_k \sigma} + \frac{k}{\rho}  \Pi_{(\alpha_1 \cdots \alpha_{k-1}}\nabla^{\sigma} \Pi_{\alpha_{k } ) \sigma} \right],
\end{align}
For $k=2$ the left hand side corresponds to the constitutive model with $\Pi_{\alpha \beta \gamma} = 0$. For odd $k$ we instead have
\begin{align}
\epsilon \mathcal{D}_1 \Pi_{\alpha_1 \cdots \alpha_{k}}  &=  - \nabla^{\sigma} \Pi_{\alpha_1 \cdots \alpha_k \sigma} - k \Biggl[ \Pi^{\sigma}_{\;\; (\alpha_1 \cdots \alpha_{k-1}} B^{\;}_{\alpha_k) \sigma } \nonumber \\
&- \frac{1}{\rho}  \Pi_{(\alpha_1 \cdots \alpha_{k-1}}\nabla^{\sigma} \Pi_{\alpha_{k } ) \sigma} - (k - 1)  \Pi_{(\alpha_1 \cdots \alpha_{k-2}} D_{\alpha_{k-1} \alpha_{k})} \Biggr].
\label{odd moments}
\end{align}
Thus we find that the correction to the evolutionary equation for the second velocity moment $\Pi_{\alpha \beta}$ is suppressed by a factor of $\epsilon^{3}$, relative to the leading order terms. Crucially this strong suppression means that Stokes numbers slightly less than one may still be well approximated by our fluid model, provided that we retain the $O(\epsilon)$ advection term ($\mathcal{D}_1 \Pi_{\alpha \beta}$) which will no longer be negligible. 

According to Equation \ref{odd moments} the evolution of the third velocity moment will depend on gradients of the 4-th velocity moment at leading order. Thus, we gain no advantages if we were to truncate the expansion at the third velocity moment over truncating at the second.

\subsubsection{Near Maxwellian ordering scheme} \label{near maxwellian scheme section}

We now wish to consider a situation where the dust distribution function is initially close to a Maxwellian velocity distribution and determine under what circumstances the departure from a Maxwellian velocity distribution remains small. Consider an asymmetric Maxwellian velocity distribution, 



\begin{equation}
f = \frac{|A|^{1/2}}{(2 \pi)^{n/2}} \exp \left( - \frac{1}{2} Q^{\alpha \beta} (V_{\alpha} - U_{\alpha}) (V_{\beta} - U_{\beta}) \right) ,
\end{equation}
with second velocity moment

\begin{equation}
W_{\alpha \beta} = \int (V_{\alpha} - U_{\alpha}) (V_{\beta} - U_{\beta}) f d^n V .
\end{equation}
This is related to $Q^{\alpha \beta}$ through $Q^{\alpha \sigma} W_{\sigma\beta} = \delta_{\beta}^{\alpha}$. More generally we define the $k$-th velocity moment for the Maxwellian velocity distribution as

\begin{equation}
W_{\alpha_1 \cdots \alpha_k} = \int (V_{\alpha_1} - U_{\alpha_1}) \cdots (V_{\alpha_k} - U_{\alpha_k}) f d^n V .
\end{equation}
For odd $k$, $W_{\alpha_1 \cdots \alpha_k} = 0$. Using standard results for Maxwellian distributions \citep[e.g.][]{Withers85} we obtain the following relationship between the $k$-th and $(k - 2)$-th velocity moment;


\begin{equation}
W_{\alpha_1 \cdots \alpha_k} = (k-1) W_{(\alpha_1 \cdots \alpha_{k-2}} W_{\alpha_{k-1}) \alpha_k} .
\label{W recurance relation}
\end{equation}
By symmetry of the velocity moments we also have $W_{\alpha_1 \cdots \alpha_k} = W_{(\alpha_1 \cdots \alpha_k)} = (k-1) W_{(\alpha_1 \cdots \alpha_{k-2}} W_{\alpha_{k-1} \alpha_k)}$. 

Starting from the assumption that the 2nd velocity moment evolves according to
\begin{equation}
D W_{\alpha_1 \alpha_2} = - 2 \left[ W^{\sigma}_{\;\; (\alpha_1} B^{\;}_{\alpha_2) \sigma} - D_{\alpha_{1} \alpha_{2}}\right] ,
\label{W2 evolution}
\end{equation}
we wish to show that the $k$-th velocity moment evolves according to 
\begin{equation}
D W_{\alpha_1 \cdots \alpha_k} = - k \left[ W^{\sigma}_{\;\; (\alpha_1 \cdots \alpha_{k-1}} B^{\;}_{\alpha_k) \sigma} - (k-1) W_{(\alpha_1 \cdots \alpha_{k-2}} D_{\alpha_{k-1} \alpha_{k)}}\right]
\label{Wk evolution}
\end{equation}
Assuming this is the case for the $(k - 2)$-th velocity moment then we can substitute Equation \ref{W recurance relation} into the above equation to obtain

\begin{align}
\begin{split}
D W_{\alpha_1 \cdots \alpha_k} &= (k-1) W_{(\alpha_{1} \alpha_2}  D W_{\alpha_3 \cdots \alpha_{k})} + (k-1) W_{(\alpha_1 \cdots \alpha_{k-2}} D  W_{\alpha_{k-1} \alpha_k)} \\
&= -(k-1) (k-2) W^{\;}_{(\alpha_1 \alpha_2} \left[ W^{\sigma}_{\;\; \alpha_3 \cdots \alpha_{k-1}} B^{\;}_{\alpha_k) \sigma} - (k-3) W^{\;}_{\alpha_3 \cdots \alpha_{k-2}} D^{\;}_{\alpha_{k-1} \alpha_{k})}\right] \\
&- 2 (k-1) W^{\;}_{(\alpha_1 \cdots \alpha_{k-2}} \left[ W^{\sigma}_{\;\;\alpha_{k-1}} B^{\;}_{\alpha_k) \sigma} - D^{\;}_{\alpha_{k-1} \alpha_{k})}\right] \\
&= -(k-1) \left[ (k-2) W^{\;}_{\sigma (\alpha_1 \cdots \alpha_{k-3}} B_{\alpha_{k-2}}^{\;\; \sigma} W^{\;}_{\alpha_{k-1} \alpha_k)} + 2 W^{\;}_{(\alpha_1 \cdots \alpha_{k-2} } W^{\sigma}_{\;\; \alpha_{k-1}} B^{\;}_{\alpha_k) \sigma} \right] \\
&+ (k-1) (k-3) \left[ (k-2) W_{(\alpha_1 \cdots \alpha_{k-4}} D_{\alpha_{k-3} \alpha_{k-2}} W_{\alpha_{k-1} \alpha_k)} + 2 W_{(\alpha_1 \cdots \alpha_{k-4}} W_{\alpha_{k-3} \alpha_{k-2}} D_{\alpha_{k-1} \alpha_k)} \right] \\
&= - k \left[ W^{\sigma}_{\;\; (\alpha_1 \cdots \alpha_{k-1}} B^{\;}_{\alpha_k) \sigma} - (k-1) W_{(\alpha_1 \cdots \alpha_{k-2}} D_{\alpha_{k-1} \alpha_{k)}}\right] .
\end{split}
\label{Wk evolution expanded}
\end{align}
Thus we see that if the $(k-2)$-th velocity moment evolves according to Equation \ref{Wk evolution}, and the 2nd velocity moment evolves according to Equation \ref{W2 evolution}, then the $k$-th velocity moment also evolves according to Equation \ref{Wk evolution}. Starting with the 4-th velocity moment we see that, given Equations \ref{W2 evolution} and \ref{Wk evolution expanded}, it evolves according to Equation \ref{Wk evolution}. We can thus proceed by induction to arbitrary $k$, and conclude that $W_{\alpha_1 \cdots \alpha_k}$ evolve according to Equation \ref{Wk evolution}.

Consider a dust fluid which varies on some short lengthscale $L_{\rm dust}$ embedded with a gas that varies on a long lengthscale $L_{\rm gas}$. This introduces a separation of scales for which we introduce $\boldsymbol{\xi}$ for coordinates describing variation on the short dust lengthscale and $\mathbf{x}$ describing variation on the gas lengthscale. Naturally, the properties of the gas depend only on $\mathbf{x}$ (and time). We propose a nearly Maxwellian dust velocity distribution with the following asymptotic scheme

\begin{align}
\Pi_{\alpha_1 \cdots \alpha_k} &= \epsilon^{k} \rho (\boldsymbol{\xi}, \mathbf{x}) W_{\alpha_1 \cdots \alpha_k} (\mathbf{x}) + \epsilon^{k + 1} \Sigma_{\alpha_1 \cdots \alpha_k} (\boldsymbol{\xi}, \mathbf{x}) , \label{p nm asymptotic} \\
U &= U_0 (\mathbf{x}) + \epsilon^{\kappa} u_0 (\boldsymbol{\xi}, \mathbf{x}) , \\
\nabla &= \epsilon^{-1} \frac{\partial}{\partial \boldsymbol{\xi}}+ \frac{\partial}{\partial \mathbf{x}} , \label{deriv nm asymptotic}
\end{align}
where $\epsilon$ is treated as a book-keeping parameters. Strictly speaking one should also expand the density, however the $O(\epsilon)$ terms due to the effects of the Non-Maxwellian velocity perturbation can be absorbed into the definition of $\Sigma_{\alpha_1 \cdots \alpha_k}$. While we can often treat $\kappa = 2$ (i.e. the part of the mean velocity that varies on the dust lengthscale is $O(\epsilon^2)$) we shall assume $\kappa = 1$ throughout as this will allow for a wider range of dust flows.

Substituting Equations \ref{p nm asymptotic}-\ref{deriv nm asymptotic} into Equation \ref{D1 k-th velocity moment}, and making use of $\mathcal{D}_1 \rho = 0$, the evolutionary equation for the k-th velocity moment becomes:

\begin{align}
 \begin{split}
 \mathcal{D}_1 \Pi_{\alpha_1 \cdots \alpha_k} &= \epsilon^{k} \rho D W_{\alpha_1 \cdots \alpha_k} + \epsilon^{k+1} u_0^{\alpha} \frac{\partial}{\partial x^{\alpha}} W_{\alpha_1 \cdots \alpha_k} + \epsilon^{k+1} \mathcal{D}_1 \Sigma_{\alpha_1 \cdots \alpha_k} \\
&= -\epsilon^{k} \left( \frac{\partial}{\partial \xi^{\sigma}} + \epsilon \frac{\partial}{\partial x^{\sigma}}  \right) (\rho W_{\alpha_1 \cdots \alpha_k \sigma}) - \epsilon^{k+1} \left( \frac{\partial}{\partial \xi^{\sigma}} + \epsilon \frac{\partial}{\partial x^{\sigma}}  \right) \Sigma_{\alpha_1 \cdots \alpha_k \sigma} \\
&- k \epsilon^{k} \Biggl [  \left(\rho W^{\sigma}_{\;\;(\alpha_1 \cdots \alpha_{k-1}} + \epsilon \Sigma^{\sigma}_{\;\; (\alpha_1 \cdots \alpha_{k-1}} \right)  B^{\;}_{\alpha_k) \sigma} \\
&- \left(\rho W_{(\alpha_1 \cdots \alpha_{k-1}} + \epsilon \Sigma_{(\alpha_1 \cdots \alpha_{k-1}} \right) \left( \frac{\partial}{\partial \xi^{\sigma}} + \epsilon \frac{\partial}{\partial x^{\sigma}}  \right) \left(\rho W_{\alpha_k) \sigma} + \epsilon \Sigma_{\alpha_k) \sigma} \right)  \\
&- (k-1) \left(\rho W_{(\alpha_1 \cdots \alpha_{k-2}} + \epsilon \Sigma_{(\alpha_1 \cdots \alpha_{k-2}} \right) D_{\alpha_{k-1} \alpha_{k})} \Biggr] ,
\end{split}
\end{align}
where, here, $D = \partial_t + U_0^{\alpha} \nabla_{\alpha}$ is the Lagrangian time derivative with respect to the leading order flow described by $U_0$.

Making use of Equation \ref{Wk evolution} for the evolution of $W_{\alpha_1 \cdots \alpha_k}$, along with the recurrence relation for $W_{\alpha_1 \cdots \alpha_k}$ (Equation \ref{W recurance relation}) and rearranging we obtain and equation for the evolution of the non-Maxwellian part of the velocity moment,

\begin{align}
 \begin{split}
\mathcal{D}_1 \Sigma_{\alpha_1 \cdots \alpha_k} &+  u_0^{\alpha} \frac{\partial}{\partial x^{\alpha}} W_{\alpha_1 \cdots \alpha_k}  + k \rho W_{\sigma (\alpha_1} \frac{\partial}{\partial x^{\sigma}}  W_{\alpha_2 \cdots \alpha_{k})}  +  \frac{\partial}{\partial \xi^{\sigma}} \Sigma_{\alpha_1 \cdots \alpha_k \sigma} \\
&+ k \Biggl [ \Sigma^{\sigma}_{\;\; (\alpha_1 \cdots \alpha_{k-1}}  B^{\;}_{\alpha_k) \sigma} - \rho W_{(\alpha_1 \cdots \alpha_{k-1}}   \frac{\partial}{\partial \xi^{\sigma}}  \Sigma_{\alpha_k) \sigma}  \\
&- \Sigma_{(\alpha_1 \cdots \alpha_{k-1}} \frac{\partial}{\partial \xi^{\sigma}} \rho W_{\alpha_k) \sigma} - (k-1)  \Sigma_{(\alpha_1 \cdots \alpha_{k-2}}  D_{\alpha_{k-1} \alpha_{k})} \Biggr ]\\
&= \epsilon \Biggl [  k \rho W_{(\alpha_1 \cdots \alpha_{k-1}} \frac{\partial}{\partial x^{\sigma}}  \Sigma_{\alpha_k) \sigma} + k \Sigma_{(\alpha_1 \cdots \alpha_{k-1}} \frac{\partial}{\partial x^{\sigma}} \left(\rho W_{\alpha_k) \sigma} + \epsilon \Sigma_{\alpha_k) \sigma}\right) \\
&+ k \Sigma_{(\alpha_1 \cdots \alpha_{k-1}} \frac{\partial}{\partial \xi^{\sigma}} \Sigma_{\alpha_k) \sigma}   - \frac{\partial}{\partial x^{\sigma}}  \Sigma_{\alpha_1 \cdots \alpha_k \sigma} \Biggr] ,
\end{split}
\end{align}
where the terms on the right hand side are all sub-leading. Dropping these subleading terms we obtain

\begin{align}
 \begin{split}
0 &= \mathcal{D}_1 \Sigma_{\alpha_1 \cdots \alpha_k} +  u_0^{\alpha} \frac{\partial}{\partial x^{\alpha}} W_{\alpha_1 \cdots \alpha_k}  + k \rho W_{\sigma (\alpha_1} \frac{\partial}{\partial x^{\sigma}}  W_{\alpha_2 \cdots \alpha_{k})}  +  \frac{\partial}{\partial \xi^{\sigma}} \Sigma_{\alpha_1 \cdots \alpha_k \sigma} \\
&+ k \Biggl [  \Sigma^{\sigma}_{\;\; (\alpha_1 \cdots \alpha_{k-1}}  B^{\;}_{\alpha_k) \sigma} - \rho W_{(\alpha_1 \cdots \alpha_{k-1}}   \frac{\partial}{\partial \xi^{\sigma}}  \Sigma_{\alpha_k) \sigma}  \\
&- \Sigma_{(\alpha_1 \cdots \alpha_{k-1}} W_{\alpha_k) \sigma} \frac{\partial}{\partial \xi^{\sigma}} \rho - (k-1)  \Sigma_{(\alpha_1 \cdots \alpha_{k-2}}  D_{\alpha_{k-1} \alpha_{k})} \Biggr ] .
\end{split}
\end{align}
This confirms that the asymptotic ordering scheme (Equations \ref{p nm asymptotic}-\ref{deriv nm asymptotic}) is self consistent and the non-Maxwellian terms are suppressed by a factor of $\epsilon \sim L_{\rm dust}/L_{\rm gas}$ relative to the Maxwellian terms. However, for the purposes of the equation of motion the pressure gradients are the more important quantity. For the nearly Maxwellian velocity distribution considered here the stress gradients are

\begin{align}
\begin{split}
\nabla_{\beta} \Pi^{\beta \alpha} &= \epsilon \left( \frac{\partial}{\partial \xi^{\beta}} + \frac{\partial}{\partial x^{\beta}} \right) \rho  W^{\beta \alpha} + \epsilon^{2} \left( \frac{\partial}{\partial \xi^{\beta}} + \frac{\partial}{\partial x^{\beta}} \right) \Sigma_{\beta \alpha} \\
 &= \epsilon W^{\beta \alpha} \frac{\partial}{\partial \xi^{\beta}} \rho + O(\epsilon^{2}) .
\end{split}
\end{align}
Thus the effects of the non-Maxwellian terms are $O(\epsilon^2)$, and are thus small relative to the acceleration and gravity, which are taken to be $O(1)$, when the dust layer is dynamically cool.

\subsubsection{Are the ordering schemes attractors?} \label{condition on decay of kth moment section}

We have two separate situations where we can truncate the moment expansion by neglecting the third (and higher) velocity moment(s). The first is when $\mathrm{St} \lesssim 1$, meaning that the dust is tightly coupled to the gas and the higher order velocity moments are suppressed by interaction with the gas. The second is for dynamically cool dust layers where $L_{\rm dust} \ll L_{\rm gas}$ (with the lengthscale typically being the dust and gas scale heights), where the non-Maxwellian velocity moments are suppressed by the confinement of the dust. This latter scenario is of interest for dust with $\mathrm{St} > 1$ in gas flows which are not strongly stirred, in the presence of vertical gravity, as these would be expected to settle into a hydrostatically supported dust layer which is much thinner than a hydrostatically supported gas flow. Of course the existence of a consistent asymptotic scaling does not guarantee that the fluid regime is an attractor. While a complete exploration of when this state becomes an attractor, and thus allow for a fluid treatment of the dust, is beyond the scope of this work, in this section we shall present an argument showing that velocity moments which start far from this asymptotic scaling are expected to damp towards this scaling, subject to the dust fluid being thermally stable.

Consider a situation situation where the either the well coupled or near-Maxwellian ordering scheme holds. We wish to explore what happens where some perturbation increases the k-th velocity moment sufficiently such that it breaks the ordering scheme. If the k-th velocity moment is large, while all other velocity moments keep the same ordering as in the fluid ordering schemes, then the only terms that are important in the evolutionary equation for k-th velocity moment are those involving $\Pi_{\alpha_1 \cdots \alpha_k}$. Thus evolution of the k-th velocity moment is approximately described by
\begin{equation}
 \mathcal{D}_1 \Pi_{\alpha_1 \cdots \alpha_k} = -k \Pi^{\sigma}_{\;\; (\alpha_1 \cdots \alpha_{k-1}} B^{\;}_{\alpha_k) \sigma} .
 \label{drag vort only eq}
\end{equation}
Defining $W_{\alpha_1 \cdots \alpha_k} = \rho^{-1} \Pi_{\alpha_1 \cdots \alpha_k}$ this simplifies to

\begin{equation}
 D W_{\alpha_1 \cdots \alpha_k} = -k W^{\sigma}_{\;\; (\alpha_1 \cdots \alpha_{k-1}} B^{\;}_{\alpha_k) \sigma} .  \label{W drag only eq}
\end{equation}
We wish to show that $W_{\alpha_1 \cdots \alpha_k}$ decays subject to certain constraints on $B_{\alpha \beta}$. To do this we make use of the adjoint problem,

\begin{equation}
 D Y^{\alpha} = B_{\beta}^{\;\; \alpha} Y^{\beta} , \label{Y adjoint eq}
\end{equation}
in order to relate the evolutionary equation for $W_{\alpha_1 \cdots \alpha_k}$ for arbitrary $k$ to that with $k = 2$. This allows us to relate the behaviour of Equation \ref{W drag only eq} to properties of the constitutive equation, in particular the thermal stability of the flow. 

Equations \ref{W drag only eq} and \ref{Y adjoint eq} are related by an invariant scalar $\chi = W_{\alpha_1 \cdots \alpha_k} Y^{\alpha_1} \cdots Y^{\alpha_k}$, with

\begin{align}
 \begin{split}
D \chi &= Y^{\alpha_{n+1}} \cdots Y^{\alpha_{k}} D Q_{\alpha_{n+1} \cdots \alpha_{k}} + (k - n) Q_{\alpha_{n+1} \cdots \alpha_{k}^{d}}Y^{\alpha_{n+1}} \cdots Y^{\alpha_{k-1}} D Y^{\alpha_{k}}   \\
&= - (k - n) Y^{\alpha_{n+1}} \cdots Y^{\alpha_{k}} Q_{\sigma \alpha_{n+1} \cdots \alpha_{k-1} } M_{\alpha_{k}}^{\;\; \sigma} + (k - n) Q_{\sigma \alpha_{n+1} \cdots \alpha_{k-1}}Y^{\alpha_{n+1}} \cdots Y^{\alpha_{k-1}} M_{\beta}^{\;\; \sigma} Y^{\beta}   \\
&= 0 \quad .
\end{split}
\end{align}

Consider now $k=2$ and define the associated scalar, $\zeta = Q_{\alpha \beta} Y^{\alpha} Y^{\beta}$, we also assume $Q_{\alpha \beta}$ is positive definite at $t = t_0$. Without loss of generality we can take $Y^{\alpha} = y_0^{\alpha}$ at $t = t_0$, where $|\mathbf{y}_0| = 1$, such that

\begin{equation}
 \zeta =\left. (Q_{\alpha \beta} Y^{\alpha} Y^{\beta} ) \right |_{t=t_0} = Q_{\alpha \beta} y_0^{\alpha} y_0^{\beta} > 0 .
\end{equation}
At $t = t_1 > t_0$ we write $Y^{\alpha} = \mathcal{Y} y^{\alpha}$, with $|\mathbf{y}| = 1$ , such that

\begin{equation}
 \zeta = \left. (Q_{\alpha \beta} Y^{\alpha} Y^{\beta} ) \right|_{t=t_1} =  \mathcal{Y}^2 Q_{\alpha \beta} y^{\alpha} y^{\beta}  .
\end{equation}
As $Q_{\alpha \beta}$ is positive semi-definite, for all $t$, $Q_{\alpha \beta} y^{\alpha} y^{\beta} \ge 0$. Using the fact that $\zeta$ is constant, we obtain

\begin{equation}
\mathcal{Y}^2 = \frac{Q_{\alpha \beta} y_0^{\alpha} y_0^{\beta}}{ Q_{\alpha \beta} y^{\alpha} y^{\beta} } ,
\end{equation}
In order for the fluid to be thermally stable $Q_{\alpha \beta}$ must ultimately decay towards zero. If this were not the case then there would exist components of $\Pi_{\alpha \beta}$ where heating by the disc turbulence is not balanced by cooling from the $\Pi^{\sigma}_{\;\; (\alpha} B_{\beta) \sigma}$ term, and would thus experience thermal runaway. Thus for $\delta > 0$, there exists a $t = t_{\rm cool} > t_0$ such that the components of $Q_{\alpha \beta}$ at $t=t_{\rm cool}$ satisfy $|Q_{\alpha \beta}| < \delta$. It should be noted that certain components of $Q_{\alpha \beta}$ can experience transient growth (e.g. due to the shearing out of the initial conditions), but must ultimately decline in order to ensure thermal stability. As $\delta$ is arbitrary we can choose $\delta$ small enough such that

\begin{equation}
 Q_{\alpha \beta} y^{\alpha} y^{\beta} \le \sum_{\alpha, \beta} |Q_{\alpha \beta}| |y^{\alpha}| |y^{\beta}|  < \delta  \sum_{\alpha, \beta} |y^{\alpha}| |y^{\beta}| < Q_{\alpha \beta} y_0^{\alpha} y_0^{\beta} ,
\end{equation}
for $t > t_{\rm cool}$. From this we can conclude that $\mathcal{Y} > 1$, for $t > t_{\rm cool}$. By choosing $t$ large enough we can make $\mathcal{Y}$ arbitrarily large. Typically one expects $t_{\rm cool}$ to be of order the cooling/settling time in the fluid as this decay is linked to the dynamical cooling of the dust fluid.

Now consider the scalar $\chi$ associated with $Q_{\alpha_{n+1} \cdots \alpha_{k}}$. As $\chi$ is constant we have,

\begin{align}
\begin{split}
\left. Q_{\alpha_{n+1} \cdots \alpha_{k}} \right |_{t=t_0} y_0^{\alpha_{n+1}} \cdots y_0^{\alpha_{k}} &= ( Q_{\alpha_{n+1} \cdots \alpha_{k}} Y^{\alpha_{n+1}} \cdots Y^{\alpha_{k}})_{t=t_0}  \\
&= ( Q_{\alpha_{n+1} \cdots \alpha_{k}} Y^{\alpha_{n+1}} \cdots Y^{\alpha_{k}})_{t=t_1} \\
&= \left. \mathcal{Y}^{k} Q_{\alpha_{n+1} \cdots \alpha_{k}} \right |_{t=t_1}  y^{\alpha_{n+1}} \cdots y^{\alpha_{k}} .
 \end{split}
\end{align}
Rearranging this we obtain,

\begin{equation}
\left. Q_{\alpha_{n+1} \cdots \alpha_{k}} \right |_{t=t_1}  y^{\alpha_{n+1}} \cdots y^{\alpha_{k}}  = \left. \mathcal{Y}^{-k}  Q_{\alpha_{n+1} \cdots \alpha_{k}} \right |_{t=t_0} y_0^{\alpha_{n+1}} \cdots y_0^{\alpha_{k}} \le \mathcal{Y}^{-k}  \sum_{\alpha_{n+1} \cdots \alpha_{k}} \left | \left. Q_{\alpha_{n+1} \cdots \alpha_{k}} \right  |_{t=t_0} \right | .
\end{equation}
Again, by choosing $t_1$ large enough we can take $\mathcal{Y}$ to be as large as we like, this means that $\left. Q_{\alpha_{n+1} \cdots \alpha_{k}} \right |_{t=t_1}  y^{\alpha_{n+1}} \cdots y^{\alpha_{k}}$ can be made arbitrarily small. As we can do this for any unit vector $y^{\alpha}$, and $Q_{\alpha_{n+1} \cdots \alpha_{k}}$ is symmetric, we conclude that the components of $Q_{\alpha_{n+1} \cdots \alpha_{k}}$ will become arbitrarily small at late times 

Thus we can conclude, from the above argument, that thermal stability of the fluid flow is a necessary, and sufficient condition for $W_{\alpha_1 \cdots \alpha_k}$ to decay. This means that the k-th velocity moment, $\Pi_{\alpha_1 \cdots \alpha_k}$, decays when its evolution can be well approximated by Equation \ref{drag vort only eq} and the fluid is thermally stable. This implies that the dust fluid ordering schemes are stable to (nonlinear) perturbations to the higher order velocity moments, which should damp until they are compatible with the fluid ordering scheme derived above on approximately the cooling/settling time of the dust fluid.

The above argument shows that thermal stability is a necessary condition for the fluid dust description to remain valid. It is not, however, a sufficient condition as the argument only applies to (nonlinear) perturbations to the k-th velocity moment in isolation. Thus, in principle, there could exist perturbations to the multiple orders of velocity moment simultaneously that can be sustained and will not damp towards the fluid ordering scheme. For now we shall work under the assumption that thermal stability is sufficient to ensure the damping of higher-order velocity moments, however the exploration of the stability of the fluid description against more general perturbations should be explored if the dust fluid model finds widespread use.

\subsection{Obtaining the dust fluid equations}

As a result of the asymptotic argument presented above, we can take $\Pi_{\alpha \beta \gamma} = 0$ and only consider the first 3 moments of the Fokker-Planck equation. This yields a continuity, momentum and constitutive relation for a 6-dimensional dust fluid. This dust fluid has a high degree of symmetry as physical properties must be independent of the gas displacement $\{\mathbf{x}_g\}$\footnote{For the Stochastic model considered here. If the fluid equations were to be derived based on a two-step Stochastic model, as outlined in \citet{Minier23}, then the dummy gas variable would influence the fluid model, which may allow the effects of spatial correlations to be included.}. One can, therefore, integrate out these redundant degrees of freedom. 

In integrating out the dummy gas degrees of freedom, we replace the connections $\nabla_{\alpha}$ with the connections $\overline{\nabla}_{\alpha}$ which ensure that the components of vectors associated with the integrated out directions remain correctly aligned. Our normalisation means that $\rho_d = \overline{\rho}_6$, we also introduce the rheological stress tensor $T_{\alpha \beta} = \overline{\Pi}_{\alpha \beta}$ and we can always choose the size of the dummy gas dimensions such that $U^{\alpha} = \overline{U^{\alpha}}$. With these choices the dust fluid equations are

\begin{align}
\overline{D} \rho_d &= - \rho_d \overline{\nabla}_{\alpha} U^{\alpha} , \label{6 d continuity aved}  \\
\rho_d \overline{D} U_{\alpha} &= \rho_d F_{\alpha} - \overline{\nabla}^{\beta} T_{\alpha \beta} - \rho_d C_{\alpha \beta} (U^{\beta} - U_g^{\beta}) , \label{6 d momentum aved} \\
\overline{\mathcal{D}}_2 T_{\alpha \beta} &= -2 \left( T^{\gamma}_{\;\; (\alpha} C^{\;}_{\beta) \gamma} - \rho_d D_{(\alpha \beta)} \right) ,
\label{6 d constituative aved}
\end{align}
where

\begin{equation}
\overline{D} = \partial_t + U^{\alpha} \overline{\nabla}_{\alpha} = \begin{pmatrix}
\partial_t + u^i_d \nabla_i & \mathbf{0} \\
\mathbf{0} & \partial_t + u^i_d \nabla_i 
\end{pmatrix},
\end{equation}
which corresponds to the usual (3-dimensional) Lagrangian time derivative with respect to the mean dust flow applied to the dust and dummy gas components of the (6-D) tensor independently. $C_{\alpha \beta}$ and $D_{\alpha \beta}$ are given by Equations \ref{C OU turb} and \ref{D OU turb} (for isotropic stochastic driving). Finally the operator $\overline{\mathcal{D}}_2$, when acting on $T_{\alpha \beta}$, is given by

\begin{equation}
\overline{\mathcal{D}}_2 T_{\alpha \beta} = \overline{D} T_{\alpha \beta}  + 2 T_{\gamma_d (\alpha} \overline{\nabla}^{\gamma_d} U_{\beta)} + T_{\alpha \beta} \overline{\nabla}_{\gamma_d} U^{\gamma_d}  .
\end{equation}

Finally to highlight the effects of the torsion we consider it's contribution to the dust fluid vorticity,

\begin{equation}
\omega^{\gamma} \epsilon_{\gamma \alpha \beta} = 2 \overline{\partial}_{[\alpha} U_{\beta]} + S^{\gamma_g}_{\alpha \beta} U_{\gamma_g} ,
\end{equation}
meaning

\begin{equation}
\omega^{\gamma} \epsilon_{\gamma \alpha_g \beta_d} = - \overline{\partial}_{\beta_d} U_{\alpha_g}  - \mathcal{T}_{\beta_d \alpha_g^{*}}^{\gamma_g^*} U_{\gamma_g} .
\end{equation}
This additional contribution to the vorticity is associated with the rotation of the gas displacement vectors.

For Cartesian coordinates ($x$, $y$, $z$, $x_g$, $y_g$, $z_g$), in Euclidean space, with $C_{\alpha \beta}$ and $D_{\alpha \beta}$ are given by Equations \ref{C OU turb} and \ref{D OU turb}, then Equations \ref{6 d continuity aved}-\ref{6 d constituative aved} are explicitly








\begin{align}
(\partial_t + U^{\alpha_d} \partial_{\alpha_d} )  \rho_d &= - \rho_d \partial_{\alpha_d} U^{\alpha_d} , \label{6 d continuity cart} \\
\rho_d (\partial_t + U^{\alpha_d} \partial_{\alpha_d} ) U_{\alpha_d} &= -\rho_d \partial_{\alpha_d} \phi - \partial_{\beta_d} T_{\alpha_d}^{\;\; \beta_d} - \frac{\rho_d}{t_s} (U_{\alpha_d}  - U_{\alpha_g})  , \\
\begin{split}
\rho_d (\partial_t + U^{\alpha_d} \partial_{\alpha_d} )  U_{\alpha_g} &= -\rho_d \partial_{\alpha_g^*} \phi - f_d \partial_{\alpha_g^*} p_g + \rho_d (U^{\beta_d} - U^{\beta_d}_{g}) \partial_{\beta_d} U_{\alpha_d}^{\rm g} \\
&- \partial_{\beta_d} T_{\alpha_g}^{\;\; \beta_d} - \frac{\rho_d}{t_c} (U_{\alpha_g} - U^g_{\alpha_g}) , 
\end{split} \\
\begin{split}
(\partial_t + U^{\alpha_d} \partial_{\alpha_d} ) T_{\alpha_d \beta_d}  &+ T^{\gamma_d}_{\;\; \alpha_d} \partial_{\gamma_d} U_{\beta_d} + T^{\gamma_d}_{\;\; \beta_d} \partial_{\gamma_d} U_{\alpha_d} + T_{\alpha_d \beta_d} \partial_{\gamma_d} U^{\gamma_d}  \\
&= -\frac{2}{t_s} T_{\alpha_d \beta_d} + \frac{1}{t_s} T_{\alpha_d \beta_d^{*}} + \frac{1}{t_s} T_{\alpha_d^{*} \beta_d}   , 
\end{split} \\
\begin{split}
(\partial_t + U^{\alpha_d} \partial_{\alpha_d} ) T_{\alpha_d \beta_g}  &+ T^{\gamma_d}_{\;\; \alpha_d} \partial_{\gamma_d} U_{\beta_g} + T^{\gamma_d}_{\;\; \beta_g} \partial_{\gamma_d} U_{\alpha_d} + T_{\alpha_d \beta_g} \partial_{\gamma_d} U^{\gamma_d}  \\
&= - \left(\frac{1}{t_c}  + \frac{1}{t_s} \right) T_{\alpha_d \beta_g} + \frac{1}{t_s}  T_{\alpha_d^* \beta_g}  , 
\end{split} \\
\begin{split}
(\partial_t + U^{\alpha_d} \partial_{\alpha_d} ) T_{\alpha_g \beta_g}  &+ T^{\gamma_d}_{\;\; \alpha_g} \partial_{\gamma_d} U_{\beta_g} + T^{\gamma_d}_{\;\; \beta_g} \partial_{\gamma_d} U_{\alpha_g} + T_{\alpha_g \beta_g} \partial_{\gamma_d} U^{\gamma_d} \\
&= - \frac{2}{t_c} \left( T_{\alpha_g \beta_g}  -  \alpha c_s^2 \rho_d \delta_{\alpha_g \beta_g} \right) . 
\end{split} \label{6 d constituative cart}
\end{align}
In the 3-dimensional picture we have $u_d^{i} = U^{i}$, $u_s^{i} = U^{i^{*}}$ are the (3D) dust velocity and fluid seen respectively and $p_{i j} = T_{i j}$, $\tau_{i j} = T_{i j^{*}}$, $\sigma_{i j} = T_{i^{*} j^{*}}$ are the dust kinetic tensor, dust-gas correlation tensor and Reynolds stress of the fluid seen respectively. Equations \ref{6 d continuity cart}-\ref{6 d constituative cart} are equivalent to

\begin{align}
(\partial_t + u_d^{i} \partial_{i} )  & \rho_d = - \rho_d \partial_{i} u_d^{i} , \\
\rho_d (\partial_t + u_d^{j} \partial_{j} ) & u^d_i = -\rho_d \partial_{i} \phi - \partial_{j} p^{j}_{\;\; i} - \frac{\rho_d}{t_s} (u^d_i  - u^s_i)  , \\
\begin{split}
\rho_d (\partial_t + u_d^{j} \partial_{j} )  & u^s_i = -\rho_d \partial_{i} \phi - f_d \partial_{i} p_g + \rho_d (u_d^{j} - u_g^j) \partial_{j} u_i^{\rm g} - \partial_{j} \tau^{j}_{\;\; i} - \frac{\rho_d}{t_c} (u^s_i - u^g_{i}) , 
\end{split} \\
\begin{split}
(\partial_t +u_d^{k} \partial_{k} ) & p_{i j}  + p^{k}_{\;\; i} \partial_{k} u^d_{j} + p^{k}_{\;\; j} \partial_{k} u^d_{i} + p_{i j} \partial_{k} u_d^{k}  = -\frac{2}{t_s} p_{i j} + \frac{1}{t_s} \tau_{i j} + \frac{1}{t_s} \tau_{j i}   , 
\end{split} \\
\begin{split}
(\partial_t + u_d^{k} \partial_{k} ) & \tau_{i j}  + p^{k}_{\;\; i} \partial_{k} u^s_{j} + \tau^{k}_{\;\; j} \partial_{k} u^d_{i} + \tau_{i j} \partial_{k} u_d^{k}  = - \left(\frac{1}{t_c}  + \frac{1}{t_s} \right) \tau_{i j} + \frac{1}{t_s}  \sigma_{i j}  , 
\end{split} \\
\begin{split}
(\partial_t + u_d^{k} \partial_{k} ) & \sigma_{i j}  + \tau^{k}_{\;\;i} \partial_{k} u^s_{j} + \tau^{k}_{\;\; j} \partial_{k} u^s_{i} + \sigma_{i  j} \partial_{k} u_d^{k} = - \frac{2}{t_c} \left( \sigma_{i  j}  -  \alpha c_s^2 \rho_d \delta_{i  j} \right) . 
\end{split} 
\end{align}
 



\section{Properties of the dust fluid model} \label{dust fluid properties}

We will now describe some key features of our dust fluid model.

\subsection{The mean gas velocity experienced by the dust is different to that experienced by the gas}

Unlike the pressureless, non-turbulent models the dust experiences a different mean gas velocity to the gas. Part of this is due to the ``crossing trajectory effect'' \citep[e.g. see][]{Minier01,Minier04,Minier14} where the mean gas velocity ``seen'' by the dust is that following the Lagrangian trajectory traced by the dust, rather than that traced by the fluid particles. In addition to this, the dust experience a subsample of the gas velocity field rather than the gas velocity field itself. This distinction is vital for producing dust dispersion by the turbulence. If the dust experienced the same gas velocity  distribution as the gas then a local dust density maxima of perfectly coupled dust would not spread in homogeneous gas turbulence. The dust to gas density ratio (in the 3D picture), in such a setup, evolves according to

\begin{align}
 \partial_t (\rho_d/\rho_g) &= - \nabla_i (\rho_d u_d^i/\rho_g) \\
&= - \nabla_i (\rho_d u_s^i/\rho_g) 
\end{align}
Thus, in order that the gas turbulence disperse the dust, we require the velocity of the fluid seen $u_s^i \ne u_g^i = 0$. The subsampling of the gas velocity distribution means the larger number of dust grains at the centre of the overdensity experience more ``draws'' from the gas velocity distribution and thus experience a greater gas dispersion (this would equally be true for ``marked'' gas fluid elements). This means the dust experiences a mean gas flow directed away from the maxima due to the resulting gradient in the cross pressure.





\subsection{Anisotropic Dust Rheological Stress Tensor}

 The most important feature of the dust fluid model is the fluid stress is not zero, and can be dynamically important. In fact one expects dust settling/drift to concentrate dust until dust stress gradients become dynamically important. Also present is a form of ``cross-pressure'', arising from correlations between the dust and gas motion, which modifies the mean-gas velocity experienced by the dust.

This rheological stress tensor is anisotropic in the presence of strong shear or rotation. In general, the gas turbulence heats the dust and isotropises the dust stress tensor on timescales longer than the correlation time. However, in strong shear flows the velocity dispersion induced in the dust by the turbulence is sheared out resulting in an anisotropic stress tensor (just as happens for the gas Reynolds tensor). The flow vorticity also provides an additional anisotropic heating term in the dust. In Section \ref{Steady state section} we explore this effect further by considering the dust stress tensor in a rotating shear flow. 
 
As we shall show in the next section, the presence of a nonzero elastic stress means the dust fluid supports waves, specifically seismic waves.


\subsection{Viscoelasticity}

The dust fluid exhibits viscoelastic behaviour (See Appendix \ref{viscoelasticity appendix}) This behaviour is easiest to see when $t_s \sim t_c = O(\mathrm{De})$, where $\mathrm{De} = t_r/t_f$ is the Deborah number of the dust fluid, which is the ratio of the characteristic relaxation time $t_r \sim t_s \sim t_c$ to the characteristic fluid timescale $t_f$. When $\mathrm{De} \gg 1$ the dust stress tensor evolves according to
  \begin{equation}
   \overline{\mathcal{D}}_2 T_{\alpha \beta} = \overline{\mathcal{D}} T_{\alpha \beta} - 2 T^{\gamma}_{\;\; (\alpha} \varepsilon^{\;}_{\beta) \gamma \sigma} \omega^{\sigma} = 0 . \label{elastic terms}
  \end{equation}
This corresponds to an elastic stress with a vortical heating \citep[or ``gyroscopic motion''][]{Gavrilyuk12} term, and evolves in an identical manner to a Reynolds stress in the absence of source terms. When $\mathrm{De} \ll 1$ the stress tensor is approximately


  \begin{equation}
    T_{\alpha \beta} =  p_d \left(1 + \frac{t_s}{t_c}\Theta^g_{\alpha \beta}\right) g_{\alpha \beta} + \frac{1}{2} p_{x} (g_{\alpha \beta^*} + g_{\alpha^* \beta}) - 2 \mu_{\alpha \beta}^{\mu \nu} \overline{\nabla}_{\mu} U_{\nu} + O(\mathrm{De}^2) .
    \label{viscous limit}
  \end{equation}
  At leading order this consists of an isotropic, isothermal, effective, dust pressure with sound speed $\sqrt{\frac{\alpha}{1 + t_s/t_c}} c_s$, a cross pressure $p_{x} = p_d$, from the dust-gas velocity correlations, and an additional pressure like contribution to the dummy gas components of the Rheological stress. The next terms in the expansion are an anisotropic viscous stress characterised by the viscosity tensor $\mu_{\alpha \beta}^{\mu \nu}$; including a ``cross'' viscosity, which likely encapsulates the the decorrelation of the dust and gas velocities in the presence of shear. Explicit expressions for $\mu_{\alpha \beta}^{\mu \nu}$ are given in Appendix \ref{viscoelasticity appendix}. For weak gas turbulence $\alpha \ll 1$ the the viscous terms, for small dust grains, are typically negligible and the dust primarily behaves like an inviscid isothermal gas with a lower temperature than than the gas phase. The difference between $U^g_{\alpha_g}$ and $U_{\alpha_d}$ (mean gas velocities experienced by the gas and dust respectively) as a result of the cross pressure term allows for dust diffusion to occur in this limit.


The local expression, Equation \ref{viscous limit}, arises due to the fact that in the $\mathrm{De} \ll 1$ limit $\mathbf{v} - \mathbf{v}^g$ and $\mathbf{v}^g - \mathbf{u}^g$, in the original stochastic-differential equations, are ``fast variables'' with no memory of the previous fluid state \citep{Minier16}. For $\tau_c \ll 1$, but $\mathrm{St} \sim 1$, only $\mathbf{v}^g - \mathbf{u}^g$ is a fast variable and we have a local closure for $T_{\alpha_d \beta_{g}}$ and $P_{\alpha_g \beta_g}$ but not $P_{\alpha_d \beta_d}$, which then has a fluid memory of order the stopping time. In the large Deborah number limit the fast terms in Equation \ref{dust particle O-U} are negligible, resulting in a fully nonlocal behaviour for $T_{\alpha \beta}$ (Equation \ref{elastic terms}).

\subsection{Eddy-Knudsen number effect}

While it might be expected that small dust grains should closely follow the gas with the dust velocity correlations being set by the gas velocity correlations. This turns out to only be the case when the dust sees the turbulence as a continuum. This is explored further in Appendix \ref{eddy knudon appendix}. Whether the dust see the turbulence as a continuum or is sensitive to individual eddies is determined by a form of ``eddy Knudsen number'':
 \begin{equation}
  \mathrm{Kn}_{\rm e} = \frac{\lambda}{L} = \frac{t_c \Delta U^{*}}{L} ,
 \end{equation}
 where $\lambda = t_c \Delta U^{*}$ is the mean free path of a dust grain in the turbulent flow representing the lengthscale a dust grain is transported by a single eddy, $\Delta U^{*}$ is a characteristic velocity difference between the dust and gas and $L$ is a characteristic lengthscale of variations in the fluid flow. 

 When $\mathrm{Kn}_{\rm e} \ll 1$ a dust grain interacts with many turbulent eddies over the lengthscale on which the dust fluid varies, meaning the dust experiences the turbulence as a continuum of stochastic perturbations. When $\mathrm{Kn}_{\rm e} \gtrsim 1$ the dust is instead strongly affected by individual eddies (in a similar manner to how weakly collisional gases can be strongly perturbed by individual collisions). Thus, in this regime, the dust is sensitive to individual eddies. In the short stopping time limit the equation for the dust stress simplify to (see Appendix \ref{eddy knudon appendix})



\begin{align}
\begin{split}
t_c \tilde{D} \left( \frac{T_{\alpha_g \beta_g}}{\rho_d} \right) &+ 2 \frac{t_c}{\rho_d} T^{\;}_{\gamma_d (\alpha_g} \nabla^{k} u^{g}_{\beta_g)} + 2 \left(\frac{T_{\alpha_g \beta_g}}{\rho_d}  - \alpha c_s^2 g_{\alpha_g \beta_g} \right) \\
&= - \mathrm{Kn}_{\rm e}  \left[ \frac{\Delta U^{\gamma}}{\Delta U^{*}} L \overline{\nabla}_{\gamma} \left( \frac{T_{\alpha_g \beta_g}}{\rho_d} \right) + 2 \frac{T_{\gamma (\alpha_g}}{\rho_d} L \overline{\nabla}^{\gamma} \frac{\Delta U_{\beta_g)}}{\Delta U^{*}} \right ] .
\end{split}
\end{align}
When $\mathrm{Kn}_{\rm e} \rightarrow 0$ this matches the equation governing the evolution of the gas velocity correlations, meaning the dust velocity correlations are indeed set by those of the gas. This is no longer the case when $\mathrm{Kn}_{\rm e} \sim 1$ and the dust velocity correlations can depart strongly from those of the gas, even when the mean velocity of the gas and dust remain tightly coupled.


\section{Hyperbolic structure and linear waves} \label{inerial hyperbolic}

In this section we will rearrange the equations into hyperbolic form, which is useful for some types of numerical solver and for understand the wave modes in the system. We wish to find a state vector $\mathbf{W}$, matrices $\mathbf{A}_i$ and source vector such that the dust-fluid equations take the form 

\begin{equation}
\frac{\partial \mathbf{W}}{\partial t} + \mathbf{A}^{\alpha} \overline{\nabla}_{\alpha} \mathbf{W} = \mathbf{S} ,
\end{equation}
To start we rearrange the equations so that all the source/sink terms are on the right hand side,

\begin{align}
\dot{\rho}_d +& U^{\alpha} \overline{\nabla}_{\alpha} \rho_d + \rho_d \overline{\nabla}_{\alpha} U^{\alpha} = 0 , \\
\dot{T}_{\alpha \beta} +& U^{\gamma} \overline{\nabla}_{\gamma} T_{\alpha \beta} + 2 T_{\gamma (\alpha} \overline{\nabla}^{\gamma} U_{\beta)} + T_{\alpha \beta} \overline{\nabla}_{\gamma} U^{\gamma} = -2 \left( T^{\gamma}_{\;\; (\alpha} C^{\;}_{\beta) \gamma} - \rho_d D_{\alpha \beta} \right) , \label{inerial constituative hyperbolic} \\
\dot{U}_{\alpha} +& U^{\beta} \overline{\nabla}_{\beta} U_{\alpha} + \frac{1}{\rho_d} \overline{\nabla}^{\beta} T_{\beta \alpha} = F_{\alpha} - C_{\alpha \beta} (U^{\beta} - U_g^{\beta}) , \label{inerial momentum hyperbolic} 
\end{align}
where we have exchanged the momentum and constitutive relation as it will make $\mathbf{A}^{\alpha}$ easier to diagonalise. The state vector for this system is

\begin{equation}
\mathbf{W} \quad = \quad \begin{pmatrix}
\rho_d \\
\mathbf{T}\\
\mathbf{U}
\end{pmatrix} ,
\end{equation}
The source vector is


\begin{equation}
\mathbf{S} \quad = \quad \begin{pmatrix}
0 \\
- \mathbf{C} \mathbf{T} - \mathbf{T} \mathbf{C}^{\mathbf{T}} + 2 \rho_d \mathbf{D} \\
\mathbf{F} - \mathbf{C} (\mathbf{U} - \mathbf{U}_g) 
\end{pmatrix} .
\end{equation}
The matrices $\mathbf{A}^{\alpha}$ are given by

\begin{equation}
\mathbf{A}^{\alpha}  \quad = \quad \begin{pmatrix}
U^{\alpha} & \mathbf{0} & \rho_d \hat{\mathbf{e}}^{\alpha}  \\
0 & U^{\alpha} \mathbf{I} & \mathbf{M}^{\alpha} \\
0 & \frac{\mathbf{I}}{\rho_d} \hat{\mathbf{e}}^{\alpha} & U^{\alpha} \mathbf{I} 
\end{pmatrix} , \label{A mat eq}
\end{equation}
where $\mathbf{I}$ denote the identity matrix and

\begin{equation}
(\mathbf{M}^{\alpha} )_{\sigma \beta \gamma} = T_{\sigma \beta} \delta_{\gamma}^{\alpha} + 2 T^{\alpha}_{\;\; ( \sigma } \delta_{\beta)}^{\gamma} ,
\end{equation}
such that 

\begin{equation}
(\mathbf{M}^{\alpha} \overline{\nabla}_{\alpha} \mathbf{U})_{\sigma \beta} = (\mathbf{M}^{\alpha} )_{\sigma \beta \gamma} \overline{\nabla}_{\alpha} U^{\gamma} ,
\end{equation}

For the system to be hyperbolic we must show that all the eigenvalues of $A^{\alpha} \hat{n}_{\alpha}$ are real for unit vector $\hat{n}_{\alpha}$, and the eigenvectors span the 28 dimensional state-space. Without loss of generality we can orient our coordinate system such that $\hat{\mathbf{n}} = \hat{\mathbf{e}}^1$ to point along the positive x-direction. Physically we must remember that the dummy gas and position dimensions are distinct, however we do not need to consider the case where $\mathbf{n}$ has non-zero components in the `dummy gas' directions as we require physical quantities to be independent of $\mathbf{x_{\rm g}}$.

It is useful to separate out the velocity into the velocity along the $x$ direction (along the direction of propagation), $U_1$, and the velocity in the other directions, $U_{\alpha}$ (where, for the rest of this section, we take the indices $\alpha,\beta$ and $\gamma$ to run over $2,\cdots,6$.). We similarly separate out the stress tensor into compression along the $x$ direction, $T_{1 1}$, shear components in the $x$ direction $T_{1 \alpha}$ and the components of the stress in other directions, $T_{\alpha \beta}$. We shall split the momentum and constitutive relation in a similar manor This results in the following state vector

\begin{equation}
\mathbf{W} \quad = \quad \begin{pmatrix}
\rho \\
T_{1 1} \\
T_{1 \alpha} \\
T_{\alpha \beta} \\
U_{1} \\
U_{\alpha}
\end{pmatrix} ,
\end{equation}


The eigenvalues, $v$, for $A^{\alpha} n_{\alpha}$ can be derived from the determinant of the following matrix

\begin{equation}
\hspace{10000pt minus 1fil}  \mathbf{A}^{1} - v \mathbf{I}  \quad = \quad \begin{pmatrix}
u^{x} - v & 0 & \mathbf{0} & \mathbf{0} & \rho_d & \mathbf{0} \\
0 & u^{x} - v & \mathbf{0} & \mathbf{0} & 3 T_{1 1} & \mathbf{0} \\
\mathbf{0} & \mathbf{0} & (u^{x} - v) \mathbf{I} & \mathbf{0} & 2 T_{1 \alpha} & T_{1 1} \mathbf{I} \\
\mathbf{0} & \mathbf{0} & \mathbf{0} & (u^{x} - v) \mathbf{I} & T_{\alpha \beta} & 2 T^{\;}_{1 (\alpha} \hat{\mathbf{e}}^{\mathbf{T}}_{\beta)} \\
0 & \frac{1}{\rho_d} & \mathbf{0} & \mathbf{0} & (u^{x} - v) & \mathbf{0} \\
\mathbf{0} & \mathbf{0} & \frac{1}{\rho_d} \mathbf{I} & \mathbf{0} & \mathbf{0} & (u^{x} - v) \mathbf{I}
\end{pmatrix} ,
\end{equation} 
which has a characteristic equation

\begin{equation}
(v - u^{x})^{16} \left( (v - u^{x})^2 - 3 \frac{T_{1 1}}{\rho_d} \right) \left( (v - u^{x})^2 - \frac{T_{1 1}}{\rho_d}  \right)^5 = 0 .
\end{equation}
This results in 16 nonpropagating (in the fluid frame) wavemodes, with wavespeed $v = u^{x}$. These consist of the entropy wave with eigenvector $\begin{pmatrix}
1 \\
0_{27}
\end{pmatrix} $
and 15 `stress' waves with eigenvectors $\begin{pmatrix}
0_7 \\
\hat{\mathbf{e}}_{\alpha \beta} \\
0_{6}
\end{pmatrix} $, where we have introduced the notation $0_n = \begin{matrix} 0 \\ \vdots \\ 0\end{matrix}$, with $n$ denoting the number of zeros in the column.


Two of the propagating waves can be identified as P-waves, with wavespeed $v = u^{x} \pm \sqrt{\frac{3 T_{1 1}}{\rho_d}}$. The P-waves are analogous to sound-waves, but with an anisotropic soundspeed, with seismic wavespeed anistropy being a well known phenomena in geophysics \citep{Thomsen86}. These wavemodes have eigenvectors,

\begin{equation}
\quad \begin{pmatrix}
\pm \rho_d \\
\pm 3 T_{1 1} \\
\pm 3 T_{1 \alpha} \\
\pm \left(T_{\alpha \beta} + \frac{2}{T_{1 1}} T_{1 (\alpha} T_{\beta) 1} \right) \\
- \sqrt{\frac{3 T_{1 1}}{\rho_d}} \\
-\sqrt{\frac{3}{\rho_d T_{1 1}}} T_{1 \alpha}
\end{pmatrix} ,
\end{equation}
Finally there are 10 propagating waves which can be identified as S-waves, with wavespeed $v = u^{x} \pm \sqrt{\frac{T_{1 1}}{\rho_d}}$. As is typical for elastic media, the S-waves have slower wavespeeds than the P-waves. These wavemodes have eigenvectors,

 \begin{equation}
\quad \begin{pmatrix}
 0 \\
 0 \\
 \pm T_{1 1} \hat{\mathbf{e}}_{\gamma} \\
 \pm 2 T^{\;}_{1 (\alpha} \delta^{\gamma}_{\beta)} \\
 0 \\
 -\sqrt{\frac{T_{1 1}}{\rho_d}} \hat{\mathbf{e}}_{\gamma} 
\end{pmatrix} .
\end{equation}
These eigenvectors span the 28-dimensional state-space of the dust fluid model.

Upon decomposing the velocity and pressure tensor, the source vector is given by 

\begin{equation}
\mathbf{S} \quad = \quad \begin{pmatrix}
 0 \\
 - 2 T^{1}_{\;\; 1} C^{\;}_{1 1} - 2 T^{\gamma}_{\;\; 1} C^{\;}_{1 \gamma}  \\
 - 2 T^{1}_{\;\; (1} C^{\;}_{\alpha) 1}  - 2 T^{\gamma}_{\;\; (1} C^{\;}_{\alpha) \gamma} \\
 - 2 T^{1}_{\;\; (\alpha} C^{\;}_{\beta) 1} - 2 T^{\gamma}_{\;\; (\alpha} C^{\;}_{\beta) \gamma} + 2 \rho_d D^{\;}_{\alpha \beta} \\
 F_{1}  - C_{1 1} (U^{1} - U_g^{1})  - C_{1 \gamma} (U^{\gamma} - U_g^{\gamma}) \\
 F_{\alpha} - C_{\alpha 1} (U^{1} - U_g^{1}) - C_{\alpha \gamma} (U^{\gamma} - U_g^{\gamma}) 
\end{pmatrix} ,
\end{equation}
Thus we see that there are no source terms for the entropy wave. Turbulent diffusion ($\mathbf{D}$) and the drag dependant coupling between pressure tensor components are sources/sinks of the stress waves. Finally the force per unit mass $\mathbf{F}$ and drag terms are sources/sinks of the P and S-waves. In practice whether the wavemodes can propagate in the dust fluid will depend on these source/sink terms as strong damping (such as by drag) may cause the waves to be evanescent in certain regions of parameter space.

While the aforementioned wavemodes represent all the waves present in the bulk. The dust fluid can support additional wavemodes when it occupies a thin layer, or other gravitationally confined structure. In such a situation the disc posses dust  breathing modes associated with periodic oscillations of the dust scale-height. These are analogous to the surface waves in seismology.


\section{Rheological Stress in a Rotating Shear Flow} \label{Steady state section}

\subsection{Steady state} \label{Steady state subsection}

In order to better understand the behaviour of the rheology, we consider the specific example of a steady rotating shear flow in the kinematic limit (i.e. we impose a rotation profile in the dust and gas and neglect the modification to the fluid flow from the resulting stress gradients). Rotating shear flows are of particular interest in astrophysics as they are important for understanding accretion discs. They are also a common setup in experimental fluid dynamics (e.g. Taylor-Couette flows). To study this problem we adopt (6-D) cylindrical polar coordinates $(R,\phi,z,R_g,\phi_g,z_g)$ with metric tensor components 

\begin{equation}
 g_{R R} = g_{R_g R_g} = g_{z z} = g_{z_g z_g} = 1, \quad g_{\phi \phi} = g_{\phi_g \phi_g} = R^2 .
\end{equation}
and connection coefficients,

\begin{equation}
 \Gamma_{\phi \phi}^{R} = \Gamma_{\phi \phi_g}^{R_g} = -R ,
\end{equation}

\begin{equation}
 \Gamma_{\phi R}^{\phi} = \Gamma_{R \phi}^{\phi}  = \Gamma_{\phi R_g}^{\phi_g}  = \Gamma_{R \phi_g}^{\phi_g}  = 1/R ,
\end{equation}
with all other components zero. The fluid flow consists of a rotating shear flow where both the dust and gas rotate on cylinders with angular velocity $\Omega = \Omega(R)$. This leads to the 6-D mean velocity of the dust fluid of

\begin{equation}
U^{\gamma} = \Omega (R) (\hat{e}^{\alpha}_{\phi} + \hat{e}^{\alpha}_{\phi_g}) .
\end{equation}
We additionally assume that the fluid is vertically homogeneous. By specifying that the dust mean velocity should exactly follow that of the gas we are implicitly taking the zero Eddy-Knudsen number limit. Thus the stress for small Stokes dusts is entirely specified by the velocity correlations in the gas.

With this geometry, and imposed velocity, the operator $\overline{\mathcal{D}}_2$ (when acting on $T_{\alpha \beta}$) is

\begin{equation}
 \overline{\mathcal{D}}_2 T_{\alpha \beta} = \overline{D} T_{\alpha \beta} + 2 T^{R}_{\;\; (\alpha} (\hat{e}_{\beta)}^{\phi} + \hat{e}_{\beta)}^{\phi_g}  ) \partial_{R} (R^2 \Omega) -2 \Omega \Gamma_{\phi (\alpha}^{\gamma} T^{\;}_{\beta) \gamma} - 2 R^2 \Omega \left(\Gamma_{\gamma (\alpha}^{\phi} + \Gamma_{\gamma (\alpha}^{\phi_g} \right) T_{\beta)}^{\;\; \gamma} .
\end{equation}
We are interested in the steady state solution to the stress tensor with $\overline{D} T_{\alpha \beta} = 0$. We thus have the following for the constitutive relation in the steady rotating shear flow,

\begin{equation}
4 R (\Omega - A) T^{R}_{\;\; (\alpha} (\hat{e}_{\beta)}^{\phi} + \hat{e}_{\beta)}^{\phi_g}  )  -2 \Omega \Gamma_{\phi (\alpha}^{\gamma} T_{\beta) \gamma} - 2 R^2 \Omega (\Gamma_{\gamma (\alpha}^{\phi} + \Gamma_{\gamma (\alpha}^{\phi_g}) T_{\beta)}^{\;\; \gamma} = -2 \left( T^{\gamma}_{\;\; (\alpha} C^{\;}_{\beta) \gamma} - \rho_d D^{\;}_{\alpha \beta} \right),
\label{rotating shear flow eq}
\end{equation}
where we have introduced Oort's first constant $A = -(R/2) \Omega_{R}$, which is a measure of the fluid shear rate. The Rayleigh stability criterion corresponds to $A/\Omega < 1$.



Explicitly, the ``dust-dust" components of Equation \ref{rotating shear flow eq}, which can be thought of as the equations governing the behaviour of the 3D dust stress, are

\begin{align}
 -\frac{4 \Omega}{R} T_{R \phi} &= -\frac{2}{t_s} (T_{R R} - T_{R R_g}) , \\
 - \frac{2 \Omega}{R} T_{\phi \phi} + 2 R (\Omega - A) T_{R R} &= -\frac{2}{t_s} T_{R \phi} + \frac{1}{t_s} (T_{R \phi_g} + T_{R_g \phi}) , \\
 4 R (\Omega - A) T_{R \phi} &= - \frac{2}{t_s} (T_{\phi \phi} - T_{\phi \phi_g}) . 
\end{align}
The ``dust-gas" components Equation \ref{rotating shear flow eq}, which govern the behaviour of the ``cross-stress" - from the cross correlation between the dust and gas velocities, are


\begin{align}
 -\frac{2 \Omega}{R} T_{R_g \phi} - \frac{\Omega}{R} (T_{R \phi} + T_{R \phi_g}) &= - \left(\frac{1}{t_s} + \frac{1}{t_c} \right) T_{R R_g} + \frac{1}{t_s} T_{R_g R_g} , \\
 -\frac{2 \Omega}{R} T_{\phi \phi_g} - \Omega R (T_{R R} - T_{R R_g}) + 2 R (\Omega - A) T_{R R} &= - \left(\frac{1}{t_s} + \frac{1}{t_c} \right) T_{R \phi_g} + \frac{1}{t_s} T_{R_g \phi_g} , \\
-\frac{\Omega}{R} (T_{\phi \phi} + T_{\phi \phi_g}) + 2 R (\Omega - A) T_{R R_g} &= - \left(\frac{1}{t_s} + \frac{1}{t_c} \right) T_{R_g \phi} + \frac{1}{t_s} T_{R_g \phi_g} , \\
2 R (\Omega - A) (T_{R \phi} + T_{R \phi_g}) + R \Omega (T_{R_g \phi} - T_{R \phi}) & = - \left(\frac{1}{t_s} + \frac{1}{t_c} \right) T_{\phi \phi_g} + \frac{1}{t_s} T_{\phi_g \phi_g} . 
\end{align}
Finally the  ``gas-gas" components of Equation \ref{rotating shear flow eq}, which describe the behaviour the (3D) gas Reynolds stress along the dust trajectory, are

\begin{align}
-\frac{2 \Omega}{R} (T_{R_g \phi_g} + T_{R_g \phi}) &= -\frac{2}{t_c} (T_{R_g R_g} - \alpha c_s^2 \rho_d) , \\
-\frac{\Omega}{R} (T_{\phi \phi_g} + T_{\phi_g \phi_g}) - \Omega R (T_{R R_g} - T_{R_g R_g}) + 2 R (\Omega - A) T_{R R_g} &= -\frac{2}{t_c} T_{R_g \phi_g} , \\
2 R \Omega T_{R_g \phi_g} + 2 R (\Omega - 2 A)  T_{R \phi_g} &= -\frac{2}{t_c} (T_{\phi_g \phi_g} - \alpha c_s^2 \rho_d R^2) .
\end{align}

It is straightforward, if rather laborious, to invert the above equations. However, the resulting expressions are somewhat cumbersome and not particularly informative. We shall instead use a symbolic algebra package to obtain expressions for the pressure tensor components that can be used in numerical computations (we provide a Mathematica script to do this in the supplementary materials). We can then numerically explore the behaviour of the dust pressure tensor.

\begin{figure}
\includegraphics[trim=100 30 220 50, clip, width=\linewidth]{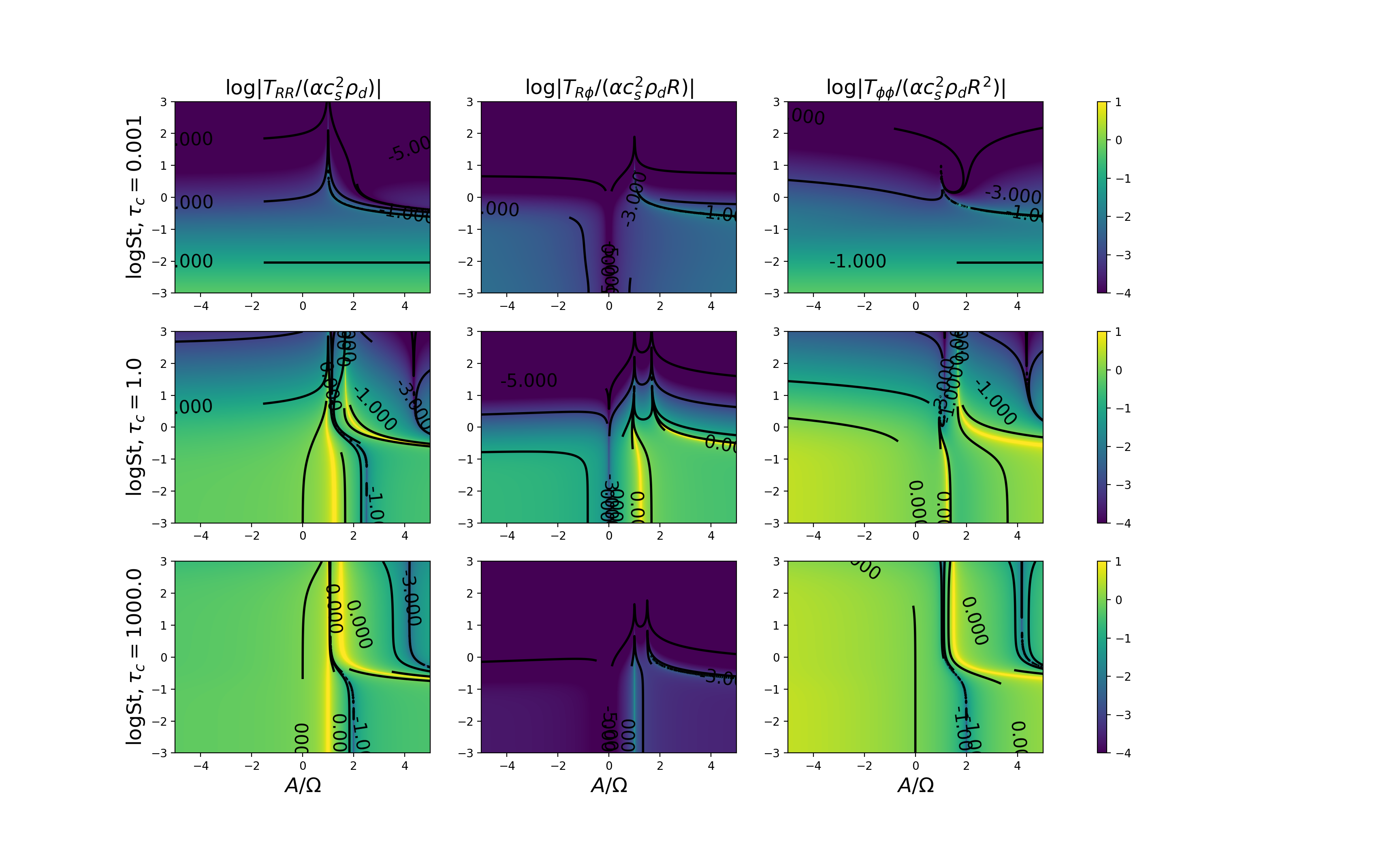}
 \caption{Horizontal stress tensor components in a rotating shear flow, as a function of $\mathrm{St}$ and $A/\Omega$, with different dimensionless correlation times}
 \label{A dependance}
\end{figure}



\begin{figure}
\centering
\begin{subfigure}{0.32 \linewidth}
\includegraphics[width=\linewidth]{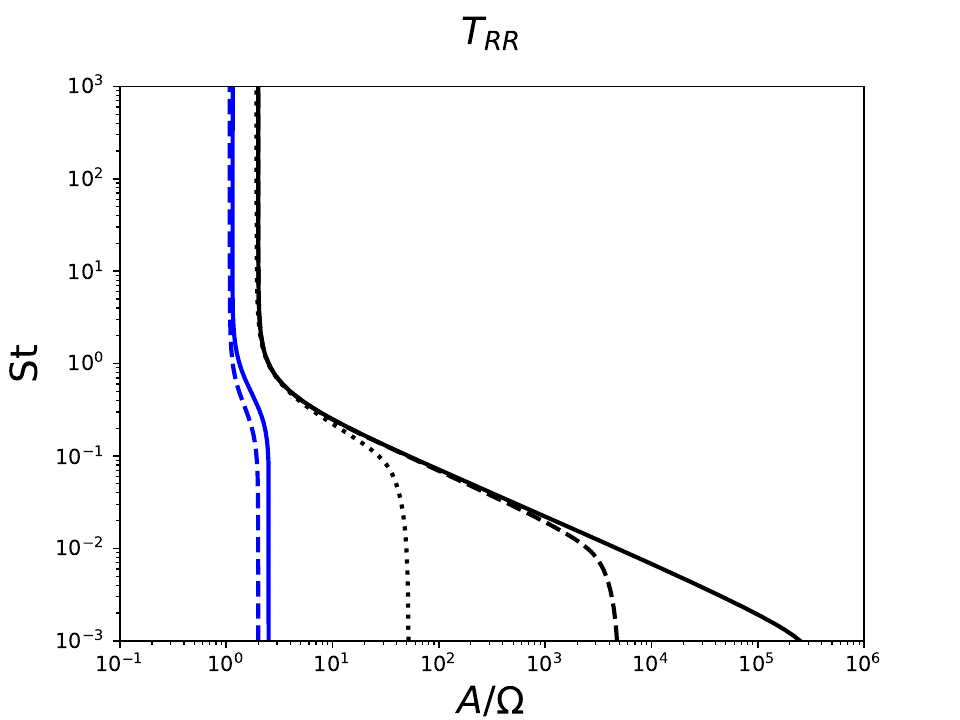}
\end{subfigure}
\begin{subfigure}{0.32 \linewidth}
\includegraphics[width=\linewidth]{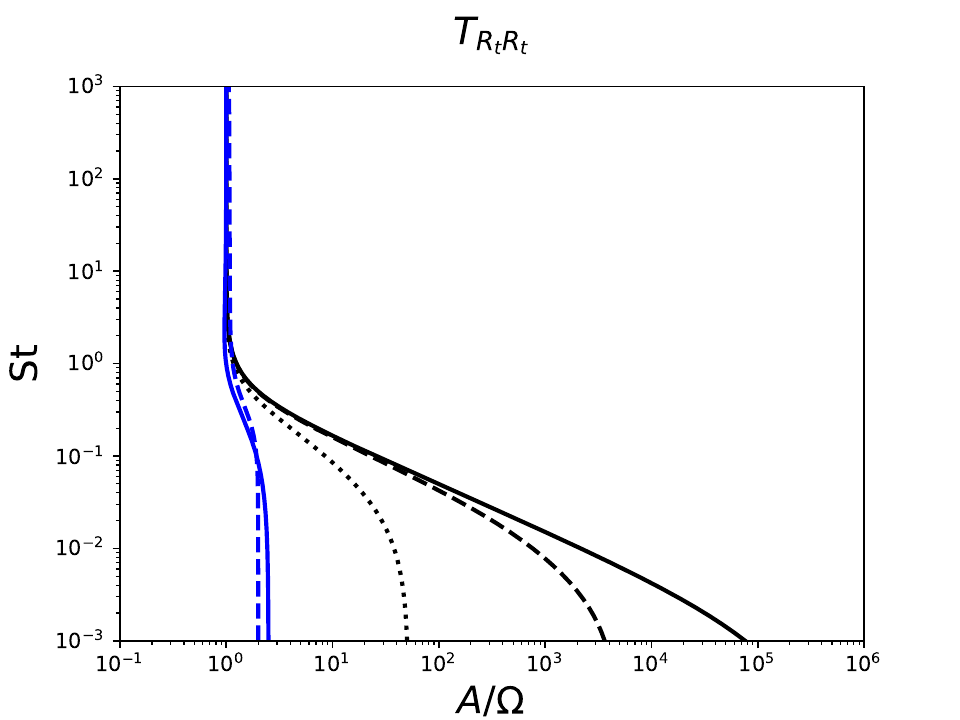}
\end{subfigure}
\begin{subfigure}{0.32 \linewidth}
\includegraphics[width=\linewidth]{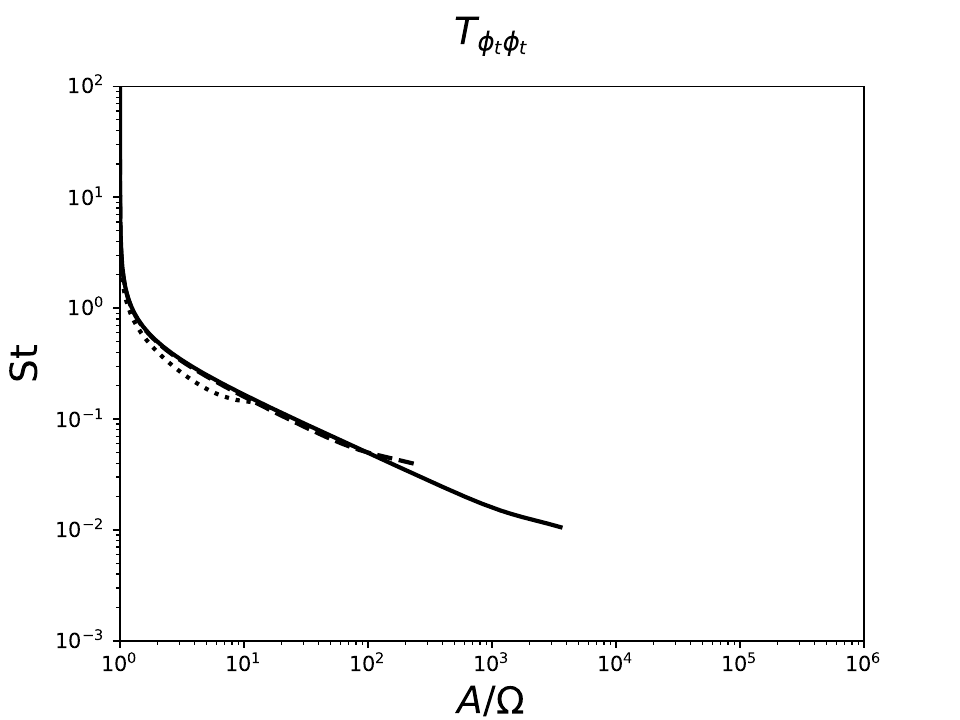}
\end{subfigure}
\caption{Locations in the parameter space of zeros of the stress tensor components, indicating a lack of steady solutions. Black Solid: $\tau_c=10^{-3}$, Black Dashed: $\tau_c=10^{-2}$, Black Dotted: $\tau_c=10^{-2}$, Blue Solid: $\tau_c=1$, Blue Dashed: $\tau_c=10^{3}$. The region of the parameter space above these lines contains no (physical) steady solutions. }
\label{pressure zeros}
\end{figure}

Figure \ref{A dependance} shows how the horizontal stress tensor components change with $A/\Omega$ and $\mathrm{St}$ for different dimensionless correlation times. Figure \ref{pressure zeros} shows the locations in the parameter space where stress tensor components pass through zero - indicating that the rotating shear flow no longer posses a steady solution. Together these shows the general behaviour of the dust stress tensor. This shows that the dust stress tends to get weaker at larger Stokes and tends to isotropy in the absence of shear $A/\Omega \rightarrow 0$. There are singularities/zeros of the pressure tensor at large $A/\Omega$ and $\mathrm{St}$ associated with the breakdown of the fluid dust description. For $\tau_c \lesssim 1$ these asymptote to the Rayleigh stability criterion for large $\mathrm{St}$, for small Stokes dust drag/cooling helps to regularise the behaviour of the pressure tensor allow for steady solutions at higher $A/\Omega$. For large $\tau_c$ small Stokes numbers are less effective at regularising the behaviour of the stress tensor and we see a zero of the stress tensor at $A/\Omega \sim 2$ for small Stokes. This occurs because of a breakdown of the turbulence model for large $\tau_c$ and $A/\Omega$ (See Appendix \ref{well mixed criteria} and \ref{Reynolds stress rotating shear flow}).



\begin{figure}
\centering
\begin{subfigure}{0.48 \linewidth}
\includegraphics[width=\linewidth]{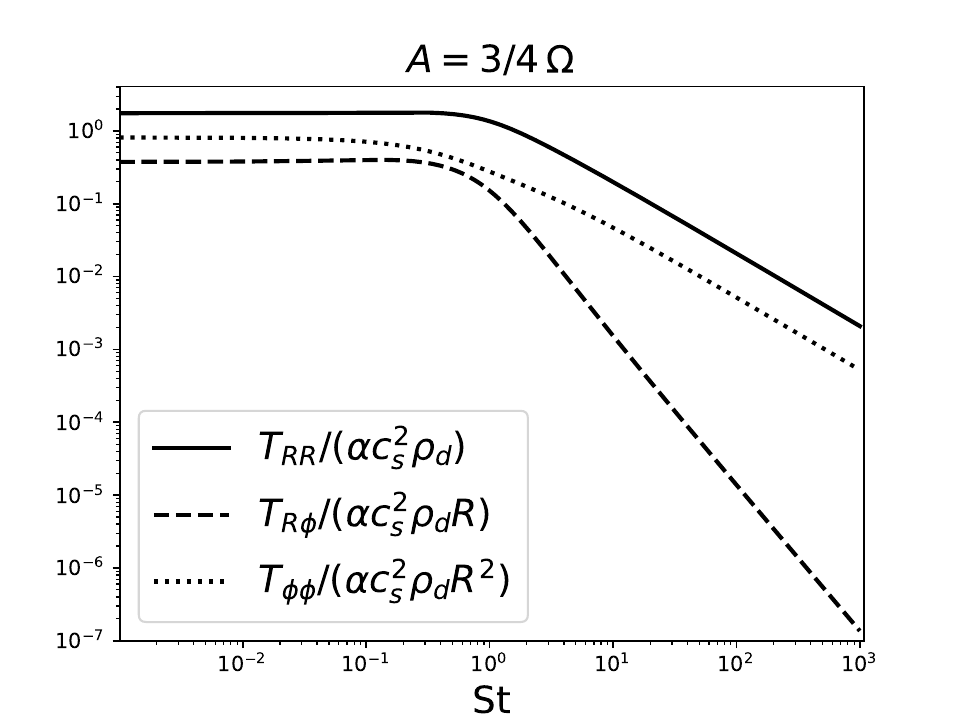}
\end{subfigure}
\begin{subfigure}{0.48 \linewidth}
\includegraphics[width=\linewidth]{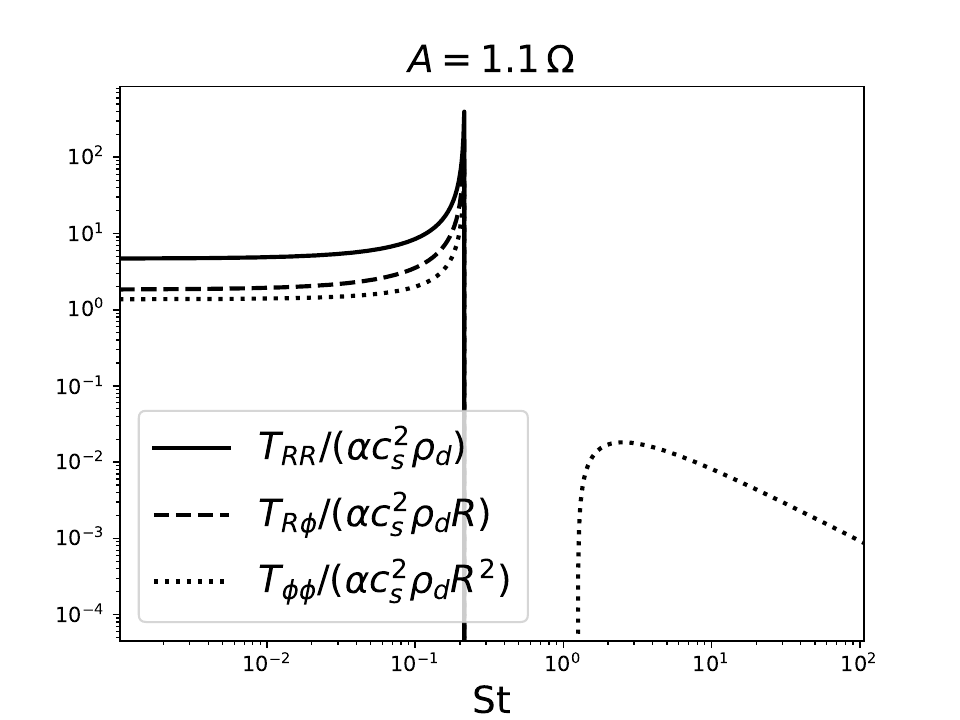}
\end{subfigure}
\caption{Horizontal stress tensor components for a fluid with a dimensionless correlation time, $\tau_c  = 1$. Left: A Rayleigh stable Keplerian shear flow where $\Omega \propto R^{-3/2}$ ($A = \frac{3}{4} \Omega$); Right: Rayleigh unstable shear flow with $A = 1.1 \Omega$. The gas Reynolds stress for these two cases is shown in Figure \ref{reynolds stress fig} in Appendix \ref{Reynolds stress rotating shear flow}. For the Keplerian shear flow the rheological stress tensor becomes increasingly anisotropic as the Stokes number increases. In the Rayleigh unstable case there is a maximum Stokes number above which the fluid dust description breaks down as the constitutive equation becomes thermally unstable.}
\label{pressure tensor fig}
\end{figure}

Figure \ref{pressure tensor fig} shows the stress tensor components for different Stokes numbers in a rotating shear flow with $\tau_c = 1$. The left plot shows the Rayleigh stable Keplerian shear flow with $A = \frac{3}{4} \Omega$ (The Reynolds stress of the gas associated with this flow is shown in the left hand plot of Figure \ref{reynolds stress fig} in Appendix \ref{Reynolds stress rotating shear flow}). For small Stokes numbers the tight coupling to the gas means the dust stress is set by the velocity correlations in the gas. As the Stokes number increases there is a competition between the isotropising effect of the turbulence and the shearing out of the radial component of the stress tensor, leading to an increasingly anisotropic flow. 

The right hand plot shows a Rayleigh unstable shear flow with $A = 1.1 \Omega$ (The Reynolds stress of the gas is shown in the right hand plot of Figure \ref{reynolds stress fig}). Again at low Stokes the dust stress is set by the gas velocity correlations. The stress tensor components diverge as the Stokes number increases, and one approaches the breaks down in the fluid description, before vanishing indicating a lack of accessible steady flow solution.




\begin{figure}
\centering
\begin{subfigure}{0.48 \linewidth}
\includegraphics[width=\linewidth]{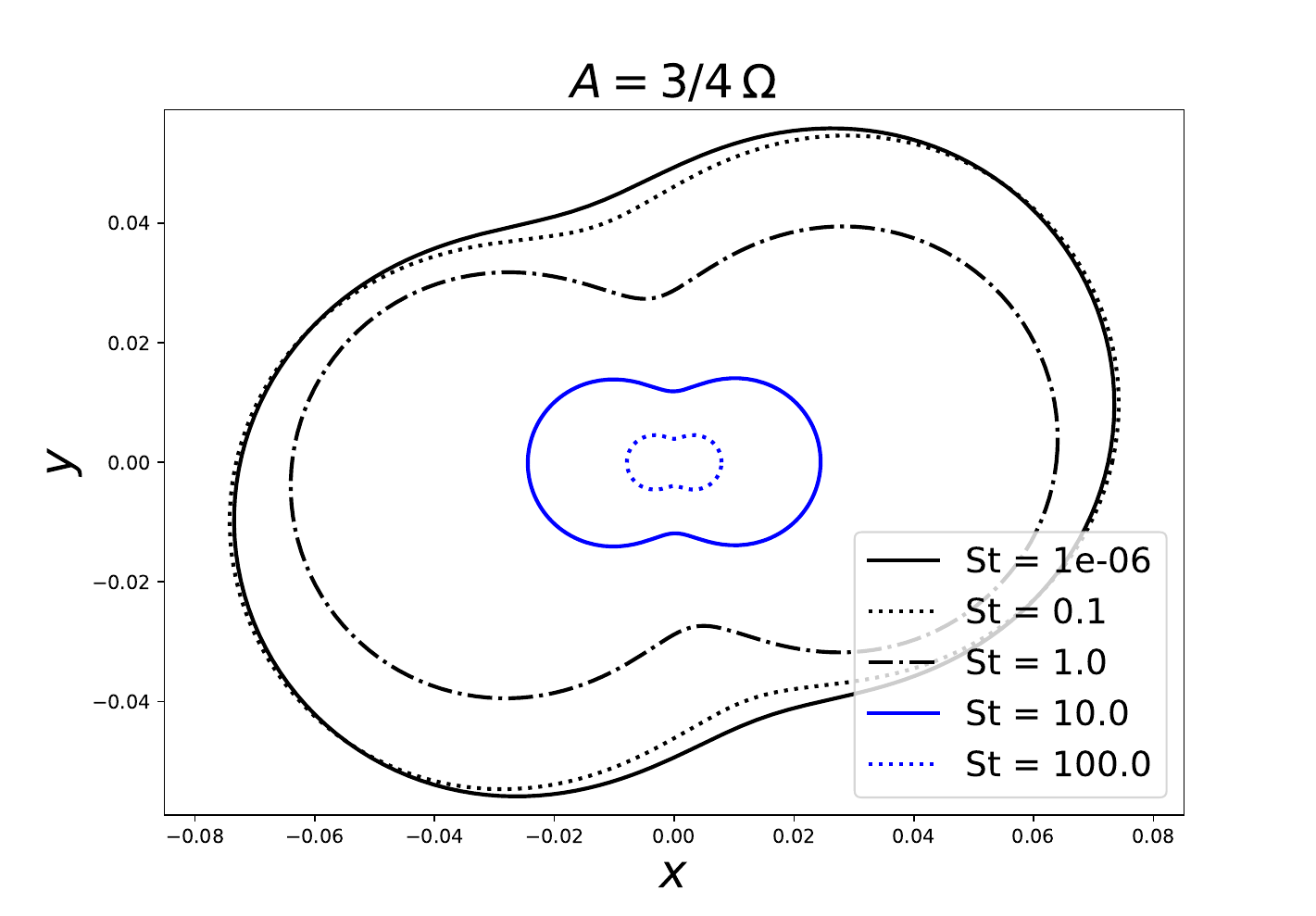}
\end{subfigure}
\begin{subfigure}{0.42 \linewidth}
\includegraphics[trim=10 10 30 30, clip,width=\linewidth]{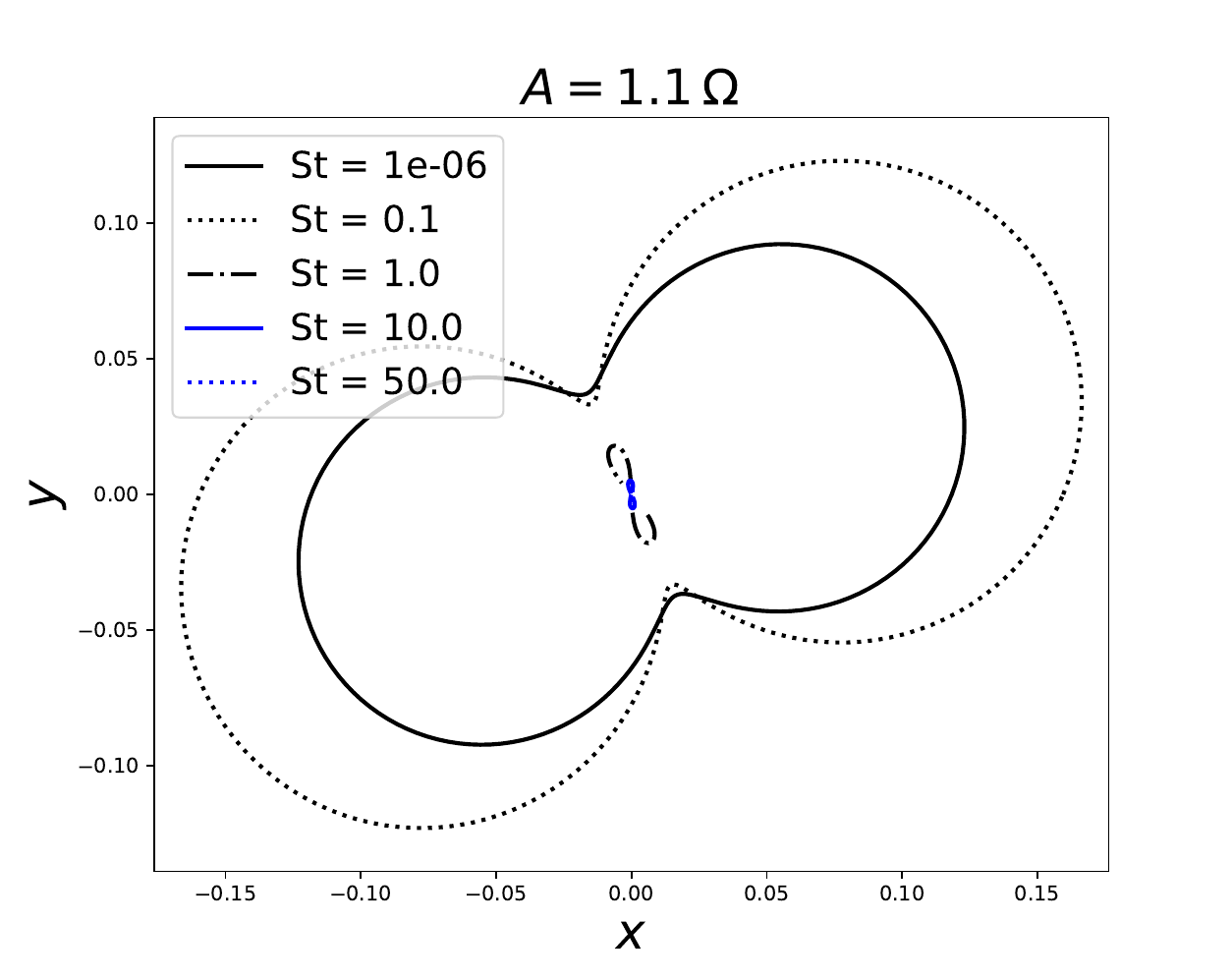}
\end{subfigure}
\caption{Illustration of the directional dependence of the P-wave velocity at different Stokes numbers for the disc considered in Figure \ref{pressure tensor fig}. The distance from the origin is proportional to the wavespeed of the P-wave, while the angle to the $x$-axis is the direction of propagation. The axes are aligned such that the x-axis is directed in the radial direction, while the the y-axis points in the direction of the fluid motion. On this figure the sound wave in the gas would be a circle of unit radius. For the Keplerian shear flow the P-waves slow down and preferentially propagate radially as the Stokes number increases. For the Rayleigh unstable flow the P-wave velocity is highly anisotropic, up to the breakdown in the fluid description where there is no longer a steady background on which the P-wave can propagate.}

\label{pwave fig}
\end{figure}

Figure \ref{pwave fig} shows how the speed of the P-wave varies with Stokes number and direction. S-waves velocities are the same as those of the P-waves but rescaled by $1/\sqrt{3}$. As the dust rheological stress becomes increasingly anisotropic the P-waves and S-waves propagate more radially than azimuthally. In the right hand plot of the Rayleigh unstable flow the P (and S) waves cannot propagate at large enough Stokes numbers - further increasing the Stokes number leads to a breakdown of the fluid description.

\subsection{Accretion flow solutions} \label{acc flow section}

Section \ref{Steady state subsection} gives the effects of the leading order velocity field on the pressure tensor in a rotating shear flow and neglects the presence of an accretion flow driven by the $R_{r\phi}$ Reynolds stress and the effects of gas pressure gradients. To study this effect we implement a 1D hydro-solver to solve the dust-fluid equations in aligned-cylindrical coordinates. We consider an axisymmetric dust flow around a Keplerian gravitational potential, neglecting the effects of vertical gravity. 

For the gas properties we solve Equations \ref{continuity total form}-\ref{constituative total form} perturbatively, with the leading order terms being that due to gravity and circular-Keplerian motion. We then consider the 1st order correction to the gas velocity due to the gas pressure and turbulence. This has the effect of driving a slow, radial accretion flow and making the gas rotation sub-Keplerian in the presence of a negative pressure gradient. We consider constant $\alpha$ and sound speed throughout.

As a starting point we consider the case of a constant gas surface density, and neglect the slow change in the gas density due to accretion. This is not self-consistent as the timescale on which the gas density evolves is typically expected to be comparable to the dust sound crossing time of interest. We also consider a more realistic example, where we solve for the steady-state, turbulent gas profile, resulting in a gas surface density of $\rho_g \propto R^{-3/2}$ Further details on the gas-flow considered are given in Appendix \ref{accretion  flow gas}.

We solve the dust fluid equations in aligned-cylindrical coordinates using an implementation of the HLL \citep{Harten83} and Roe \citep{Roe81} approximate-Reimann solvers, and constant reconstruction. We use an operator-split Van-Leer integrator \citep{vanLeer06,Toro99}, and use an RK(2)3 integrator to integrate the source terms. 

We take units such that the radius of the inner boundary is 1 and $GM=1$, resulting in the circular Keplerian frequency on the inner boundary also being unity. The domain spans $R \in [1,5]$. In these units we consider a gas disc with constant sound speed $c_=0.2$, and turbulence with $\alpha = 0.02$ and $\tau_c = 0.1$ or $\tau_c = 0.01$. This is adopted for computational convenience (principally difficulties with ensuring positivity of the stress tensor and so that the dust sound crossing time is not too long), and is not reflective of realistic disc turbulence (particularly for dust hosting discs). We consider dust with stopping time $t_s=0.01$ at the reference orbit $R=1$ and reference gas density $\rho_g=1$. For the constant gas density case we start with a constant dust density $\rho_d = 1$, while for the steady state we start with a step profile,

\begin{equation}
 \rho_d = 0.1 + 0.9 [\tanh(2 R + 1) + 1 ] .
\end{equation}
In both cases we take the initial dust velocity to be equal to the gas velocity and an isotropic dust stress with $T_{\alpha \beta} = \alpha c_s^2 g_{\alpha \beta}$. Notably this initial dust stress does not correspond to the anisotropic stress expected in a steady rotating shear flow (as shown in Section \ref{Steady state subsection}). We adopt zero-gradient boundary conditions with a diode inner boundary for $U^{R}$, i.e. we set $U^{R} (1) = 0$ whenever it is positive thereby only allowing an outflow on the inner boundary. We add wavekilling zones to our simulations, applying a large artificial viscosity near the boundary, which decreases to zero within a distance of 0.2 of the boundary. This is done to stop grid scale oscillations being excited by the boundary, particularly when using the Roe solver. For the constant gas density case we integrate for 1000 inner orbits, while for the steady state accretion flow we integrate for 3000 inner orbits. In both cases this does not reach a steady state due to the long relaxation time in the outer disc.


\begin{figure}
\centering
\begin{subfigure}{0.48 \linewidth}
\includegraphics[width=\linewidth]{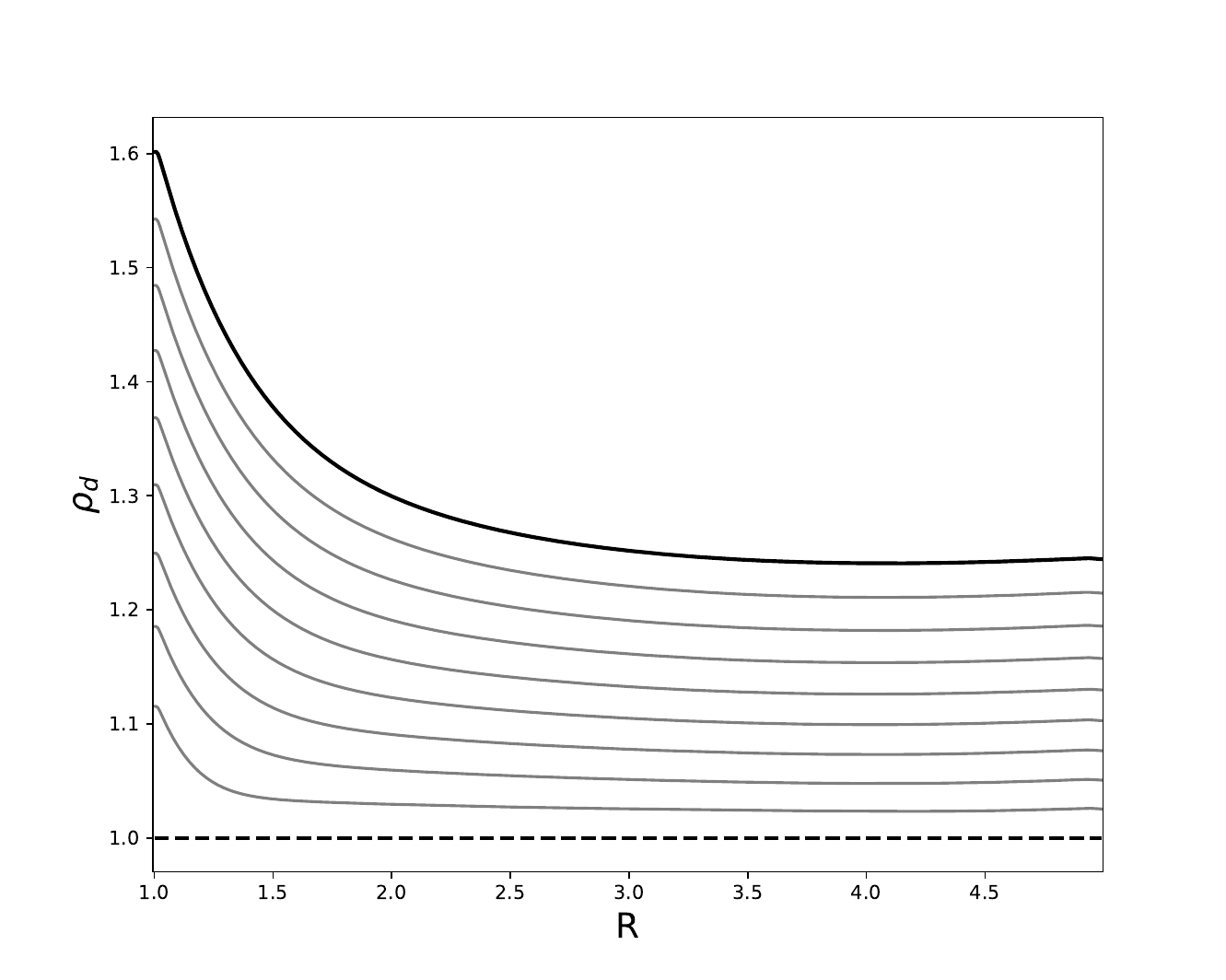}
\end{subfigure}
\begin{subfigure}{0.48 \linewidth}
\includegraphics[width=\linewidth]{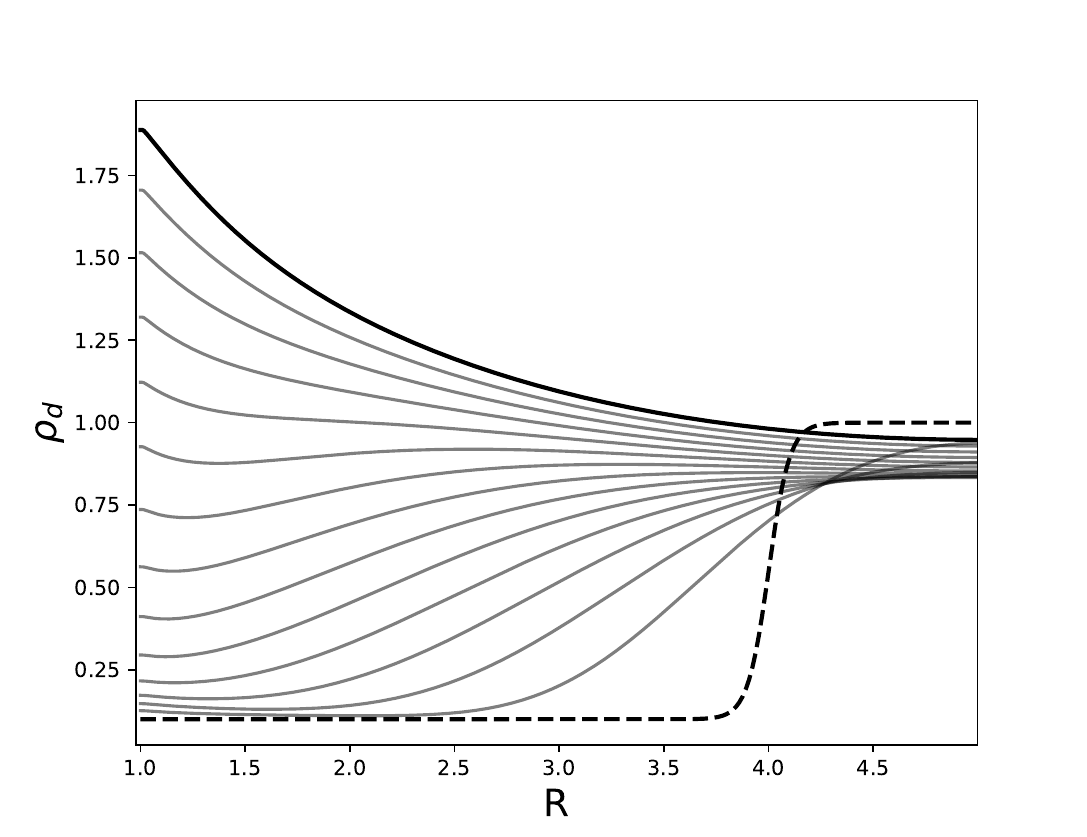}
\end{subfigure}
\caption{Density plots for the rotating shearflow, (left) constant density flow, (right) steady state accretion flow. Both are for Keplerian shear flows, solved with the HLL solver with $\tau_c=0.1$. The dashed line indicates the initial density profile, while the solid black line shows the final density profile. Grey lines show the evolution of the density profile, spaced every 100 inner orbits (left), and every 200 inner orbits (right).}
\label{den plot rot shear}
\end{figure}

\begin{figure}
\centering
\begin{subfigure}{0.48 \linewidth}
\includegraphics[width=\linewidth]{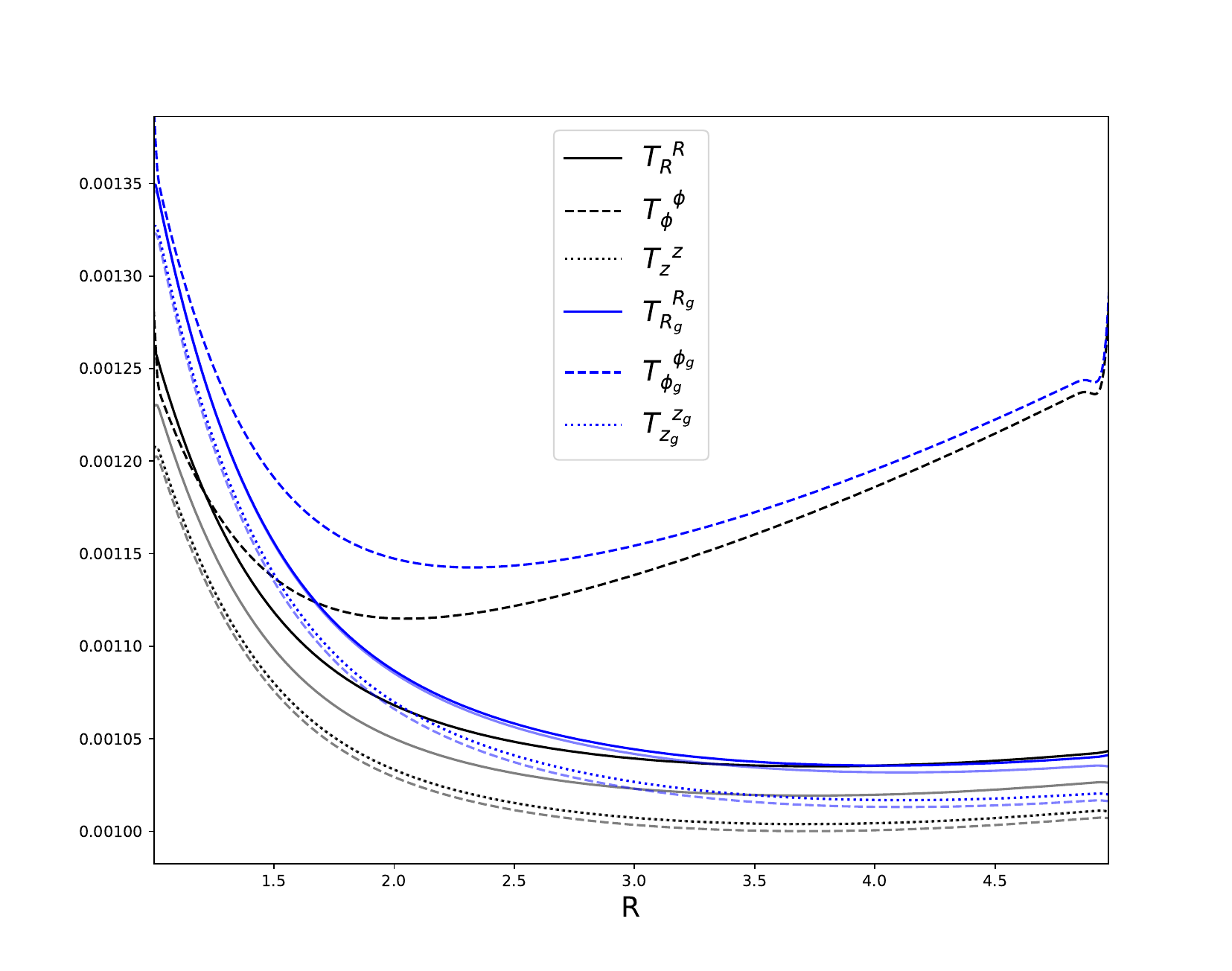}
\end{subfigure}
\begin{subfigure}{0.48 \linewidth}
\includegraphics[width=\linewidth]{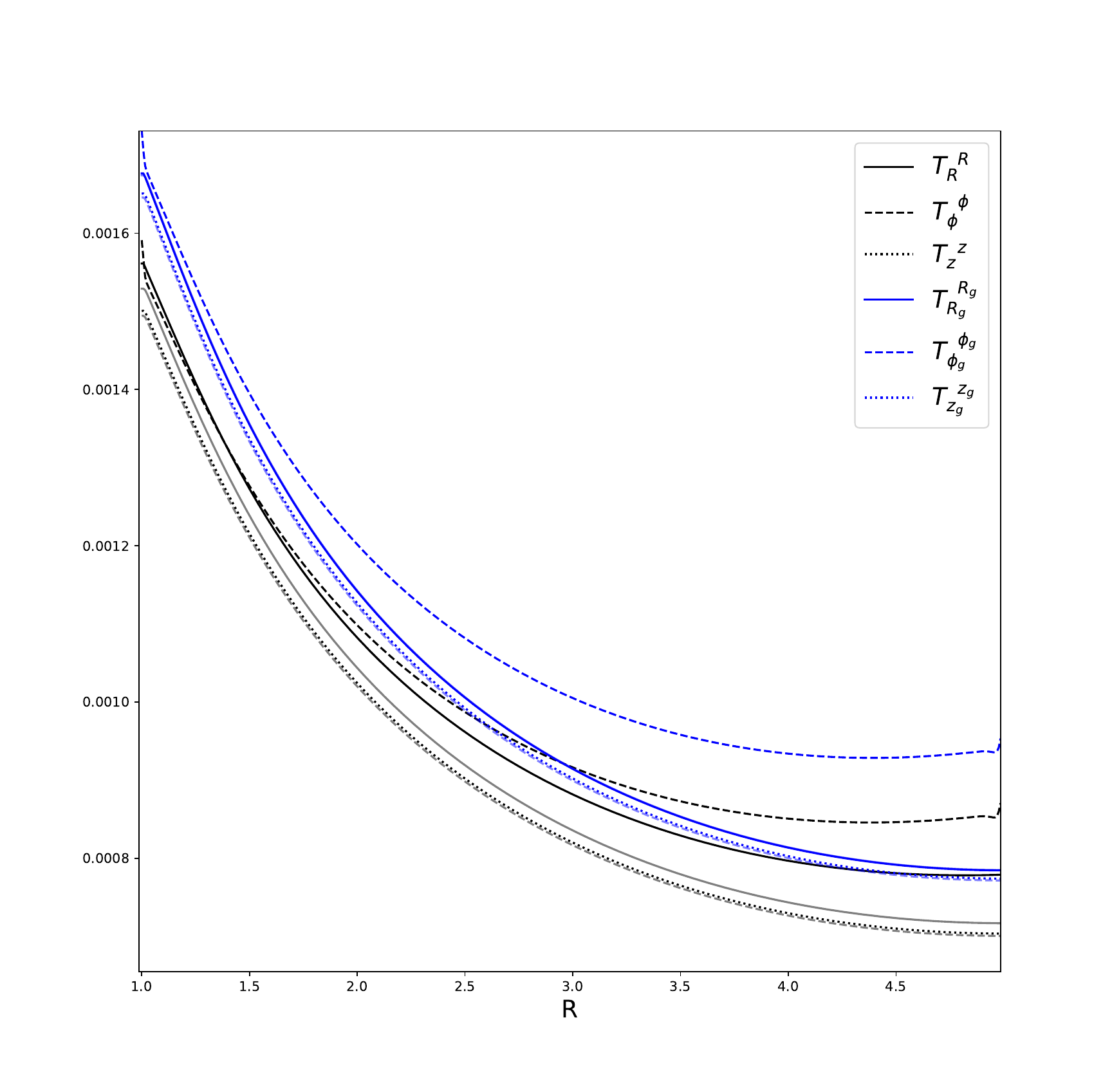}
\end{subfigure}
\caption{Diagonal dust-stress components for the flows considered in Figure \ref{den plot rot shear}. Full colour lines are the simulation, while greyed out/transparent lines are the solutions from Section \ref{Steady state subsection}. These are closer in the inner disc where the velocity is closer to Keplerian motion. The outer disc in the constant gas density case is still far from the steady state profile.}
\label{stress plot rot shear}
\end{figure}

Figure \ref{den plot rot shear} shows the density of both cases at different times, integrated with the HLL solver for $\tau_c = 0.1$. In the steady state case the initial dust step in the outer disc has drifted in due to the drag from the sub-Keplerian gas, before the dust density start to relax towards the steady state for the induced drift velocity Figure \ref{stress plot rot shear} shows the dust stress tensor for the final output of both these simulations, compared against the steady state solutions of Section \ref{Steady state subsection}. The Roe solver is not able to reach as large a correlation time as the HLL solver (which is more diffusive), however we carry out the same simulations with a correlation time of $\tau_c=0.01$ where we find agreement between the two solvers.

The simulations exhibit several of the expected feature of the dust fluid in an accretion disc. In the steady accretion flow, Figure \ref{den plot rot shear} show the initial step profile drifting inwards due to the action of gas drag with the sub-Keplerian gas flow. This is well established behaviour for dust in accretion flows and occurs even in the absent of dust-pressure. For the constant gas density case the dust is dragged inwards by the accretion flow and builds up on the boundary. This appears to be due to the adopted boundary conditions being partially permeable to the dust, with the zero-gradient in the radial velocity resulting in a lower outflow velocity than expected for a continuation of the accretion flow. The presence of this partial obstruction is then communicated into the disc by diffusion of the dust due to the gas turbulence. Preliminary tests simulating the dust fluid in accretion flows with a gas-pressure maximum suggests, as expected, the gas pressure maximum is no longer a dust trap. This differs from the behaviour found in pressureless dust models, with the dust now able to reach the inner boundary given enough time, This effect will have major implications for the transport of solids in protoplanetary discs and needs to be explored more thoroughly in the future.

The numerical implementation of the dust-fluid solver, presented here, is far from a practical implementation and requires numerous improvements to be useful. For most purposes the HLL solver is too diffusive and one should either further develop the Roe solver to be more stable at large correlation times or implement a HLLC \citep{Toro94} type solver. Currently the solver struggles to maintain positivity of the stress tensor, particularly $T_{\phi \phi}$,  at larger correlation times, with the correlation time reachable by the HLL solver being $\tau_c = 0.1$, rather than the $\tau_c=1-5$ expected of realistic disc turbulence. One possible reason for this is using the total second velocity moment, $T_{\alpha \beta} + \rho_d U_{\alpha} U_{\beta}$ as a conservative variable. This leads to errors due to the large difference in scale between the orbital motion, $\rho_d U_{\phi} U_{\phi}$ and the dust stress tensor component $T_{\phi \phi}$. On obvious improvement would be to implement a FARGO \citep{Masset00} like advection scheme and subtract off the orbital motion, to reduce the error on $T_{\phi \phi}$. Is is also possible that a finite difference scheme will perform better than the finite-volume Reimann solvers employed here when the orbital motion is treated in this manor \citep{BenitezLlambay16}. Finally much work needs to be done to determine appropriate boundary conditions, as the current zero-gradient boundaries excite grid scale oscillations that are currently dealt with by adding viscosity close to the boundary. This probably indicates that such a choice of boundary are ill posed in the dusty, rotating shear flow setting. 

\section{Discussion} \label{discussion}

In this paper we have chosen not to include the back reaction of the dust on the gas, despite it's importance even in dust poor flows. One can include back reaction in an ad-hoc manner by adding a dust drag term to $f_i^{\rm g}$,

\begin{equation}
 f_i^{\rm g} = - \nabla_i \phi - \frac{1}{\rho_g} \nabla_i p_g - \frac{1}{t_s} f_d (u_i^{\rm g} - u_i^{\rm d}) , \label{mean flow back reaction}
\end{equation}
where $u_i^{\rm d}$ is the mean dust velocity and $f_d = \rho_d/\rho_g$ is the dust to gas density ratio. Here the back reaction on the gas is only between the mean velocities and does not account for the effect of the dust on the gas turbulence, through corresponding source terms in the evolutionary equation for the gas Reynolds stress. A more self consistent approach would be to allow back reaction with the stochastic gas and dust velocities, which leads to the obvious improvement on Equation \ref{mean flow back reaction} by replacing the mean flow back reaction term with $- \frac{1}{t_s} f_d (v_i^{\rm g} - v_i^{\rm d})$ \citep[e.g. as done in][]{Minier01b,Minier04,Minier15}. However, $\mathbf{v}^{\rm g}$ is the velocity of a fluid element (seen), while $\mathbf{v}^{\rm d}$ is the velocity of an individual dust particle, and one expects there to be multiple dust particles within a given fluid element - it is thus not clear whether this modification is self consistent \citep[see also the discussion in][]{Minier01b}.


As above, in addition to adding in the dust drag, dust loading can affect the fluid turbulence. An alternative way to include this effect is make $t_c$ and $\alpha$ dependant on collective dust properties, the most important effect likely being a dependence on the dust to gas ratio $f_d = \rho_d/\rho_g$.  It would, thus, be useful to have a more rigorous treatment of back reaction, this would also be important for ensuring total energy in the gas+dust system is conserved (particularly to ensure the turbulence doesn't act as an infinite source of energy).



A rigorous treatment of energy conservation allows for energy to be exchanged between the gas turbulence, gas thermal energy and dust mechanical potential energy, along with the mean flow of both phases. This is particularly important in the compressible setting as the pressure is dynamical, and affected by the gas temperature, rather than being a Lagrange multiplier enforcing incompressibility \citep[the coupling has been considered in the incompressible setting e.g. by][]{Fox14}. This coupling naturally leads to the damping of the gas turbulence due to dust loading as energy is transformed from the turbulent fluctuations into heating the dust and is then transformed to the gas thermal energy due to gas drag leading to turbulent gas motions being converted to heat on $\sim t_s f_d^{-1}$. Finally this more complete modelling of the energy exchange between the three stores of energy may allow for more complex phenomena like intermittence and predator-prey dynamics which are known to be important in many turbulence processes \citep[e.g.][]{Diamond94}. 


Our model considers a situation where dust-dust collisions are rare due to the low number density of the dust relative to the gas. As the dust density increases, dust-dust collisions can become important, particularly for small grains. Inclusion of dust-dust collisions would allow contact between the dust fluid theory and existing work on dust pressure in weakly collisionless dust systems \citep{Goldreich78,Borderies85,Araki86,Latter08,Larue23}. The inclusion of dust collisions in a SDE model for particle laden flows, and associate moment closure, has been studied in \citet{Innocenti19,Innocenti21} and \citet{Capecelatro16a,Capecelatro16b}. This includes a separation of the dust Reynolds stress from the pressure tensor, which is important when dust-dust collisions are included as the dust collision  velocity is principally sensitive to the particle velocity dispersion rather than the turbulent velocity dispersion \citep{Fox14,Capecelatro16b}. Gas kinetic effects can also be important for smaller dust grains in regions of low gas density, where finite Knudsen number effects become important. Here dust gas collisions are infrequent enough, and impart sufficient momentum on the dust grain, that are an additional source of stochasticity on top of the gas turbulence which will act to heat the dust. Both dust-dust collisions and kinetic gas effects are likely important in systems with very large dust to gas ratios - particularly when the gas is produced by sublimating/colliding dust. 



Finally more sophisticated models of gas turbulence's \citep[e.g.][]{Sawford91,Pope02} include two timescales (correlation time and  Kolmogorov time) and maybe used to derive finite Reynolds number effects (formally our model is for turbulence with an infinite Reynolds number). As Reynolds numbers in astrophysical (and geophysical) gasses are very large this effect is likely only important in a limited region of parameter space - but maybe needed to obtain the correct collisional velocity for small grains \citep[e.g. to reproduce the results of][]{Ormel07}, or for the experimental verification of the model.

\section{Conclusion} \label{conclusion}

In this paper we have derived a fluid model for collisionless dust in a turbulent gas, starting from a system of stochastic differential equations describing the motion of a single dust grain. To allow for the coordinate systems and geometries common to astrophysics, we have adopted a covariant form for our dust-fluid model. We show that the continuum mechanics properties of dust in a turbulent gas corresponds to a 6-dimensional anisotropic Maxwell fluid with a dynamically important rheological stress tensor. The 6-dimensional formulation keeps the dust and fluid seen velocities, and their respective moments, on the same footing. The coupling between the dust kinetic tensor, dust-gas cross pressure and fluid seen Reynolds stress are obtained from the advection of the 6-dimensional dust stress tensor by the fluid flow.

In summery our conclusions are:

\begin{itemize}
 \item We have developed a dust fluid model, using a closure valid in the accretion disc context, and demonstrated that the self-consistency of the moment truncation used to obtain the fluid description is closely related to the thermals stability of the fluid.
 \item Collisionless dust has a non-zero anisotropic rheological stress which can be dynamically important, such as in dusty atmospheres where the dust is in hydrostatic equilibrium between the dust stress, vertical gravity and the gas Reynolds stress.
  \item The dust can support seismic (P and S) waves 
 \item Whether velocity correlations of small dust grains are set by the gas velocity correlations is determined by a form of Eddy-Knudsen number. Which can lead to small dust grains not being well mixed with the gas.
\end{itemize}

We have suggested several potential extension to our model some of which we intend to pursue in future work. 



\backsection[Acknowledgements]{The authors wish to thank Richard Booth and Micha\"{e}l Bourgoin for invaluable help with the literature along with Henrik Latter, Tobias Heinamann, Geoffroy Lesur and Francesco Lovascio for discussions which greatly clarified the physical interpretation of our model. We also thank Cathie Clarke, Andrew Sellek and the group of the CRAL for useful discussions on this work. We thank all three referees for a thorough review, bringing several suggestions that put the manuscript into it's present form.}

\backsection[Funding]{ The authors would like to thank the European Research Council (ERC). This research was supported by the ERC through the CoG project PODCAST No 864965. This project has received funding from the European Union’s Horizon 2020 research and innovation programme under the Marie Skłodowska-Curie grant agreement No 823823. }

\backsection[Declaration of interests]{The authors report no conflict of interest.}

\backsection[Data availability statement]{The data underlying this article will be shared on reasonable request to the corresponding author.}



\appendix

\section{Model for the gas} \label{gas model appendix}

\subsection{Formulation}


In our model for the turbulent gas an individual fluid element evolves according to the following set of stochastic differential equations,

\begin{align}
d \mathbf{x} &= \mathbf{v} d t , \\
d \mathbf{v} &= \mathbf{f}_{g} d t - \frac{1}{t_c} (\mathbf{v} - \mathbf{u}) d t + \sqrt{\frac{2 \alpha}{t_c}} c_s d \mathbf{W} ,
\end{align}
where ($\mathbf{x}$, $\mathbf{v}$) are the position and velocity of the fluid elements, $\mathbf{F}$ is the force per unit mass on the gas in the absence of turbulence and the turbulence results in an Ornstein-Uhlenbeck walk around the mean fluid flow with correlation time $t_c$. $\mathbf{W}$ is a Wiener process, with $c_s$ the gas sound speed and $\alpha$ a dimensionless measure of the strength of the turbulence. The fluid flow is a member of a statistical ensemble of similar flows with each fluid element following a single realisation of the flow \citep{Pope85,Thomson87}.

The Fokker-Planck equation associated with these equations can be derived in a similar way to that of the dust,

\begin{equation}
 \frac{\partial p}{\partial t} + \frac{\partial}{\partial x^{i}} [v^i p] + \frac{\partial}{\partial v^{i}} \left[p f_g^{i} - \frac{1}{t_c} (v^i - u^i) p \right] = \frac{\alpha c_s^2}{t_c} \frac{\partial^2 p}{\partial v^2} .
\end{equation}
Taking the 0th, 1st and 2nd velocity moments of this equation we arrive at

\begin{align}
 \frac{\partial \rho}{\partial t} &+ \frac{\partial}{\partial x^i} [ u_i \rho] = 0 , \\
 \frac{\partial}{\partial t} [\rho  u_i] &+ \frac{\partial}{\partial x^j} [R_{i j} + \rho u_i u_j ] - \rho f^{g}_{i} = 0 ,\\
\frac{\partial}{\partial t} [R_{i j} + \rho u_i u_j ] &+ \frac{\partial}{\partial x^k} [R_{i j k} + 3 u_{(i} R_{j k)} + \rho  u_i  u_j u_k ] - 2 \rho u^{\;}_{(i} f^{g}_{j )}  + \frac{2}{t_c} R_{i j} = 2 \frac{\alpha c_s^2}{t_c} \rho g_{i j} .
\end{align}
These can be rearranged to obtain

\begin{align}
 D \rho & = -\rho \nabla_i u^i , \\
 \rho D u_i &= - \nabla^j R_{i j}  + \rho f^{g}_{i} , \\
(D + \nabla_k u^k) R_{i j} + 2 R_{k (i} \nabla^k u_{j)} &= - \nabla^{k} R_{i j k} -\frac{2}{t_c} \left ( R_{i j} - \alpha c_s^2 \rho g_{i j} \right)  , 
\end{align}
where, in this Appendix, $D = \partial_t + u^i \nabla_i$ is the Lagrangian derivative with respect to the gas flow. Our closure scheme for this model assumes $R_{i j k} = 0$. We shall show, in the next section, this can be justified on the basis of a near-Maxwellian ordering scheme for the velocity moments, similar to the dust. Finally using a similar argument to that presented in Appendix \ref{realisability of P}, for the dust, we can show that (when $R_{i j k} = 0$) $R_{i j}$ is positive semi-definite for positive semidefinite initial conditions. Thus the equations to be solved for the gas phase are


\begin{align}
 D \rho & = -\rho \nabla_i u^i , \label{continuity total form} \\
 \rho D u_i &= - \nabla^j R_{i j}  + \rho f^{g}_{i} , \\
(D + \nabla_k u^k) R_{i j} + 2 R_{k (i} \nabla^k u_{j)} &= -\frac{2}{t_c} \left ( R_{i j} - \alpha c_s^2 \rho g_{i j} \right)  . \label{constituative total form}
\end{align}

Equivalently, one can perform a Reynolds decomposition of this flow resulting in the equivalent set of equations, which are closer to the formulation of \citet{Thomson87},

\begin{align}
d \mathbf{x} &= (\mathbf{v}_t + \mathbf{u}) d t , \label{turbulent gas possition evo} \\
d \mathbf{v}_t &=- \frac{1}{t_c} (\mathbf{v}_t - \mathbf{v}_{\rm hs}) d t + \sqrt{\frac{2 \alpha}{t_c}} c_s d \mathbf{W} , \label{turbulent gas velocity evo}
\end{align}
where the total gas velocity is equal to the sum of the mean velocity $\mathbf{u}$ and turbulent velocity $\mathbf{v}_t$, $\mathbf{v} = \mathbf{u} + \mathbf{v}_t$. The mean velocity obeys the usual Reynolds averaged equation

\begin{equation}
 \rho D \mathbf{u} = \rho \mathbf{f}_g - \nabla \cdot \mathbf{R} ,
 \label{mean momentum mean turb split}
\end{equation}
and we have introduced $\mathbf{v}_{\rm hs} = \frac{t_c}{\rho} \nabla \cdot \mathbf{R} - t_c \mathbf{v}_t \cdot \nabla \mathbf{u}$. \citet{Minier14} has shown that these two formulations are equivalent.



\subsection{Justification of the closure scheme} \label{well mixed criteria}


In this section we will determining a closure for the gas phase in our model. This will exploit the separation of scales between the hypersonic background motion and the highly subsonic turbulence and show that there exist a well defined asymptotic scaling in which the departure from an anisotropic Maxwellian velocity distribution is small. This is similar to the near-Maxwellian ordering scheme for the dust fluid considered in Section \ref{near maxwellian scheme section}. As we did in the dust fluid one can obtain an evolutionary equation for the k-th velocity moment of the gas turbulence,





\begin{align}
 \begin{split}
 \mathcal{D}_1 R_{i_1 \cdots i_k} + k R_{j (i_1 \cdots i_{k-1}} \nabla^{j} u_{i_k)} &= - \nabla^{j} R_{i_1 \cdots i_{k} j} +\frac{k}{\rho} R_{(i_1 \cdots i_{k-1} } \nabla^{j} R_{i_k) j} \\
&- \frac{k}{t_c} \left( R_{i_1 \cdots i_k}  - (k - 1) \alpha c_s^2 R_{(i_1 \cdots i_{k-2}} g_{i_{k-1} i_{k} )} \right) , \label{R k-th moment}
 \end{split}
\end{align}
where we have introduced the differential operator $\mathcal{D}_1 = D + \nabla_i u^i$. 

Consider a high Mach-number gas flow with subsonic turbulence and introduce two (potentially) small parameters $\delta$, which is of order $1/\mathcal{M}$ with $\mathcal{M}$ the Mach number, and $\alpha < 1$ which is a measure of the strength of the turbulence. We introduce two lengthscales a long lengthscale $L= O(1)$ (with long lengthscale variable $\mathbf{x}$), and `short' lengthscale $l = O(\delta)$ (with short lengthscale variable $\boldsymbol{\xi}$). To leading order the gas has a Maxwellian velocity distribution where the mean velocity is $O(1)$ and the gas sound speed is $O(\delta)$. This ordering scheme is subtly different to the near-Maxwellian ordering scheme of the dust fluid as we generally have $L \le l \le L_{\rm dust}$. The small parameter $\epsilon$ in the dust fluid problem is $O(\alpha^{1/2} \delta)$ in the gas ordering scheme. Our ordering scheme will be valid provided that the turbulent velocities are small compared with the the typical fluid velocities and correspond to $\epsilon \ll 1$. This can either be due to the flow having a high Mach-number $(\delta \ll 1)$, as is typical in astrophysics, or when the turbulence is weak ($\alpha \ll 1$).

At leading order we consider a gas with a Maxwellian velocity distribution which varies on the long lengthscale $L$, but having a gas density that can vary on the short lengthscale $l$. At higher order the distribution function has a non-Maxwellian velocity component, which is allowed to vary on the short lengthscale. The nearly Maxwellian asymptotic scheme is

\begin{align}
R_{i_1 \cdots i_k} &= \alpha^{\mathrm{ceil}(k/2)} \delta^{k} \rho (\boldsymbol{\xi}, \mathbf{x}) W_{i_1 \cdots i_k} (\mathbf{x}) + \alpha^{\mathrm{ceil}((k + 1)/2)} \delta^{k+1} \Sigma_{i_1 \cdots i_k} (\boldsymbol{\xi}, \mathbf{x}) , \label{R gas ordering scheme} \\
\mathbf{u} &= \mathbf{u}_0 (\mathbf{x}) + \alpha^{2} \delta^{2} \mathbf{u}_1 (\boldsymbol{\xi}, \mathbf{x}) , \\
\nabla &= \delta^{-1} \frac{\partial}{\partial \boldsymbol{\xi}}+ \frac{\partial}{\partial \mathbf{x}}  , \label{delta gas ordering scheme}
\end{align}
where $W_{i_1 \cdots i_k}$ are the Maxwellian velocity correlations and have the same properties as to there dust counterparts and evolve according to

\begin{equation}
 D W_{i_1 \cdots i_k} = - k \left[ W^{j}_{ \;\; (i_1 \cdots i_{k-1}} B^{\;}_{i_{k} ) j} - (k - 1) W_{(i_1 \cdots i_{k-2}} D_{i_{k-1} i_{k})}  \right] . \label{kth maxwellian gas corr}
\end{equation}
As we did with the dust, we can absorb perturbations to the gas density, from the non-Maxwellian terms, into the definition of $\Sigma_{i_1 \cdots i_k}$. 

Substituting the ordering scheme (Equations \ref{R gas ordering scheme}-\ref{delta gas ordering scheme}) into Equation \ref{R k-th moment} we arrive at

\begin{align}
 \begin{split}
 \mathcal{D}_1 R_{i_1 \cdots i_k} &= \alpha^{\mathrm{ceil}(k/2)} \delta^{k} \rho D W_{i_1 \cdots i_k} + \alpha^{\mathrm{ceil}(k/2) + 2} \delta^{k + 2} u_1^{j} \frac{\partial}{\partial x^{j}} W_{i_1 \cdots i_k} + \alpha^{\mathrm{ceil}((k + 1)/2)} \delta^{k+1} \mathcal{D}_1 \Sigma_{i_1 \cdots i_k} \\
&= -\alpha^{\mathrm{ceil}((k+1)/2)} \delta^{k} \left( \frac{\partial}{\partial \xi^{j}} + \delta \frac{\partial}{\partial x^{j}}  \right) (\rho W_{i_1 \cdots i_k j}) -  \alpha^{\mathrm{ceil}(k/2) + 1} \delta^{k+1} \left( \frac{\partial}{\partial \xi^{j}} + \delta \frac{\partial}{\partial x^{j}}  \right) \Sigma_{i_1 \cdots i_k j} \\
&- k \delta^{k} \Biggl [  \alpha^{\mathrm{ceil}((k+1)/2)} \rho W_{i_1 \cdots i_{k}} + \alpha^{\mathrm{ceil}(k/2) + 1} \delta \Sigma^{\sigma}_{i_1 \cdots i_{k}}  \\
&- \left(\alpha^{\mathrm{ceil}((k+1)/2)}  W_{(i_1 \cdots i_{k-1}} + \alpha^{\mathrm{ceil}(k/2) + 1} \delta \rho^{-1} \Sigma_{(i_1 \cdots i_{k-1}} \right) \left( \frac{\partial}{\partial \xi^{j}} + \delta \frac{\partial}{\partial x^{j}}  \right) \left( \rho W_{i_k) j} + \alpha \delta \Sigma_{i_k) j} \right)  \\
&- (k-1) \left(\alpha^{\mathrm{ceil}(k/2)} \rho W_{(i_1 \cdots i_{k-2}} + \alpha^{\mathrm{ceil}((k + 1)/2)} \delta \Sigma_{(i_1 \cdots i_{k-2}} \right) g_{i_{k-1} i_{k})} \Biggr] ,
\end{split}
\end{align}
where, here, $D = \partial_t + u_0^{i} \nabla_{i}$ and we have made use of $\mathcal{D}_1 \rho_{g} = (D + \nabla_{i} u_0^{i}) \rho_{g} = 0$.

Making use of Equation \ref{kth maxwellian gas corr} for the evolution of $W_{i_1 \cdots i_k}$, along with the recurrence relation for $W_{i1 \cdots i_k}$ and rearranging we obtain and equation for the evolution of the non-Maxwellian part of the turbulent velocity moment,

\begin{align}
 \begin{split}
  \alpha^{\mathrm{ceil}((k + 1)/2)} &\mathcal{D}_1 \Sigma_{i_1 \cdots i_k} + \alpha^{\mathrm{ceil}(k/2) + 1} \frac{\partial}{\partial \xi^{j}}  \Sigma_{i_1 \cdots i_k j}  + k \Biggl [ \alpha^{\mathrm{ceil}(k/2) + 1} \Sigma^{\sigma}_{i_1 \cdots i_{k}} \\
&- \alpha^{\mathrm{ceil}((k+1)/2) + 1}  W_{(i_1 \cdots i_{k-1}} \frac{\partial}{\partial \xi^{j}} \Sigma_{i_k) j}  + \alpha^{\mathrm{ceil}((k+1)/2)} \rho W_{j (i_k} \frac{\partial}{\partial x^{j}} W_{i_1 \cdots i_{k-1})} \\
&- \alpha^{\mathrm{ceil}(k/2) + 1} \rho^{-1} \Sigma_{(i_1 \cdots i_{k-1}} \frac{\partial}{\partial \xi^{j}} \left( \rho W_{i_k) j} \right)  - (k-1) \alpha^{\mathrm{ceil}((k + 1)/2)} \Sigma_{(i_1 \cdots i_{k-2}} g_{i_{k-1} i_{k})} \Biggr ] \\
&= \delta \Biggl [ k \alpha^{\mathrm{ceil}((k+1)/2) + 1}  W_{(i_1 \cdots i_{k-1}} \frac{\partial}{\partial x^{j}} \Sigma_{i_k) j}  + k \alpha^{\mathrm{ceil}(k/2) + 1} \rho^{-1} \Sigma_{(i_1 \cdots i_{k-1}} \frac{\partial}{\partial x^{j}} \left( \rho W_{i_k) j} \right)  \\
&+ k \alpha^{\mathrm{ceil}(k/2) + 2} \rho^{-1} \Sigma_{(i_1 \cdots i_{k-1}} \left( \frac{\partial}{\partial \xi^{j}} + \delta \frac{\partial}{\partial x^{j}}  \right) \Sigma_{i_k) j} \\
&+\alpha^{\mathrm{ceil}(k/2) + 2} u_1^{j} \frac{\partial}{\partial x^{j}} W_{i_1 \cdots i_k}  +  \alpha^{\mathrm{ceil}(k/2) + 1}  \frac{\partial}{\partial x^{\sigma}} \Sigma_{\alpha_1 \cdots \alpha_k \sigma} \Biggr ] .
\end{split}
\end{align}
Keeping only leading order terms in $\delta$ then, for even $k$, we have

\begin{align}
 \begin{split}
  \mathcal{D}_1 & \Sigma_{i_1 \cdots i_k} + \frac{\partial}{\partial \xi^{j}}  \Sigma_{i_1 \cdots i_k j}  + k \Biggl [ \Sigma^{\sigma}_{i_1 \cdots i_{k}} - \rho^{-1} \Sigma_{(i_1 \cdots i_{k-1}} \frac{\partial}{\partial \xi^{j}} \left( \rho W_{i_k) j} \right)  - (k-1) \Sigma_{(i_1 \cdots i_{k-2}} g_{i_{k-1} i_{k})} \Biggr ] \\
&-\delta \Biggl [  k \rho^{-1} \Sigma_{(i_1 \cdots i_{k-1}} \frac{\partial}{\partial x^{j}} \left( \rho W_{i_k) j} \right) + \frac{\partial}{\partial x^{\sigma}} \Sigma_{\alpha_1 \cdots \alpha_k \sigma} \Biggr ] \\
&= \delta \alpha \Biggl [ k W_{(i_1 \cdots i_{k-1}} \frac{\partial}{\partial x^{j}} \Sigma_{i_k) j} + k \rho^{-1} \Sigma_{(i_1 \cdots i_{k-1}} \left( \frac{\partial}{\partial \xi^{j}} + \delta \frac{\partial}{\partial x^{j}}  \right) \Sigma_{i_k) j} + u_1^{j} \frac{\partial}{\partial x^{j}} W_{i_1 \cdots i_k}  \Biggr ]  ,
\end{split}
\end{align}
while for odd $k$, we have

\begin{align}
 \begin{split}
  \mathcal{D}_1 & \Sigma_{i_1 \cdots i_k} + \alpha \frac{\partial}{\partial \xi^{j}}  \Sigma_{i_1 \cdots i_k j}  + k \Biggl [ \alpha \Sigma^{\sigma}_{i_1 \cdots i_{k}} - \alpha  W_{(i_1 \cdots i_{k-1}} \frac{\partial}{\partial \xi^{j}} \Sigma_{i_k) j}  +  \rho W_{j (i_k} \frac{\partial}{\partial x^{j}} W_{i_1 \cdots i_{k-1})} \\
&- \alpha \rho^{-1} \Sigma_{(i_1 \cdots i_{k-1}} \frac{\partial}{\partial \xi^{j}} \left( \rho W_{i_k) j} \right)  - (k-1) \Sigma_{(i_1 \cdots i_{k-2}} g_{i_{k-1} i_{k})} \Biggr ] \\
&= \delta \alpha \Biggl [ k \rho^{-1} \Sigma_{(i_1 \cdots i_{k-1}} \frac{\partial}{\partial x^{j}} \left( \rho W_{i_k) j} \right) + k \alpha \rho^{-1} \Sigma_{(i_1 \cdots i_{k-1}} \left( \frac{\partial}{\partial \xi^{j}} + \delta \frac{\partial}{\partial x^{j}}  \right) \Sigma_{i_k) j}  + \frac{\partial}{\partial x^{\sigma}} \Sigma_{\alpha_1 \cdots \alpha_k \sigma} \Biggr ] .
\end{split}
\end{align}
In both cases the right-hand side can be dropped at leading order if either $\alpha$ or $\delta$ are small. This confirms that the asymptotic ordering scheme (Equations \ref{R gas ordering scheme}-\ref{delta gas ordering scheme}) is self-consistent and the non-Maxwellian terms are suppressed by a factor of $\sim \alpha/\mathcal{M}$ relative to the Maxwellian terms. For the purposes of the effect on the gas equation of motion one must consider the effect on the Reynolds stress gradients. For the nearly Maxwellian velocity distribution considered the gradients of the Reynolds Stress are

\begin{equation}
\nabla_{j} R^{i j} = \alpha \delta \left( \frac{\partial}{\partial \xi^{j}} + \frac{\partial}{\partial x^{j}} \right) \rho  W^{i j} + \alpha^{2} \delta^{2} \left( \frac{\partial}{\partial \xi^{j}} + \frac{\partial}{\partial x^{j}} \right) \Sigma_{i j} .
\end{equation}
Thus the effects of the non-Maxwellian terms are $O(\alpha^{2} \delta^2)$ and are thus small relative to the acceleration and gravity, which are taken to be $O(1)$, or the pressure gradients which are $O(\delta)$

As with the dust fluid,  the existence of a consistent asymptotic scaling does not guarantee that the it is an attractor. We shall not repeat the argument here, but one can follow a similar line of reasoning to Section \ref{condition on decay of kth moment section} to demonstrate that turbulent velocity moments which start far from the asymptotic scaling are expected to damp towards the scaling, subject to the equation governing the evolution of the Reynolds stress having a stable equilibrium. As with the dust fluid, this is not sufficient to completely show that the near-Maxwellian ordering is an attractor as it does not account for the possibility of (nonlinear) perturbations to multiple velocity moments mutually supporting each other against decay. As with the dust fluid a more complete analysis of when the near-Maxwellian ordering scheme acts as an attractor must be left for future work.

\subsection{Reynolds stress in a rotating shear flow} \label{Reynolds stress rotating shear flow}

In this section we shall derive the steady state Reynolds stress in a rotating shear flow. This will aid our discussion of the behaviour of the dust rheological stress in a rotating shear flow in Section \ref{Steady state section} along with illustrating some key properties of our turbulence model. The steady state Reynolds stress in the gas satisfies the following equation,

\begin{equation}
 \nabla_k \left(u^k  R_{i j} \right) + 2 R_{k ( i} \nabla^{k} u_{j)} = -\frac{2}{t_c} (R_{i j} - \alpha c_s^2 \rho_g g_{i j}) . 
 \label{steady state reynolds stress}
\end{equation}

\begin{figure}
\centering
\begin{subfigure}{0.48 \linewidth}
\includegraphics[trim = 0 0 50 0, clip,width=\linewidth]{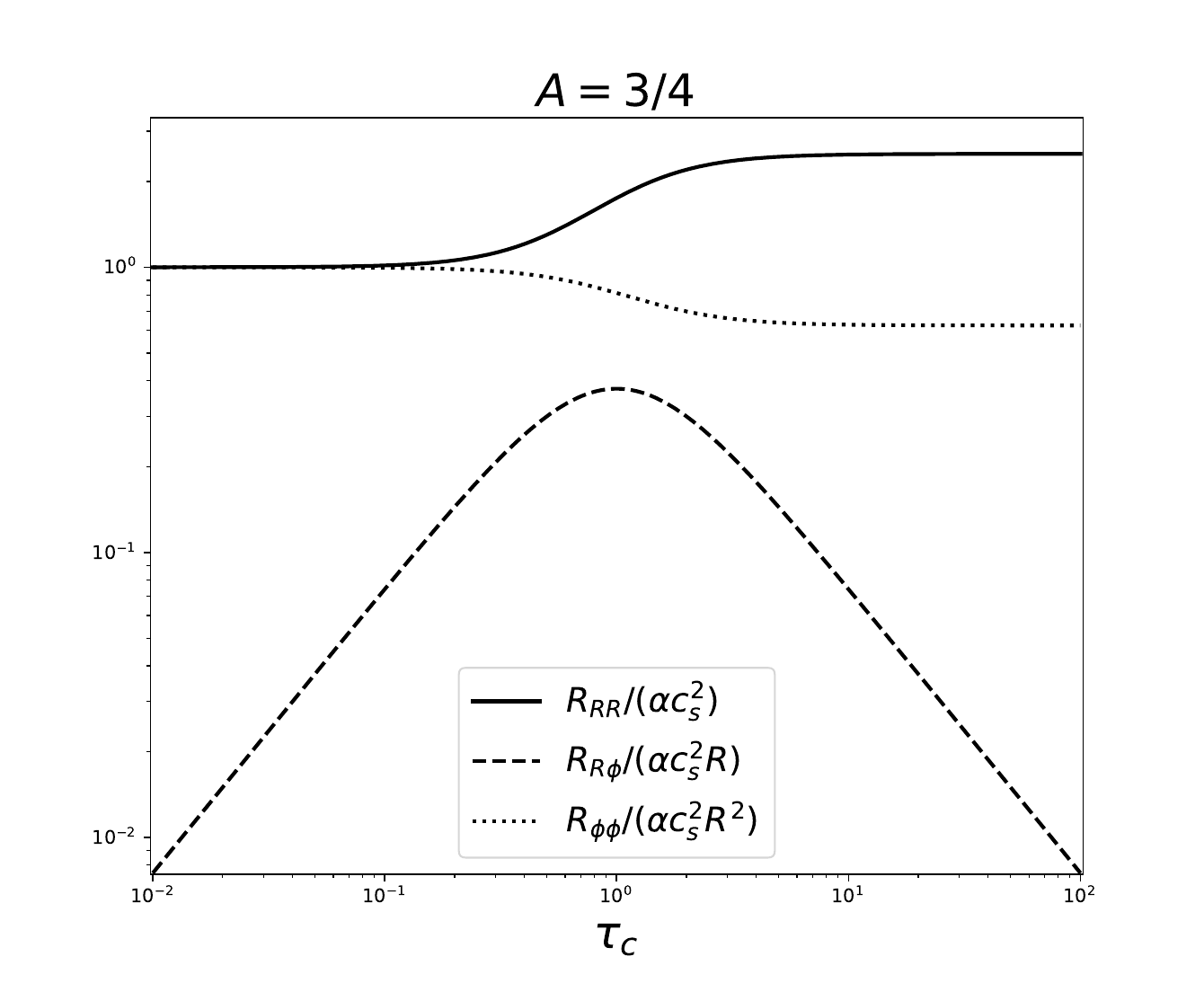}
\end{subfigure}
\begin{subfigure}{0.48 \linewidth}
\includegraphics[trim = 50 0 50 0, clip,width=\linewidth]{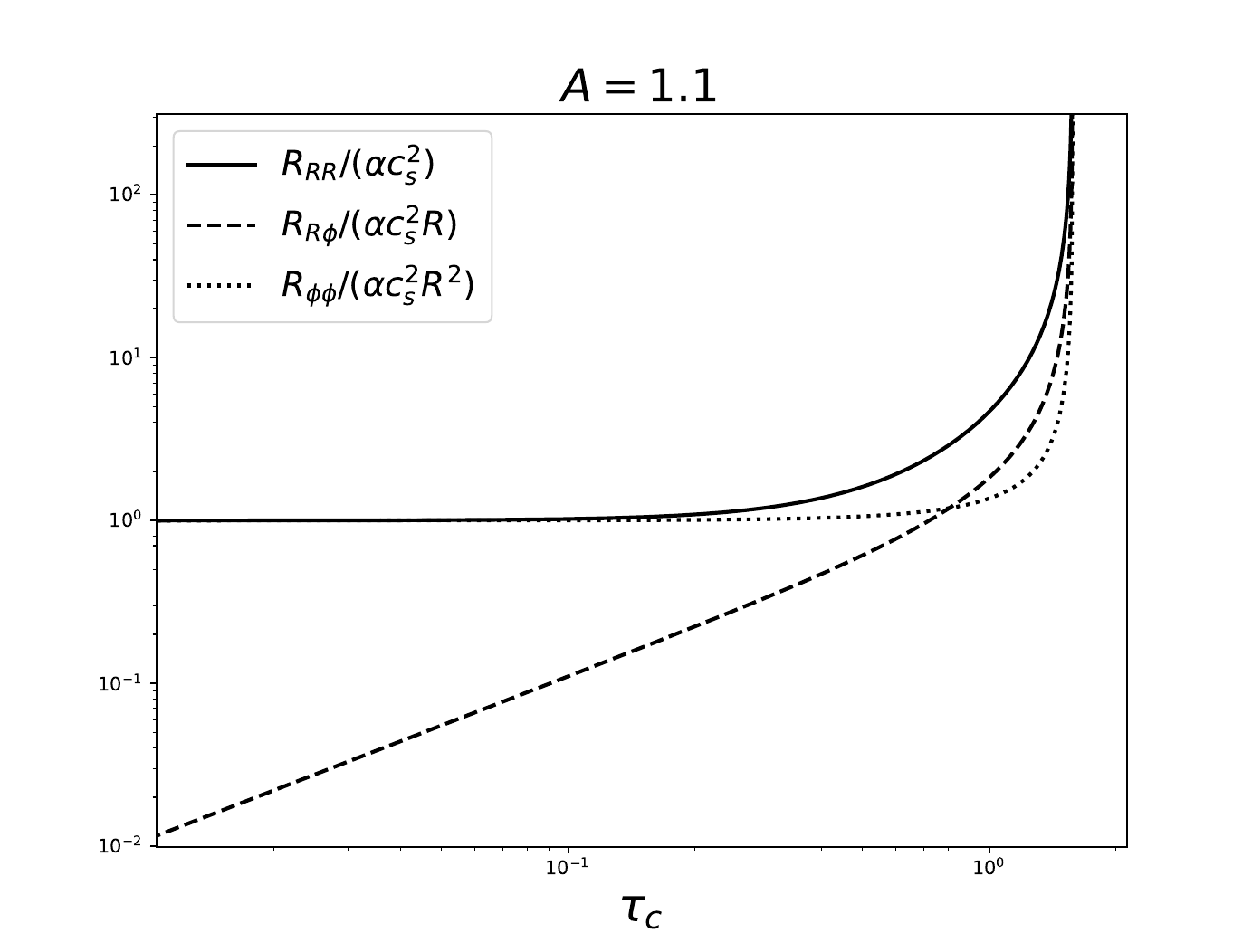}
\end{subfigure}
\caption{Reynolds stress components (in units of $\alpha c_s^2$) for two different rotating shear flows. The left is for the, Rayleigh stable, circular Keplerian rotational profile where $\Omega \propto R^{-3/2}$. Here the Reynolds stress becomes increasingly anisotropic as $\tau_c$ increases, although the cross term $R_{R \phi}$ is maximum near $\tau_c = 1$. The right is for a marginally Rayleigh unstable flow where $A = 1.1 \Omega$. Here we see that the Reynolds stress diverges as $\tau_c \rightarrow \sqrt{5/2}$, where the moment expansion used to derive the turbulent stress model breaks down.}
\label{reynolds stress fig}
\end{figure}

Adopting cylindrical polar coordinates $(R,\phi,z)$, with the usual expressions for $g_{i j}$ and $\Gamma_{i j}^k$. We consider a rotating shear flow with velocity, $u^{k} = \Omega (R) \hat{e}_{\phi}^k$. Substituting this into Equation \ref{steady state reynolds stress}, and neglecting gradients in $R_{i j}$ at leading order, we have

\begin{equation}
 - 2 \Omega \Gamma_{\phi (i}^{s} R^{\;}_{j) s} + 4 R (\Omega - A) R^{R}_{\;\; (i} \hat{e}_{j)}^{\phi}  - 2 R^{k}_{\; \; ( i} \Gamma_{j) k}^{\phi} R^2 \Omega = -\frac{2}{t_c} (R_{i j} - \alpha c_s^2 \rho_g g_{i j}) , 
\end{equation}
where we have in introduced Oort's constant $A = -\frac{R}{2} \Omega_R$. Explicitly the components of the Reynolds stress equations are 

\begin{align}
 - \frac{4 \Omega}{R} R_{R \phi} &= -\frac{2}{t_c} (R_{R R} - \alpha c_s^2 \rho_g ) , \\
 - \frac{2 \Omega}{R} R_{\phi \phi} + 2 R (\Omega - A) R_{R R} &= -\frac{2}{t_c} R_{R \phi} , \\
 4 R (\Omega - A) R_{R \phi}  &= -\frac{2}{t_c} (R_{\phi \phi} - \alpha c_s^2 \rho_g R^2) . 
\end{align}
We can rearrange these to obtain the following expressions for the Reynolds stress components in the rotating shear flow,

\begin{align}
 R_{R R} &=   \alpha c_s^2 \rho_g \frac{1 + 2 \tau_c^2 \left(2 - \frac{A}{\Omega}\right) }{1 + 4 \tau_c^2 \left(1 - \frac{A}{\Omega}\right)  }   , \label{Reynolds RR} \\
 R_{R \phi} &= \alpha c_s^2 R \left( \frac{A}{\Omega} \right) \frac{\tau_c}{1 + 4 \tau_c^2 \left(1 - \frac{A}{\Omega}\right)  } \rho_g  , \\
 R_{\phi \phi}  &= \alpha c_s^2  \rho_g R^2 \frac{1 + 2 \tau_c^2\left(1 - \frac{A}{\Omega}\right) \left(2 - \frac{A}{\Omega} \right) }{1 + 4 \tau_c^2 \left(1 - \frac{A}{\Omega}\right)  } . \label{Reynolds phi phi} \\
\end{align}
This solution breaks down in the presence of strong shear, when $\frac{A}{\Omega} \ge 1 + \frac{1}{4 \tau_c^2}$, as either $R_{R R}$ or $R_{\phi \phi}$ will be negative. This means there is no stable equilibrium for the Reynolds stress. As discussed in the previous section, this will also result in a breakdown of the moment expansion used to derive the equation governing the evolution of $R_{i j}$ as higher order moments can grow to become important. In Figure \ref{reynolds stress fig} we show the Reynolds stress components for the Rayleigh stable Keplerian shear flow and a Rayleigh unstable flow with $A = 1.1 \Omega$.

\subsection{Accretion flow solutions - gas} \label{accretion  flow gas}

In this section we derive the background gas solutions used in the numerical modelling of Section \ref{acc flow section}. Consider a Keplerian shear flow where the, circular, Keplerian motion is take to be order 1 and $u^{R}$, $\mathbf{R}$, $p$ and partial time derivative $\partial_t$ are $O(\epsilon)$ , for some small parameter $\epsilon$. We will neglect vertical gravity throughout, such that the Keplerian potential $\Phi = -1/R$ is a function of the Cylindrical radius, R, only (where we have adopted units in which $GM = 1$).  We consider velocity, in cylindrical coordinates, of the form

\begin{align}
 u^{R} &= \epsilon \tilde{u}^{R} , \\
 u^{\phi} &= \Omega_K +  \epsilon \tilde{u}^{\phi} , \\
 u^{z} &= 0 .
\end{align}
The gas equations for an axisymmetric, vertically invariant flow, neglecting vertical gravity are

\begin{align}
\dot{\rho} + R^{-1} \partial_{R} (R \rho_g \tilde{u}^{R}) &= 0 , \\
  2 \rho R \Omega_K \tilde{u}^{\phi} &= \partial_{R} p + (\nabla \cdot \mathbf{R})_{R} , \\
 \frac{1}{2} \rho \tilde{u}^{R} R \Omega_K   &= - (\nabla \cdot \mathbf{R})_{\phi} .
\end{align}
The vertical momentum equation is trivially solved by $u^{z} = 0$. The only time derivative remaining is that in the continuity equation, responsible for the slow accretion of the gas. At this order the Reynolds stress is given by Equation \ref{Reynolds RR}-\ref{Reynolds phi phi} and $R_{z z} = \alpha c_s^2 \rho_g$. For the Keplerian shear flow we take $A/\Omega = 3/4$ and $\alpha$, $c_s$ and $\tau_c$ to be constants.  Substituting these into the dust stress gradients we have

\begin{align}
 (\nabla \cdot \mathbf{R})_{R} &=  \frac{1}{2} \alpha \frac{ 2 + 5 \tau_c^2}{1 + \tau_c^2   }  \partial_{R} p  + \frac{15}{8}  \frac{p}{R} \frac{ \alpha \tau_c^2 }{1 + \tau_c^2 } , \\ 
 (\nabla \cdot \mathbf{R})_{\phi} &= \frac{3}{4} \frac{\alpha \tau_c}{1 + \tau_c^2 } R^{-1} \partial_{R} \left[ p R^2   \right] , 
\end{align}
noting that at this order $R_{R z} = R_{\phi z} = 0$. This results in the following corrections to the gas velocity,

\begin{align}
 \tilde{u}^{\phi} &= \frac{1}{2 \rho R \Omega_K} \left[\left(1 + \frac{1}{2} \alpha \frac{ 2 + 5 \tau_c^2}{1 + \tau_c^2   } \right) \partial_{R} p  + \frac{15}{8}  \frac{p}{R} \frac{ \alpha \tau_c^2 }{1 + \tau_c^2 }  \right] , \\
 \tilde{u}^{R}    &= - \frac{3}{2 \rho R^2 \Omega_K} \frac{\alpha \tau_c}{1 + \tau_c^2 } \partial_{R} \left[ p R^2   \right] .
\end{align}
We consider two scenarios. One where we neglect the gas continuity equation and consider a fixed dust density. This approximation can be reasonable if the dust drift timescale or the characteristic lengthscale of the dust fluid is short, however we adopt it here mostly for illustrative purposes. The second scenario is to solve for a steady accretion flow with $\dot{\rho} = 0$ this leads to a gas pressure profile of

\begin{equation}
  p = -\frac{4}{3} \mathcal{F} \frac{1 + \tau_c^2 }{\alpha \tau_c}  \Omega_K   + \mathcal{C} R^{-2},
\end{equation}
where $\mathcal{F}$ and $\mathcal{C}$ are constants. For our dust-fluid simulations in Section \ref{acc flow section} we take $\rho \propto R^{-3/2}$, which is compatible with the steady state pressure profile above.

\section{Properties of the Dust Fluid Model}

\subsection{Realisability} \label{realisability of P}

A necessary property of the constitutive relation is that the stress tensor be realisable from a second velocity moment of some distribution function. A similar property must hold for the dust Reynolds stress, the proof of which proceeds the same as the proof for the dust rheological stress tensor - to avoid repeating our selves we shall only cover the latter. As $\Pi_{\alpha \beta} = \int p (V_{\alpha} - U_{\alpha}) (V_{\beta} - U_{\beta})  d^6 \mathbf{V}$, $\Pi_{\alpha \beta}$ must be positive semi-definite. Thus for all positive semidefinite initial conditions $\Pi_{\alpha \beta} (0)$ our constitutive model most conserve the positive semi-definite character of $\Pi_{\alpha \beta}$. This is similar to the requirements for constitutive models of the MRI \citep{Ogilvie03,Lynch21}.
 

Following \citep{Lynch21} we introduce the quadratic form $Q = \Pi_{\alpha \beta} Y^{\alpha} Y^{\beta}$, if the stress tensor is positive semi-definite then $Q \ge 0$ for all vectors $Y^{\alpha}$ at all points in the fluid. We will show by contradiction that an initially positive semi-definite $\Pi_{\alpha \beta}$ cannot evolve into one that is not positive semi-definite. Suppose, to the contrary, that some point in the flow $Q < 0$ for some vector $X^{\alpha}$ at some time after the initial state. Let us consider a smooth, evolving vector field $Y^{\alpha}$ that matches the vector $X^{\alpha}$ at the given point and time. The corresponding quadratic form $Q$ is then a scalar field that evolves according to

\begin{align}
\begin{split}
\mathcal{D} Q &= Y^{\alpha} Y^{\beta} \mathcal{D}_2 \Pi_{\alpha \beta} + \Pi_{\alpha \beta} \mathcal{D}_2 (Y^{\alpha} Y^{\beta}) \\
&= - 2 Y^{\alpha} Y^{\beta} \left( \Pi^{\gamma}_{\;\; \alpha} \bar{A}_{\beta \gamma} - \rho_d D_{\alpha \beta} \right) + \Pi_{\alpha \beta} \mathcal{D}_2 Y^{\alpha} Y^{\beta} \quad , \\
\end{split}
\end{align}
where, when operating on $Y^{\alpha}$, the differential operator $\mathcal{D}_2$ is

\begin{equation}
\mathcal{D}_2 Y^{\alpha} = D Y^{\alpha} - Y^{\gamma} \nabla^{\alpha} U_{\gamma} - \frac{1}{2} Y^{\alpha} \nabla_{\gamma} U^{\gamma} .
\end{equation}

By assumption, $Q$ is initially positive and evolves continuously to a negative value at the given later time. Therefore $Q$ must pass through zero at some intermediate time, which we denote by $t = 0$ without loss of generality. We can also assume, without loss of generality, that the vector field evolves according to $\mathcal{D}_2 Y^{\alpha} = 0$, which means that it is advected by the flow. The equation for $Q$ then becomes

\begin{equation}
D Q = - 2 Y^{\alpha} Y^{\beta} \left( \Pi^{\gamma}_{\;\; \alpha} C_{\beta \gamma} - \rho_d D_{\alpha \beta} \right)\quad,
\end{equation}
At $t = 0$, $Q = 0$ and, as $\Pi_{\alpha \beta}$ is positive semi-definite, one can show that $Y^{\alpha} \Pi_{\gamma \alpha} = 0$ so that the time derivative of $Q$ is given by


\begin{equation}
D Q |_{t = 0} = 2 Y^{\alpha} Y^{\beta} \rho_d D_{\alpha \beta} \ge 0 \quad ,
\end{equation}
provided that $D_{\alpha \beta}$ is also positive semidefinite (which is guaranteed as $D_{\alpha \beta} = \frac{1}{2} g^{\mu \nu} \sigma_{\alpha \mu} \sigma_{\beta \nu}$). This contradicts the assumption that $Q$ passes through zero from positive to negative at $t=0$. We thus conclude that $\Pi_{\alpha \beta}$ remains positive semi-definite provided it is initially.

\subsection{Viscoelasticity} \label{viscoelasticity appendix}



In this appendix we shall explore the viscoelastic behaviour of the dust rheological stress. It's easiest to see the viscoelastic behaviour of the model when $t_s \sim t_c$. Introducing a characteristic relaxation time of the dust fluid $t_r \sim t_s \sim t_c$, and a characteristic fluid timescale $t_f$. We introduce the Deborah number $\mathrm{De} = t_r/t_f$, with $C_{\alpha \beta}$ and $D_{\alpha \beta}$ being $O(\mathrm{De}^{-1})$ while $\rho_d$ and $\overline{\mathcal{D}}_2$ are $O(1)$. We can rewrite the constitutive relation as


\begin{equation}
\overline{\mathcal{D}}_2 T_{\alpha \beta} = - \frac{2}{\mathrm{De}} \left( T^{\gamma}_{\;\; (\alpha} C^{\;}_{\beta) \gamma} - \rho_d D^{\;}_{\alpha \beta}\right) ,
\label{constituative with Deborah}
\end{equation}
where we now treat $\mathrm{De}$ as a bookkeeping parameter to keep track of terms in an expansion in Deborah number. The constitutive model now has a similar form to classic viscoelastic models (e.g. the Oldroyd-B model, \citet{Oldroyd50}). The elastic limit can be recovered by taking $\mathrm{De} \rightarrow \infty$ leading to,

\begin{equation}
 \overline{\mathcal{D}}_2 T_{\alpha \beta} = \overline{\mathcal{D}} T_{\alpha \beta} - 2 T^{\gamma}_{\;\; (\alpha} \varepsilon^{\;}_{\beta) \gamma \sigma} \omega^{\sigma} = 0 .
\end{equation}
This corresponds to an elastic flow, with a source term from the flow vorticity. It is equivalent to the evolution of the Reynolds stress in the absence of source terms \citep[e.g. see][]{Gavrilyuk12}. 

If we instead take the short Deborah number limit we can develop a series solution to Equation \ref{constituative with Deborah} \citep[similar to][]{Lynch21}, which takes the form

\begin{equation}
 T_{\alpha \beta} = \sum_{n = 0}^{\infty} \mathrm{De}^{n} T^{(n)}_{\alpha \beta} ,
\end{equation}
where

\begin{equation}
T^{(0) \gamma}_{\;\;\;\; (\alpha} C^{\;}_{\beta) \gamma} = \rho_d D_{\alpha \beta}
\end{equation}
and

\begin{equation}
 T^{(n) \gamma}_{\;\;\;\; (\alpha} C^{\;}_{\beta) \gamma} = -\frac{1}{2} \overline{\mathcal{D}}_2 T_{\alpha \beta}^{(n-1)} \quad , n > 0 . 
\end{equation}
Both these equations take the form $T^{(n) \gamma}_{\;\;\;\; (\alpha} C_{\beta) \gamma} = Q^{(n)}_{\alpha \beta}$, which can be inverted to obtain


\begin{align}
 T^{(n)}_{\alpha_d \beta_d} &= t_s Q^{(n)}_{\alpha_d \beta_d} + \frac{t_s t_c}{t_s + t_c} \left(  Q^{(n)}_{(\alpha_d \beta_d^{*})} + Q^{(n)}_{(\alpha_d^{*} \beta_d)} + \frac{t_c}{t_s} Q^{(n)}_{\alpha_d^{*} \beta_d^{*}} \right) \label{P(n) d d} \\
 T^{(n)}_{\alpha_d \beta_g}  &= \frac{t_s t_c}{t_s + t_c} \left( 2 Q^{(n)}_{(\alpha_d \beta_g)} + \frac{t_c}{t_s} Q^{(n)}_{\alpha_d^{*} \beta_g} \right) , \\
 T^{(n)}_{\alpha_g \beta_g}  &= t_c Q^{(n)}_{\alpha_g \beta_g} \label{P(n) t t} .
\end{align}
Substituting $Q^{(0)}_{\alpha \beta} = D_{\alpha \beta}$, and making use of the properties of $D_{\alpha \beta}$ we obtain

\begin{align}
T^{(0)}_{\alpha_d \beta_d} &= \frac{t_c^2}{t_s + t_c} D_{\alpha_d^{*} \beta_d^{*}} = \frac{t_c}{t_s + t_c} \alpha c_s^2 \rho_d g_{\alpha_d \beta_d}  \\
 T^{(0)}_{\alpha_d \beta_g}  &= \frac{t_c^2}{t_s + t_c} D_{\alpha_d^{*} \beta_g} = \frac{t_c}{t_s + t_c} \alpha c_s^2 \rho_d g_{\alpha_d^{*} \beta_g} , \\
 T^{(0)}_{\alpha_g \beta_g}  &= t_c D_{\alpha_t \beta_t} = \alpha c_s^2 \rho_d g_{\alpha_g \beta_g} .
\end{align}
To calculate $T^{(1)}_{\alpha \beta}$ we first need to calculate $\overline{\mathcal{D}}_2 T^{(0)}_{\alpha \beta}$. For simplicity we shall assume that $\alpha$, $c_s$, $t_s$, $t_c$ are constant, and that the metric tensor is time independent then  we obtain


\begin{align}
 (\overline{\mathcal{D}}_2 \mathbf{T}^{(0)})_{\alpha_d \beta_d} &= \frac{2 t_c}{t_s + t_c} \alpha c_s^2 \rho_d \bar{\nabla}_{(\alpha_d} U_{\beta_d)} , \\
 (\overline{\mathcal{D}}_2 \mathbf{T}^{(0)})_{\alpha_d \beta_g} &=  \frac{t_c}{t_s + t_c} \alpha c_s^2 \rho_d \left( \bar{\nabla}_{\alpha_d} U_{\beta_g} + \bar{\nabla}_{\beta_g^{*}} U_{\alpha_d} \right) \\
 (\overline{\mathcal{D}}_2 \mathbf{T}^{(0)})_{\alpha_g \beta_g}  &=  \frac{t_c}{t_s + t_c} \alpha c_s^2 \rho_d \left( \bar{\nabla}_{\alpha_g^{*}} U_{\beta_g} + \bar{\nabla}_{\beta_g^{*}} U_{\alpha_g} \right) .
\end{align}
Substituting this into Equations \ref{P(n) d d}-\ref{P(n) t t} we obtain

\begin{align}
 T^{(1)}_{\alpha_d \beta_d} &= -\frac{1}{2} \frac{t_c^2}{t_s + t_c} \alpha c_s^2 \rho_d \left( 2 \frac{t_s}{t_c} \frac{t_s + 2 t_c}{t_s + t_c} \bar{\nabla}_{(\alpha_d} U_{\beta_d)} + \bar{\nabla}_{\alpha_d} U_{\beta_d^{*}} + \bar{\nabla}_{\beta_d} U_{\alpha_d^{*}} \right) \\
 T^{(1)}_{\alpha_d \beta_g}  &= -\frac{1}{2} \frac{t_s t_c^2}{(t_s + t_c)^2} \alpha c_s^2 \rho_d \left[ \left( 2 + \frac{t_c}{t_s} \right) \bar{\nabla}_{\alpha_d} U_{\beta_g} + 2 \bar{\nabla}_{\beta_g^{*}} U_{\alpha_d} + \frac{t_c}{t_s}  \bar{\nabla}_{\beta_g^{*}} U_{\alpha_d^{*}} \right] , \\
 T^{(1)}_{\alpha_g \beta_g}  &= -\frac{1}{2} \frac{t_c^2}{t_s + t_c} \alpha c_s^2 \rho_d \left( \bar{\nabla}_{\alpha_g^{*}} U_{\beta_g} + \bar{\nabla}_{\beta_g^{*}} U_{\alpha_g} \right) .
\end{align}
This results in a rheological stress tensor of the form,

\begin{equation}
 T_{\alpha \beta} = p_d \left(1 + \frac{t_s}{t_c}\Theta^g_{\alpha \beta}\right) g_{\alpha \beta} + \frac{1}{2} p_{x} (g_{\alpha \beta^*} + g_{\alpha^* \beta}) - 2 \mu_{\alpha \beta}^{\mu \nu} \overline{\nabla}_{\mu} U_{\nu} + O(\mathrm{De}^2)
\end{equation}
where we have introduced the dust pressure, $p_d$, and cross pressure, $p_x$, with $p_d = p_x = \alpha c_s^2 \frac{t_c}{t_s + t_c} \rho_d$; and for convenience we have defined $\Theta^g_{\alpha \beta}$, with $\Theta^g_{\alpha_d \beta_d} = \Theta^g_{\alpha_d \beta_g} = \Theta^g_{\alpha_g \beta_d} = 0$ and $\Theta^g_{\alpha_g \beta_g} = 1$. We have also introduced an anisotropic viscosity tensor, $\mu_{\alpha \beta}^{\mu \nu}$, given by,

\begin{equation}
 \mu_{\alpha \beta}^{\mu \nu}  = \mu_d \delta_{(\alpha}^{\mu} \delta_{\beta)}^{\nu} + \mu_{x} \left( \delta_{(\alpha}^{\mu} \delta_{\beta^{*})}^{\nu} + \delta_{(\alpha^{*}}^{\mu} \delta_{\beta)}^{\nu}\right) + \eta_{\alpha \beta}^{\mu \nu} ,
\end{equation}
with

\begin{align}
 \mu_d &= \frac{1}{2} \frac{t_s t_c (t_s + 2 t_c)}{(t_s + t_c)^2} \alpha c_s^2 \rho_d , \\
 \mu_x &= \frac{1}{2} \frac{t_c^2}{t_s + t_c} \alpha c_s^2 \rho_d ,
\end{align}
and

\begin{align}
\eta_{\alpha_d \beta_d}^{\mu \nu} &= \eta_{\alpha_g \beta_g}^{\mu \nu} = 0,\\
\eta_{\alpha_d \beta_g}^{\mu \nu} &=  \frac{1}{2} \mu_x \left[ \frac{t_c^2 - t_s^2}{t_c} \delta_{\alpha_d}^{\mu} \delta_{\beta_g}^{\nu} + (t_s - t_c) \delta_{\beta_g^{*}}^{\mu} \delta_{\alpha_d}^{\nu} - (t_s + t_c) \delta_{\alpha_d}^{\mu} \delta_{\beta_g^{*}}^{\nu} + \frac{t_c}{t_s} \delta_{\beta_g^{*}}^{\mu} \delta_{\alpha_g^{*}}^{\nu} \right].
\end{align}

\subsection{Dust velocity correlations in the short stopping time limit.} \label{eddy knudon appendix}

In this section we consider the short stopping time behaviour of the rheological stress tensor. Naively one might expect the dust velocity correlations to match the gas velocity correlations due to the tight coupling between the gas and dust. This would mean the dust stress tensor would be given by $T_{\alpha_d \beta_d} = f_d R_{\alpha_d^{*} \beta_d^{*}}$ where $f_d$ is the dust to gas ratio. We shall show that this is only the case when the dust experiences the turbulence as a continuum where the dust interacts with many turbulent eddies over the lengthscale on which the dust fluid varies. If, however an individual eddy transports a dust particle a significant distance in the fluid then the dust velocity correlations can deviate strongly from those of the gas.


We wish to compare the evolutionary equations for the gas Reynolds stress to that of the dust stress tensor. The gas Reynolds stress evolves according to

\begin{equation}
 \tilde{D} R_{i j} + 2 R^{\;}_{k (i} \nabla^{k} u^{g}_{j)} + R_{i j} \nabla_k u_{g}^k = -\frac{2}{t_c} (R_{i j} - \rho_g D_{i j}) ,
\end{equation}
where $\tilde{D}$ is the Lagrangian time derivative with respect to the mean gas flow, $\mathbf{u}^g$. The dust stress tensor evolves according to

\begin{equation}
 \overline{D} T_{\alpha \beta} + 2 T_{\gamma (\alpha}  \overline{\nabla}^{\gamma} U_{\beta)} + T_{\alpha \beta}  \overline{\nabla}_{\gamma} U^{\gamma} = -2 \left(T^{\gamma}_{\;\; (\alpha} C^{\;}_{\beta) \gamma} - \rho_d D_{\alpha \beta} \right) ,
\end{equation}
where $\overline{D}$ is the Lagrangian time derivative with respect to the mean dust flow.

Because of the factor of the dust to gas ratio between the dust rheological and the gas Reynolds stress, it is more convenient to work with the respective velocity correlation tensors. The dust velocity correlation tensor $W_{\alpha \beta} = T_{\alpha \beta}/\rho_d$ which evolves according to 

\begin{equation}
 \overline{D} W_{\alpha \beta} + 2 W_{\gamma (\alpha} \overline{\nabla}^{\gamma} U_{\beta)} = -2 \left(W^{\gamma}_{\;\; (\alpha} C^{\;}_{\beta) \gamma} - D_{\alpha \beta} \right) ,
 \label{dust velocity correlation evo}
\end{equation}
We can also write the the evolutionary equation for the gas velocity correlation tensor, $R_{i j}/\rho_g$, in the form



\begin{equation}
 \tilde{\mathcal{D}}_2 (R_{i j}/\rho_g) = -\frac{2}{t_c} \left((R_{i j}/\rho_g) - D_{i j} \right) ,
 \label{gas velocity corr equation}
\end{equation}
where we have introduce the differential operator $\tilde{\mathcal{D}}_2$ which, when acting on $R_{i j}/\rho_g$, is given by

\begin{equation}
\tilde{\mathcal{D}}_2 (R_{i j}/\rho_g) =  \tilde{D} (R_{i j}/\rho_g) + (2/\rho_g) R^{\;}_{k (i} \nabla^{k} u^{g}_{j)} .
\end{equation}

We now consider small dust grains $(t_s \rightarrow 0)$ embedded in a gas flow with mean velocity $u_i^g$. The `dust' components of the dust fluid momentum equation, in the limit $t_s \rightarrow 0$, simplify to

\begin{equation}
 U^{\alpha_d} = U^{\alpha_d^{*}} ,
\end{equation}
while the `dummy gas' components are

\begin{equation}
 \rho_d \overline{D} U_{\alpha_g} = \rho_d F_{\alpha_g} - \overline{\nabla}^{\beta_d} T_{\alpha_g \beta_d} - \frac{1}{t_c} \rho_d \left(U_{\alpha_g} - U^{g}_{\alpha_g} \right) .
\end{equation}
While the mean dust velocity is tightly coupled to the mean gas velocity $(U^{\alpha_d} = U^{\alpha_g^{*}})$ the mean gas velocity as experience by the dust is not generally the same as the mean gas velocity experience by the gas, $u_i^g$. This is because the dust is effectively a subsample of the gas velocity field and can experience a mean gas velocity relative to the gas frame due to correlation in the gas turbulence. This distinction is vital for allowing zero stopping time particles to diffuse in gas turbulence. 

We can write the 6D dust velocity as follows

\begin{equation}
 U^{\alpha_d} = u_g^{\alpha_d} + \Delta U^{\alpha_d} , \quad  U_{\;}^{\alpha_g} = u_g^{\alpha_g^*} + \Delta U_{\;}^{\alpha_g^{*}} ,
\end{equation}
where $\Delta U^i$ is the relative velocity with respect to the mean gas flow experienced by the gas, which need not be small. With this velocity for the dust flow the Lagrangian time derivative, $\overline{D}$, can be related to $\tilde{D}$ by

\begin{equation}
 \overline{D} = \tilde{D} + \Delta U^{\gamma} \overline{\nabla}_{\gamma} .
\end{equation}

Substituting this velocity into Equation \ref{dust velocity correlation evo} and separating the dust and dummy gas components of the dust constitutive relation we get

\begin{align}
\begin{split}
 \tilde{D} W_{\alpha_d \beta_d} &+ 2 W^{\;}_{\gamma (\alpha_d} \overline{\nabla}^{\gamma} u^{g}_{\beta_d)}  + \Delta U^{\gamma} \overline{\nabla}_{\gamma} W_{\alpha_d \beta_d}  + 2 W_{\gamma (\alpha_d} \overline{\nabla}^{\gamma} \Delta U_{\beta_d)}  = -\frac{1}{t_s} \left(2 W_{\alpha_d \beta_d} - W_{\alpha_d^{*} \beta_d} - W_{\alpha_d \beta_d^{*}} \right) , \label{dust velocity corr d d} 
\end{split} \\
\begin{split}
 \tilde{D} W_{\alpha_d \beta_g} &+ W_{\gamma \alpha_d} \overline{\nabla}^{\gamma} \left(u^{g}_{\beta_g^{*}} + \Delta U_{\beta_g^{*}} \right) + W_{\gamma \beta_g} \overline{\nabla}^{\gamma} \left( u^{g}_{\alpha_d} + \Delta U_{\alpha_d} \right) \\
&+ \Delta U^{\gamma} \overline{\nabla}_{\gamma} W_{\alpha_d \beta_g}  = -\left(\frac{1}{t_s} + \frac{1}{t_c} \right) W_{\alpha_d \beta_g} + \frac{1}{t_s} W_{\alpha_d^{*} \beta_g}   , \label{dust velocity corr d t} 
\end{split}\\
\begin{split}
 \tilde{D} W_{\alpha_g \beta_g} &+ 2 W^{\;}_{\gamma (\alpha_g} \overline{\nabla}^{\gamma} u^{g}_{\beta_g)^{*}} + \Delta U^{\gamma} \overline{\nabla}_{\gamma} W_{\alpha_g \beta_g} + 2 W_{\gamma (\alpha_g} \overline{\nabla}^{\gamma} \Delta U_{\beta_g)^{*}}  = -\frac{2}{t_c} \left(W_{\alpha_g \beta_g}  - \alpha c_s^2 g_{\alpha_g \beta_g}\right) . \label{dust velocity corr t t} 
\end{split}
\end{align}

Taking the short stopping time limit of Equations \ref{dust velocity corr d d} and \ref{dust velocity corr d t} leads to

\begin{equation}
W_{\alpha_d \beta_d} = \frac{1}{2} (W_{\alpha_d^{*} \beta_d} + W_{\alpha_d \beta_d^{*}}) ,
\end{equation}
and

\begin{equation}
W_{\alpha_d \beta_g} = W_{\alpha_d^{*} \beta_g} .
\end{equation}
We can use these relations to simplify the `dummy gas' components (Equation \ref{dust velocity corr t t}), which can be rearranged to obtain

\begin{equation}
t_c \tilde{\mathcal{D}}_2 W_{\alpha_g \beta_g} + 2(W_{\alpha_g \beta_g}  - \alpha c_s^2 g_{\alpha_g \beta_g}) = -t_c \left( \Delta U^{\gamma} \overline{\nabla}_{\gamma} W_{\alpha_g \beta_g} + 2 W_{\gamma (\alpha_g} \overline{\nabla}^{\gamma} \Delta U_{\beta_g)} \right ) .
\label{t t constituative relation}
\end{equation}
If $\Delta U^{*}$ is the characteristic scale of the relative velocity $\Delta U^{\alpha}$ and $L$ is a characteristic lengthscale of variations in the fluid flow then we can introduce an Eddy-Knudsen number,

\begin{equation}
 \mathrm{Kn}_{\rm e} = \frac{\lambda}{L} = \frac{t_c \Delta U^{*}}{L} ,
\end{equation}
where $\lambda = t_c \Delta U^{*}$ is the mean free path of a dust grain in the turbulent flow representing the lengthscale a dust grain is transported by a single eddy. When $ \mathrm{Kn}_{\rm e} \ll 1$ the dust experiences the gas turbulence as a continuum, interacting with a large number of turbulent eddies on the lengthscale of the fluid. When $\mathrm{Kn}_{\rm e}  \gtrsim 1$ the dynamics of a dust grain is dominated by the last eddy that it interacted with - in a similar manor to the effects of individual particle collisions in weakly collisional gasses. Rescaling the right-hand side of Equation \ref{t t constituative relation} and making use of the Eddy-Knudsen number we arrive at

\begin{equation}
t_c \tilde{\mathcal{D}}_2 W_{\alpha_g \beta_g} + 2(W_{\alpha_g \beta_g}  - \alpha c_s^2 g_{\alpha_g \beta_g}) = - \mathrm{Kn}_{\rm e}  \left( \frac{\Delta U^{\gamma}}{\Delta U^{*}} L \overline{\nabla}_{\gamma} W_{\alpha_g \beta_g} + 2 W_{\gamma (\alpha_g} L \overline{\nabla}^{\gamma} \frac{\Delta U_{\beta_g)}}{\Delta U^{*}} \right ) .
\end{equation}
In the limit $\mathrm{Kn}_{\rm e} \rightarrow 0$ this matches Equation \ref{gas velocity corr equation} for the turbulent gas velocity correlations. Thus we conclude that in limit $t_s, \, \mathrm{Kn}_{e} \rightarrow 0$ the dust velocity correlations are set by the gas velocity correlations. However when $\mathrm{Kn}_{\rm e} \gtrsim 1$ the dust velocity correlations no longer match those of the gas as the dust velocity correlations are strongly affected by individual eddies.

\section{Higher moments of the Fokker-Planck equation terms} \label{Higher moment F-P terms}

The individual terms in the Fokker-Planck equation satisfy the follow relations, which are important for deriving the evolutionary equations for the higher velocity moments,

\begin{equation}
 \int (V_{\alpha_1} - U_{\alpha_1}) \cdots (V_{\alpha_k} - U_{\alpha_k})  \frac{\partial p}{\partial t} d^6 \mathbf{V} = \frac{\partial \Pi_{\alpha_1 \cdots \alpha_{k}}}{\partial t} + k \Pi_{(\alpha_1 \cdots \alpha_{k-1}} \frac{\partial}{\partial t} U_{\alpha_k)} ,
\end{equation}

\begin{align}
\begin{split}
 \int  &(V_{\alpha_1} - U_{\alpha_1}) \cdots (V_{\alpha_k} - U_{\alpha_k})  \frac{\partial}{\partial X_{\sigma}} [V_{\sigma} p] d^6 \mathbf{V}  \\
&= \frac{\partial}{\partial X_{\sigma}} \left( \Pi_{\alpha_1 \cdots \alpha_{k} \sigma} + U_{\sigma} \Pi_{\alpha_1 \cdots \alpha_{k}} \right) + k \Pi^{\sigma}_{\;\; (\alpha_1 \cdots \alpha_{k-1}} \frac{\partial}{\partial X^{\sigma}} U_{\alpha_k )} + k U^{\sigma} \Pi_{(\alpha_1 \cdots \alpha_{k-1}} \frac{\partial }{\partial X^{\sigma}} U_{\alpha_k )} ,
\end{split}
\end{align}

\begin{align}
\begin{split}
 \int  (V_{\alpha_1} - U_{\alpha_1}) &\cdots (V_{\alpha_k} - U_{\alpha_k})  \frac{\partial}{\partial V_{\sigma}} \left[ p \nabla_{\sigma} \Phi + p C_{\sigma \gamma} (V^{\gamma} - U_g^{\sigma}) \right] d^6 \mathbf{V}  \\
&= - k \Pi^{\;}_{(\alpha_1 \cdots \alpha_{k-1}} \left[ \nabla^{\;}_{\alpha_k)} \phi  + C_{\alpha_n)}^{\;\; \gamma} (U_{\gamma} - U_{\gamma}^{g} )\right] - k \Pi^{\gamma}_{\;\; (\alpha_1 \cdots \alpha_{k-1}} C^{\;}_{\alpha_n) \gamma}
\end{split}
\end{align}

\begin{equation}
 \int  (V_{\alpha_1} - U_{\alpha_1}) \cdots (V_{\alpha_k} - U_{\alpha_k})  D_{\mu \nu} \frac{\partial^2 p}{\partial V_{\mu} \partial V_{\nu}} d^6 \mathbf{V} = k (k-1) \Pi_{(\alpha_1 \cdots \alpha_{k-2}} D_{\alpha_{k-1} \alpha_{k-2})} .
\end{equation}

\bibliographystyle{jfm}
\bibliography{dust_fluid_JFM}

\begin{thebibliography}{89}
\expandafter\ifx\csname natexlab\endcsname\relax\def\natexlab#1{#1}\fi
\def\au#1{#1} \def\ed#1{#1} \def\yr#1{#1}\def\at#1{#1}\def\jt#1{\textit{#1}}
  \def\bt#1{#1}\def\bvol#1{\textbf{#1}} \def\vol#1{#1} \def\pg#1{#1}
  \def\publ#1{#1}\def\arxiv#1{#1}\def\org#1{#1}\def\st#1{\textit{#1}}

\bibitem[{Araki} \& {Tremaine}(1986)]{Araki86}
{\sc \au{{Araki}, S.} \& \au{{Tremaine}, S.}} \yr{1986}  \at{{The dynamics of
  dense particle disks}}.  \jt{\icarus}  \bvol{65}~(1),  \pg{83--109}.

\bibitem[Armitage(2020)]{Armitage20}
{\sc \au{Armitage, Philip~J.}} \yr{2020} {\em Astrophysics of Planet
  Formation\/}, 2nd edn.  \publ{Cambridge University Press}.

\bibitem[{Bai} \& {Stone}(2010)]{Bai10}
{\sc \au{{Bai}, Xue-Ning} \& \au{{Stone}, James~M.}} \yr{2010}
  \at{{Particle-gas Dynamics with Athena: Method and Convergence}}.  \jt{\apjs}
   \bvol{190}~(2),  \pg{297--310},  \arxiv{arXiv: 1005.4980}.

\bibitem[{Baines} {\em et~al.\/}(1965){Baines}, {Williams} \&
  {Asebiomo}]{Baines65}
{\sc \au{{Baines}, M.~J.}, \au{{Williams}, I.~P.} \& \au{{Asebiomo}, A.~S.}}
  \yr{1965}  \at{{Resistance to the motion of a small sphere moving through a
  gas}}.  \jt{\mnras}  \bvol{130},  \pg{63}.

\bibitem[{Balbus} \& {Hawley}(1991)]{Balbus91}
{\sc \au{{Balbus}, S.~A.} \& \au{{Hawley}, J.~F.}} \yr{1991}  \at{{A powerful
  local shear instability in weakly magnetized disks. I - Linear analysis. II -
  Nonlinear evolution}}.  \jt{\apj}  \bvol{376},  \pg{214--233}.

\bibitem[{Balsara} {\em et~al.\/}(2009){Balsara}, {Tilley}, {Rettig} \&
  {Brittain}]{Balsara09}
{\sc \au{{Balsara}, Dinshaw~S.}, \au{{Tilley}, David~A.}, \au{{Rettig},
  Terrence} \& \au{{Brittain}, Sean~D.}} \yr{2009}  \at{{Dust settling in
  magnetorotationally driven turbulent discs - I. Numerical methods and
  evidence for a vigorous streaming instability}}.  \jt{\mnras}
  \bvol{397}~(1),  \pg{24--43},  \arxiv{arXiv: 0810.0246}.

\bibitem[{Barker} \& {Ogilvie}(2014)]{Barker14}
{\sc \au{{Barker}, A.~J.} \& \au{{Ogilvie}, G.~I.}} \yr{2014}
  \at{{Hydrodynamic instability in eccentric astrophysical discs}}.
  \jt{\mnras}  \bvol{445},  \pg{2637--2654},  \arxiv{arXiv: 1409.6488}.

\bibitem[{Barri{\`e}re-Fouchet} {\em et~al.\/}(2005){Barri{\`e}re-Fouchet},
  {Gonzalez}, {Murray}, {Humble} \& {Maddison}]{Barriere05}
{\sc \au{{Barri{\`e}re-Fouchet}, L.}, \au{{Gonzalez}, J.~F.}, \au{{Murray},
  J.~R.}, \au{{Humble}, R.~J.} \& \au{{Maddison}, S.~T.}} \yr{2005}  \at{{Dust
  distribution in protoplanetary disks. Vertical settling and radial
  migration}}.  \jt{\aap}  \bvol{443}~(1),  \pg{185--194},  \arxiv{arXiv:
  astro-ph/0508452}.

\bibitem[{Ben{\'\i}tez-Llambay} \& {Masset}(2016)]{BenitezLlambay16}
{\sc \au{{Ben{\'\i}tez-Llambay}, Pablo} \& \au{{Masset}, Fr{\'e}d{\'e}ric~S.}}
  \yr{2016}  \at{{FARGO3D: A New GPU-oriented MHD Code}}.  \jt{\apjs}
  \bvol{223}~(1),  \pg{11},  \arxiv{arXiv: 1602.02359}.

\bibitem[{Bi} {\em et~al.\/}(2021){Bi}, {Lin} \& {Dong}]{Bi21}
{\sc \au{{Bi}, Jiaqing}, \au{{Lin}, Min-Kai} \& \au{{Dong}, Ruobing}} \yr{2021}
   \at{{Puffed-up Edges of Planet-opened Gaps in Protoplanetary Disks. I.
  Hydrodynamic Simulations}}.  \jt{\apj}  \bvol{912}~(2),  \pg{107},
  \arxiv{arXiv: 2103.09254}.

\bibitem[{Binkert}(2023)]{Binkert23}
{\sc \au{{Binkert}, Fabian}} \yr{2023}  \at{{Beyond diffusion: a generalized
  mean-field theory of turbulent dust transport in protoplanetary discs}}.
  \jt{\mnras}  \bvol{525}~(3),  \pg{4299--4320},  \arxiv{arXiv: 2306.06103}.

\bibitem[Bobylev(1982)]{Bobylev82}
{\sc \au{Bobylev, Aleksandr~Vasil’evich}} \yr{1982} The chapman-enskog and
  grad methods for solving the boltzmann equation.  \bt{In {\em Akademiia Nauk
  SSSR Doklady\/}}, ,  \vol{vol. 262},  \pg{pp. 71--75}.

\bibitem[Bobylev(2018)]{Bobylev17}
{\sc \au{Bobylev, A.~V.}} \yr{2018}  \at{Boltzmann equation and hydrodynamics
  beyond navier–stokes}.  \jt{Philosophical Transactions of the Royal Society
  A: Mathematical, Physical and Engineering Sciences}  \bvol{376}~(2118),
  \pg{20170227},  \arxiv{arXiv:
  https://royalsocietypublishing.org/doi/pdf/10.1098/rsta.2017.0227}.

\bibitem[{Booth} \& {Clarke}(2021)]{Booth21}
{\sc \au{{Booth}, Richard~A.} \& \au{{Clarke}, Cathie~J.}} \yr{2021}
  \at{{Modelling the delivery of dust from discs to ionized winds}}.
  \jt{\mnras}  \bvol{502}~(2),  \pg{1569--1578},  \arxiv{arXiv: 2101.04121}.

\bibitem[{Borderies} {\em et~al.\/}(1985){Borderies}, {Goldreich} \&
  {Tremaine}]{Borderies85}
{\sc \au{{Borderies}, N.}, \au{{Goldreich}, P.} \& \au{{Tremaine}, S.}}
  \yr{1985}  \at{{A granular flow model for dense planetary rings}}.
  \jt{\icarus}  \bvol{63}~(3),  \pg{406--420}.

\bibitem[Capecelatro {\em et~al.\/}(2016{\natexlab{{\em a\/}}})Capecelatro,
  Desjardins \& Fox]{Capecelatro16a}
{\sc \au{Capecelatro, Jesse}, \au{Desjardins, Olivier} \& \au{Fox, Rodney~O}}
  \yr{2016{\natexlab{{\em a\/}}}}  \at{Strongly coupled fluid-particle flows in
  vertical channels. i. reynolds-averaged two-phase turbulence statistics}.
  \jt{Physics of Fluids}  \bvol{28}~(3).

\bibitem[Capecelatro {\em et~al.\/}(2016{\natexlab{{\em b\/}}})Capecelatro,
  Desjardins \& Fox]{Capecelatro16b}
{\sc \au{Capecelatro, Jesse}, \au{Desjardins, Olivier} \& \au{Fox, Rodney~O}}
  \yr{2016{\natexlab{{\em b\/}}}}  \at{Strongly coupled fluid-particle flows in
  vertical channels. ii. turbulence modeling}.  \jt{Physics of Fluids}
  \bvol{28}~(3).

\bibitem[{Carballido} {\em et~al.\/}(2006){Carballido}, {Fromang} \&
  {Papaloizou}]{Carballido06}
{\sc \au{{Carballido}, Augusto}, \au{{Fromang}, S{\'e}bastien} \&
  \au{{Papaloizou}, John}} \yr{2006}  \at{{Mid-plane sedimentation of large
  solid bodies in turbulent protoplanetary discs}}.  \jt{\mnras}
  \bvol{373}~(4),  \pg{1633--1640},  \arxiv{arXiv: astro-ph/0610075}.

\bibitem[Chapman \& Cowling(1990)]{Chapman90}
{\sc \au{Chapman, Sydney} \& \au{Cowling, Thomas~George}} \yr{1990} {\em The
  mathematical theory of non-uniform gases: an account of the kinetic theory of
  viscosity, thermal conduction and diffusion in gases\/}.  \publ{Cambridge
  university press}.

\bibitem[{Commer{\c{c}}on} {\em et~al.\/}(2023){Commer{\c{c}}on}, {Lebreuilly},
  {Price}, {Lovascio}, {Laibe} \& {Hennebelle}]{Commercon23}
{\sc \au{{Commer{\c{c}}on}, B.}, \au{{Lebreuilly}, U.}, \au{{Price}, D.~J.},
  \au{{Lovascio}, F.}, \au{{Laibe}, G.} \& \au{{Hennebelle}, P.}} \yr{2023}
  \at{{Dynamics of dust grains in turbulent molecular clouds. Conditions for
  decoupling and limits of different numerical implementations}}.  \jt{\aap}
  \bvol{671},  \pg{A128},  \arxiv{arXiv: 2301.04946}.

\bibitem[Csanady(1963)]{Csanady63}
{\sc \au{Csanady, GT}} \yr{1963}  \at{Turbulent diffusion of heavy particles in
  the atmosphere}.  \jt{Journal of Atmospheric Sciences}  \bvol{20}~(3),
  \pg{201--208}.

\bibitem[Diamond {\em et~al.\/}(1994)Diamond, Liang, Carreras \&
  Terry]{Diamond94}
{\sc \au{Diamond, P.~H.}, \au{Liang, Y.-M.}, \au{Carreras, B.~A.} \& \au{Terry,
  P.~W.}} \yr{1994}  \at{Self-regulating shear flow turbulence: A paradigm for
  the l to h transition}.  \jt{Phys. Rev. Lett.}  \bvol{72},  \pg{2565--2568}.

\bibitem[{Dubrulle} {\em et~al.\/}(1995){Dubrulle}, {Morfill} \&
  {Sterzik}]{Dubrulle95}
{\sc \au{{Dubrulle}, B.}, \au{{Morfill}, G.} \& \au{{Sterzik}, M.}} \yr{1995}
  \at{{The dust subdisk in the protoplanetary nebula.}}  \jt{\icarus}
  \bvol{114}~(2),  \pg{237--246}.

\bibitem[Eleuterio(1999)]{Toro99}
{\sc \au{Eleuterio, F~Toro}} \yr{1999} Riemann solvers and numerical methods
  for fluid dynamics.

\bibitem[{Epstein}(1924)]{Epstein24}
{\sc \au{{Epstein}, Paul~S.}} \yr{1924}  \at{{On the Resistance Experienced by
  Spheres in their Motion through Gases}}.  \jt{Physical Review}
  \bvol{23}~(6),  \pg{710--733}.

\bibitem[{Flock} {\em et~al.\/}(2017){Flock}, {Nelson}, {Turner}, {Bertrang},
  {Carrasco-Gonz{\'a}lez}, {Henning}, {Lyra} \& {Teague}]{Flock17}
{\sc \au{{Flock}, Mario}, \au{{Nelson}, Richard~P.}, \au{{Turner}, Neal~J.},
  \au{{Bertrang}, Gesa H.~M.}, \au{{Carrasco-Gonz{\'a}lez}, Carlos},
  \au{{Henning}, Thomas}, \au{{Lyra}, Wladimir} \& \au{{Teague}, Richard}}
  \yr{2017}  \at{{Radiation Hydrodynamical Turbulence in Protoplanetary Disks:
  Numerical Models and Observational Constraints}}.  \jt{\apj}  \bvol{850}~(2),
   \pg{131},  \arxiv{arXiv: 1710.06007}.

\bibitem[Fox(2014)]{Fox14}
{\sc \au{Fox, Rodney~O.}} \yr{2014}  \at{On multiphase turbulence models for
  collisional fluid–particle flows}.  \jt{Journal of Fluid Mechanics}
  \bvol{742},  \pg{368–424}.

\bibitem[{Fromang} \& {Papaloizou}(2006)]{Fromang06}
{\sc \au{{Fromang}, S.} \& \au{{Papaloizou}, J.}} \yr{2006}  \at{{Dust settling
  in local simulations of turbulent protoplanetary disks}}.  \jt{\aap}
  \bvol{452}~(3),  \pg{751--762},  \arxiv{arXiv: astro-ph/0603153}.

\bibitem[Gavrilyuk \& Gouin(2012)]{Gavrilyuk12}
{\sc \au{Gavrilyuk, S.} \& \au{Gouin, H.}} \yr{2012}  \at{Geometric evolution
  of the reynolds stress tensor}.  \jt{International Journal of Engineering
  Science}  \bvol{59},  \pg{65--73}, the Special Issue in honor of VICTOR L.
  BERDICHEVSKY.

\bibitem[{Goldreich} \& {Tremaine}(1978)]{Goldreich78}
{\sc \au{{Goldreich}, P.} \& \au{{Tremaine}, S.}} \yr{1978}  \at{{The
  excitation and evolution of density waves.}}  \jt{\apj}  \bvol{222},
  \pg{850--858}.

\bibitem[Grad(1948)]{Grad48}
{\sc \au{Grad, Harold}} \yr{1948} {\em Approximation to the Boltzmann equation
  by moments\/}.  \publ{New York University}.

\bibitem[Grad(1949)]{Grad49}
{\sc \au{Grad, Harold}} \yr{1949}  \at{On the kinetic theory of rarefied
  gases}.  \jt{Communications on pure and applied mathematics}  \bvol{2}~(4),
  \pg{331--407}.

\bibitem[Harten {\em et~al.\/}(1983)Harten, Lax \& Leer]{Harten83}
{\sc \au{Harten, Amiram}, \au{Lax, Peter~D} \& \au{Leer, Bram~van}} \yr{1983}
  \at{On upstream differencing and godunov-type schemes for hyperbolic
  conservation laws}.  \jt{SIAM review}  \bvol{25}~(1),  \pg{35--61}.

\bibitem[Hawley \& Balbus(1991)]{Hawley91}
{\sc \au{Hawley, John~F} \& \au{Balbus, Steven~A}} \yr{1991}  \at{A powerful
  local shear instability in weakly magnetized disks. ii. nonlinear evolution}.
   \jt{The Astrophysical Journal}  \bvol{376},  \pg{223}.

\bibitem[{Hawley} {\em et~al.\/}(1995){Hawley}, {Gammie} \& {Balbus}]{Hawley95}
{\sc \au{{Hawley}, J.~F.}, \au{{Gammie}, C.~F.} \& \au{{Balbus}, S.~A.}}
  \yr{1995}  \at{{Local Three-dimensional Magnetohydrodynamic Simulations of
  Accretion Disks}}.  \jt{\apj}  \bvol{440},  \pg{742}.

\bibitem[Hobson {\em et~al.\/}(2006)Hobson, Efstathiou \& Lasenby]{Hobson06}
{\sc \au{Hobson, Michael~Paul}, \au{Efstathiou, George~P} \& \au{Lasenby,
  Anthony~N}} \yr{2006} {\em General relativity: an introduction for
  physicists\/}.  \publ{Cambridge University Press}.

\bibitem[Innocenti {\em et~al.\/}(2019)Innocenti, Fox \& Chibbaro]{Innocenti19}
{\sc \au{Innocenti, Alessio}, \au{Fox, R.} \& \au{Chibbaro, Sergio}} \yr{2019}
  \at{A lagrangian probability-density-function model for collisional turbulent
  fluid–particle flows}.  \jt{Journal of Fluid Mechanics}  \bvol{862},
  \pg{449}.

\bibitem[Innocenti {\em et~al.\/}(2021)Innocenti, Fox \& Chibbaro]{Innocenti21}
{\sc \au{Innocenti, Alessio}, \au{Fox, Rodney~O.} \& \au{Chibbaro, Sergio}}
  \yr{2021}  \at{{A Lagrangian probability-density-function model for turbulent
  particle-laden channel flow in the dense regime}}.  \jt{Physics of Fluids}
  \bvol{33}~(5),  \pg{053308},  \arxiv{arXiv:
  https://pubs.aip.org/aip/pof/article-pdf/doi/10.1063/5.0045690/16085748/053308\_1\_online.pdf}.

\bibitem[{Laibe} {\em et~al.\/}(2020){Laibe}, {Br{\'e}hier} \&
  {Lombart}]{Laibe20}
{\sc \au{{Laibe}, Guillaume}, \au{{Br{\'e}hier}, Charles-Edouard} \&
  \au{{Lombart}, Maxime}} \yr{2020}  \at{{On the settling of small grains in
  dusty discs: analysis and formulae}}.  \jt{\mnras}  \bvol{494}~(4),
  \pg{5134--5147},  \arxiv{arXiv: 2004.03689}.

\bibitem[{Laibe} \& {Price}(2012{\natexlab{{\em a\/}}})]{Laibe12a}
{\sc \au{{Laibe}, Guillaume} \& \au{{Price}, Daniel~J.}}
  \yr{2012{\natexlab{{\em a\/}}}}  \at{{Dusty gas with smoothed particle
  hydrodynamics - I. Algorithm and test suite}}.  \jt{\mnras}  \bvol{420}~(3),
  \pg{2345--2364},  \arxiv{arXiv: 1111.3090}.

\bibitem[{Laibe} \& {Price}(2012{\natexlab{{\em b\/}}})]{Laibe12b}
{\sc \au{{Laibe}, Guillaume} \& \au{{Price}, Daniel~J.}}
  \yr{2012{\natexlab{{\em b\/}}}}  \at{{Dusty gas with smoothed particle
  hydrodynamics - II. Implicit timestepping and astrophysical drag regimes}}.
  \jt{\mnras}  \bvol{420}~(3),  \pg{2365--2376},  \arxiv{arXiv: 1111.3089}.

\bibitem[{Laibe} \& {Price}(2014)]{Laibe14}
{\sc \au{{Laibe}, Guillaume} \& \au{{Price}, Daniel~J.}} \yr{2014}  \at{{Dusty
  gas with one fluid}}.  \jt{\mnras}  \bvol{440}~(3),  \pg{2136--2146},
  \arxiv{arXiv: 1402.5248}.

\bibitem[{Larue} {\em et~al.\/}(2023){Larue}, {Latter} \& {Rein}]{Larue23}
{\sc \au{{Larue}, R{\'e}my}, \au{{Latter}, Henrik} \& \au{{Rein}, Hanno}}
  \yr{2023}  \at{{Thermal hysteresis and front propagation in dense planetary
  rings}}.  \jt{\mnras}  \bvol{520}~(1),  \pg{1128--1145},  \arxiv{arXiv:
  2301.03289}.

\bibitem[{Latter} \& {Ogilvie}(2008)]{Latter08}
{\sc \au{{Latter}, Henrik~N.} \& \au{{Ogilvie}, Gordon~I.}} \yr{2008}
  \at{{Dense planetary rings and the viscous overstability}}.  \jt{\icarus}
  \bvol{195}~(2),  \pg{725--751},  \arxiv{arXiv: 0803.4123}.

\bibitem[{Latter} \& {Papaloizou}(2017)]{Latter17}
{\sc \au{{Latter}, Henrik~N.} \& \au{{Papaloizou}, John}} \yr{2017}  \at{{Local
  models of astrophysical discs}}.  \jt{\mnras}  \bvol{472}~(2),
  \pg{1432--1446},  \arxiv{arXiv: 1708.09285}.

\bibitem[{Lesur} {\em et~al.\/}(2022){Lesur}, {Ercolano}, {Flock}, {Lin},
  {Yang}, {Barranco}, {Benitez-Llambay}, {Goodman}, {Johansen}, {Klahr},
  {Laibe}, {Lyra}, {Marcus}, {Nelson}, {Squire}, {Simon}, {Turner}, {Umurhan}
  \& {Youdin}]{Lesur22}
{\sc \au{{Lesur}, G.}, \au{{Ercolano}, B.}, \au{{Flock}, M.}, \au{{Lin},
  M.~K.}, \au{{Yang}, C.~C.}, \au{{Barranco}, J.~A.}, \au{{Benitez-Llambay},
  P.}, \au{{Goodman}, J.}, \au{{Johansen}, A.}, \au{{Klahr}, H.}, \au{{Laibe},
  G.}, \au{{Lyra}, W.}, \au{{Marcus}, P.}, \au{{Nelson}, R.~P.}, \au{{Squire},
  J.}, \au{{Simon}, J.~B.}, \au{{Turner}, N.}, \au{{Umurhan}, O.~M.} \&
  \au{{Youdin}, A.~N.}} \yr{2022}  \at{{Hydro-, Magnetohydro-, and Dust-Gas
  Dynamics of Protoplanetary Disks}}.  \jt{arXiv e-prints}  \pg{p.
  arXiv:2203.09821},  \arxiv{arXiv: 2203.09821}.

\bibitem[Levermore(1996)]{Levermore96}
{\sc \au{Levermore, C~David}} \yr{1996}  \at{Moment closure hierarchies for
  kinetic theories}.  \jt{Journal of statistical Physics}  \bvol{83},
  \pg{1021--1065}.

\bibitem[{Lin}(2019)]{Lin19}
{\sc \au{{Lin}, Min-Kai}} \yr{2019}  \at{{Dust settling against hydrodynamic
  turbulence in protoplanetary discs}}.  \jt{\mnras}  \bvol{485}~(4),
  \pg{5221--5234},  \arxiv{arXiv: 1903.03620}.

\bibitem[{Lin} \& {Youdin}(2015)]{Lin15}
{\sc \au{{Lin}, Min-Kai} \& \au{{Youdin}, Andrew~N.}} \yr{2015}  \at{{Cooling
  Requirements for the Vertical Shear Instability in Protoplanetary Disks}}.
  \jt{\apj}  \bvol{811}~(1),  \pg{17},  \arxiv{arXiv: 1505.02163}.

\bibitem[{Lin} \& {Youdin}(2017)]{Lin17}
{\sc \au{{Lin}, Min-Kai} \& \au{{Youdin}, Andrew~N.}} \yr{2017}  \at{{A
  Thermodynamic View of Dusty Protoplanetary Disks}}.  \jt{\apj}
  \bvol{849}~(2),  \pg{129},  \arxiv{arXiv: 1708.02945}.

\bibitem[{Lynch} \& {Ogilvie}(2021)]{Lynch21}
{\sc \au{{Lynch}, Elliot~M.} \& \au{{Ogilvie}, Gordon~I.}} \yr{2021}
  \at{{Importance of magnetic fields in highly eccentric discs with
  applications to tidal disruption events}}.  \jt{\mnras}  \bvol{501}~(4),
  \pg{5500--5516},  \arxiv{arXiv: 2101.01221}.

\bibitem[{Masset}(2000)]{Masset00}
{\sc \au{{Masset}, F.~S.}} \yr{2000} {FARGO: A Fast Eulerian Transport
  Algorithm for Differentially Rotating Disks}.  \bt{In {\em Disks,
  Planetesimals, and Planets\/} (ed. \ed{G.~{Garz{\'o}n}, C.~{Eiroa}, D.~{de
  Winter} \& T.~J. {Mahoney}})},  \st{Astronomical Society of the Pacific
  Conference Series},  \vol{vol. 219},  \pg{p.~75}.

\bibitem[{Mignone} {\em et~al.\/}(2019){Mignone}, {Flock} \&
  {Vaidya}]{Mignone19}
{\sc \au{{Mignone}, A.}, \au{{Flock}, M.} \& \au{{Vaidya}, B.}} \yr{2019}
  \at{{A Particle Module for the PLUTO Code. III. Dust}}.  \jt{\apjs}
  \bvol{244}~(2),  \pg{38},  \arxiv{arXiv: 1908.10793}.

\bibitem[Minier(2001)]{Minier01}
{\sc \au{Minier, Jean-Pierre}} \yr{2001}  \at{Probabilistic approach to
  turbulent two-phase flows modelling and simulation: theoretical and numerical
  issues}  \bvol{7}~(3-4),  \pg{295--310}.

\bibitem[Minier(2015)]{Minier15}
{\sc \au{Minier, Jean-Pierre}} \yr{2015}  \at{On lagrangian stochastic methods
  for turbulent polydisperse two-phase reactive flows}.  \jt{Progress in Energy
  and Combustion Science}  \bvol{50},  \pg{1--62}.

\bibitem[Minier(2016)]{Minier16}
{\sc \au{Minier, jean-pierre}} \yr{2016}  \at{Statistical descriptions of
  polydisperse turbulent two-phase flows}.  \jt{Physics Reports}  \bvol{665
  (2016)},  \pg{1--122}.

\bibitem[Minier {\em et~al.\/}(2014)Minier, Chibbaro \& Pope]{Minier14}
{\sc \au{Minier, Jean-Pierre}, \au{Chibbaro, Sergio} \& \au{Pope, Stephen~B}}
  \yr{2014}  \at{Guidelines for the formulation of lagrangian stochastic models
  for particle simulations of single-phase and dispersed two-phase turbulent
  flows}.  \jt{Physics of Fluids}  \bvol{26}~(11),  \pg{113303}.

\bibitem[{Minier} \& {Henry}(2023)]{Minier23}
{\sc \au{{Minier}, Jean-Pierre} \& \au{{Henry}, Christophe}} \yr{2023}
  \at{{The dynamics of discrete particles in turbulent flows: open issues and
  current challenges in statistical modeling}}.  \jt{arXiv e-prints}  \pg{p.
  arXiv:2311.01921},  \arxiv{arXiv: 2311.01921}.

\bibitem[Minier \& Peirano(2001)]{Minier01b}
{\sc \au{Minier, Jean-Pierre} \& \au{Peirano, Eric}} \yr{2001}  \at{The pdf
  approach to turbulent polydispersed two-phase flows}.  \jt{Physics Reports}
  \bvol{352}~(1),  \pg{1--214}.

\bibitem[Minier {\em et~al.\/}(2004)Minier, Peirano \& Chibbaro]{Minier04}
{\sc \au{Minier, Jean-Pierre}, \au{Peirano, Eric} \& \au{Chibbaro, Sergio}}
  \yr{2004}  \at{Pdf model based on langevin equation for polydispersed
  two-phase flows applied to a bluff-body gas-solid flow}.  \jt{Physics of
  fluids}  \bvol{16}~(7),  \pg{2419--2431}.

\bibitem[{Nelson} {\em et~al.\/}(2013){Nelson}, {Gressel} \&
  {Umurhan}]{Nelson13}
{\sc \au{{Nelson}, Richard~P.}, \au{{Gressel}, Oliver} \& \au{{Umurhan},
  Orkan~M.}} \yr{2013}  \at{{Linear and non-linear evolution of the vertical
  shear instability in accretion discs}}.  \jt{\mnras}  \bvol{435}~(3),
  \pg{2610--2632},  \arxiv{arXiv: 1209.2753}.

\bibitem[{Ogilvie}(2003)]{Ogilvie03}
{\sc \au{{Ogilvie}, G.~I.}} \yr{2003}  \at{{On the dynamics of
  magnetorotational turbulent stresses}}.  \jt{\mnras}  \bvol{340}~(3),
  \pg{969--982},  \arxiv{arXiv: astro-ph/0212442}.

\bibitem[{Ogilvie} \& {Barker}(2014)]{Ogilvie14}
{\sc \au{{Ogilvie}, G.~I.} \& \au{{Barker}, A.~J.}} \yr{2014}  \at{{Local and
  global dynamics of eccentric astrophysical discs}}.  \jt{\mnras}  \bvol{445},
   \pg{2621--2636},  \arxiv{arXiv: 1409.6487}.

\bibitem[{Ogilvie} \& {Latter}(2013{\natexlab{{\em a\/}}})]{Ogilvie13b}
{\sc \au{{Ogilvie}, G.~I.} \& \au{{Latter}, H.~N.}} \yr{2013{\natexlab{{\em
  a\/}}}}  \at{{Hydrodynamic instability in warped astrophysical discs}}.
  \jt{\mnras}  \bvol{433},  \pg{2420--2435},  \arxiv{arXiv: 1303.0264}.

\bibitem[{Ogilvie} \& {Latter}(2013{\natexlab{{\em b\/}}})]{Ogilvie13a}
{\sc \au{{Ogilvie}, G.~I.} \& \au{{Latter}, H.~N.}} \yr{2013{\natexlab{{\em
  b\/}}}}  \at{{Local and global dynamics of warped astrophysical discs}}.
  \jt{\mnras}  \bvol{433},  \pg{2403--2419},  \arxiv{arXiv: 1303.0263}.

\bibitem[{Oldroyd}(1950)]{Oldroyd50}
{\sc \au{{Oldroyd}, J.~G.}} \yr{1950}  \at{{On the Formulation of Rheological
  Equations of State}}.  \jt{Proceedings of the Royal Society of London Series
  A}  \bvol{200}~(1063),  \pg{523--541}.

\bibitem[{Ormel} \& {Cuzzi}(2007)]{Ormel07}
{\sc \au{{Ormel}, C.~W.} \& \au{{Cuzzi}, J.~N.}} \yr{2007}  \at{{Closed-form
  expressions for particle relative velocities induced by turbulence}}.
  \jt{\aap}  \bvol{466}~(2),  \pg{413--420},  \arxiv{arXiv: astro-ph/0702303}.

\bibitem[{Ormel} \& {Liu}(2018)]{Ormel18}
{\sc \au{{Ormel}, Chris~W.} \& \au{{Liu}, Beibei}} \yr{2018}  \at{{Catching
  drifting pebbles. II. A stochastic equation of motion for pebbles}}.
  \jt{\aap}  \bvol{615},  \pg{A178},  \arxiv{arXiv: 1803.06150}.

\bibitem[{Papaloizou}(2005{\natexlab{{\em a\/}}})]{Papaloizou05b}
{\sc \au{{Papaloizou}, J.~C.~B.}} \yr{2005{\natexlab{{\em a\/}}}}  \at{{Global
  numerical simulations of differentially rotating disks with free
  eccentricity}}.  \jt{\aap}  \bvol{432}~(3),  \pg{757--769}.

\bibitem[{Papaloizou}(2005{\natexlab{{\em b\/}}})]{Papaloizou05a}
{\sc \au{{Papaloizou}, J.~C.~B.}} \yr{2005{\natexlab{{\em b\/}}}}  \at{{The
  local instability of steady astrophysical flows with non circular streamlines
  with application to differentially rotating disks with free eccentricity}}.
  \jt{\aap}  \bvol{432}~(3),  \pg{743--755},  \arxiv{arXiv: astro-ph/0412587}.

\bibitem[Peirano {\em et~al.\/}(2006)Peirano, Chibbaro, Pozorski \&
  Minier]{Peirano06}
{\sc \au{Peirano, E.}, \au{Chibbaro, S.}, \au{Pozorski, J.} \& \au{Minier,
  J.-P.}} \yr{2006}  \at{Mean-field/pdf numerical approach for polydispersed
  turbulent two-phase flows}.  \jt{Progress in Energy and Combustion Science}
  \bvol{32}~(3),  \pg{315--371}.

\bibitem[Pope(1985)]{Pope85}
{\sc \au{Pope, S.B.}} \yr{1985}  \at{Pdf methods for turbulent reactive flows}.
   \jt{Progress in Energy and Combustion Science}  \bvol{11}~(2),
  \pg{119--192}.

\bibitem[Pope(1987)]{Pope87}
{\sc \au{Pope, SB}} \yr{1987}  \at{Consistency conditions for random-walk
  models of turbulent dispersion}.  \jt{The Physics of fluids}  \bvol{30}~(8),
  \pg{2374--2379}.

\bibitem[Pope(2000)]{Pope00}
{\sc \au{Pope, Stephen~B.}} \yr{2000} {\em Turbulent Flows\/}.  \publ{Cambridge
  University Press}.

\bibitem[Pope(2002)]{Pope02}
{\sc \au{Pope, Stephen~B}} \yr{2002}  \at{A stochastic lagrangian model for
  acceleration in turbulent flows}.  \jt{Physics of Fluids}  \bvol{14}~(7),
  \pg{2360--2375}.

\bibitem[Roe(1981)]{Roe81}
{\sc \au{Roe, Philip~L}} \yr{1981}  \at{Approximate riemann solvers, parameter
  vectors, and difference schemes}.  \jt{Journal of computational physics}
  \bvol{43}~(2),  \pg{357--372}.

\bibitem[Sawford(1991)]{Sawford91}
{\sc \au{Sawford, BL}} \yr{1991}  \at{Reynolds number effects in lagrangian
  stochastic models of turbulent dispersion}.  \jt{Physics of Fluids A: Fluid
  Dynamics}  \bvol{3}~(6),  \pg{1577--1586}.

\bibitem[{Svanberg} {\em et~al.\/}(2022){Svanberg}, {Cui} \&
  {Latter}]{Svanberg22}
{\sc \au{{Svanberg}, Eleonora}, \au{{Cui}, Can} \& \au{{Latter}, Henrik~N.}}
  \yr{2022}  \at{{Wavelike nature of the vertical shear instability in global
  protoplanetary discs}}.  \jt{\mnras}  \bvol{514}~(3),  \pg{4581--4587},
  \arxiv{arXiv: 2206.03840}.

\bibitem[{Testi} {\em et~al.\/}(2014){Testi}, {Birnstiel}, {Ricci}, {Andrews},
  {Blum}, {Carpenter}, {Dominik}, {Isella}, {Natta}, {Williams} \&
  {Wilner}]{Testi14}
{\sc \au{{Testi}, L.}, \au{{Birnstiel}, T.}, \au{{Ricci}, L.}, \au{{Andrews},
  S.}, \au{{Blum}, J.}, \au{{Carpenter}, J.}, \au{{Dominik}, C.}, \au{{Isella},
  A.}, \au{{Natta}, A.}, \au{{Williams}, J.~P.} \& \au{{Wilner}, D.~J.}}
  \yr{2014} {Dust Evolution in Protoplanetary Disks}.  \bt{In {\em Protostars
  and Planets VI\/} (ed. \ed{Henrik {Beuther}, Ralf~S. {Klessen}, Cornelis~P.
  {Dullemond} \& Thomas {Henning}})},  \pg{pp. 339--361},  \arxiv{arXiv:
  1402.1354}.

\bibitem[Thomsen(1986)]{Thomsen86}
{\sc \au{Thomsen, Leon}} \yr{1986}  \at{Weak elastic anisotropy}.
  \jt{Geophysics}  \bvol{51}~(10),  \pg{1954--1966}.

\bibitem[Thomson(1987)]{Thomson87}
{\sc \au{Thomson, DJ}} \yr{1987}  \at{Criteria for the selection of stochastic
  models of particle trajectories in turbulent flows}.  \jt{Journal of fluid
  mechanics}  \bvol{180},  \pg{529--556}.

\bibitem[Toro {\em et~al.\/}(1994)Toro, Spruce \& Speares]{Toro94}
{\sc \au{Toro, Eleuterio~F}, \au{Spruce, Michael} \& \au{Speares, William}}
  \yr{1994}  \at{Restoration of the contact surface in the hll-riemann solver}.
   \jt{Shock waves}  \bvol{4},  \pg{25--34}.

\bibitem[Van~Leer(2006)]{vanLeer06}
{\sc \au{Van~Leer, Bram}} \yr{2006} Upwind and high-resolution methods for
  compressible flow: From donor cell to residual-distribution schemes.  \bt{In
  {\em 16th aiaa computational fluid dynamics conference\/}},  \pg{p. 3559}.

\bibitem[{Whipple}(1972)]{Whipple72}
{\sc \au{{Whipple}, F.~L.}} \yr{1972} {On certain aerodynamic processes for
  asteroids and comets}.  \bt{In {\em From Plasma to Planet\/} (ed. \ed{Aina
  {Elvius}})},  \pg{p. 211}.

\bibitem[Withers(1985)]{Withers85}
{\sc \au{Withers, C.S.}} \yr{1985}  \at{The moments of the multivariate
  normal}.  \jt{Bulletin of the Australian Mathematical Society}
  \bvol{32}~(1),  \pg{103–107}.

\bibitem[{Yang} \& {Johansen}(2016)]{Yang16b}
{\sc \au{{Yang}, Chao-Chin} \& \au{{Johansen}, Anders}} \yr{2016}
  \at{{Integration of Particle-gas Systems with Stiff Mutual Drag
  Interaction}}.  \jt{\apjs}  \bvol{224}~(2),  \pg{39},  \arxiv{arXiv:
  1603.08523}.

\bibitem[{Youdin} \& {Johansen}(2007)]{Youdin07b}
{\sc \au{{Youdin}, A.} \& \au{{Johansen}, A.}} \yr{2007}  \at{{Protoplanetary
  Disk Turbulence Driven by the Streaming Instability: Linear Evolution and
  Numerical Methods}}.  \jt{\apj}  \bvol{662}~(1),  \pg{613--626},
  \arxiv{arXiv: astro-ph/0702625}.

\bibitem[{Youdin} \& {Lithwick}(2007)]{Youdin07}
{\sc \au{{Youdin}, Andrew~N.} \& \au{{Lithwick}, Yoram}} \yr{2007}
  \at{{Particle stirring in turbulent gas disks: Including orbital
  oscillations}}.  \jt{\icarus}  \bvol{192}~(2),  \pg{588--604},  \arxiv{arXiv:
  0707.2975}.

\bibitem[{Zhu} {\em et~al.\/}(2014){Zhu}, {Stone}, {Rafikov} \& {Bai}]{Zhu14}
{\sc \au{{Zhu}, Zhaohuan}, \au{{Stone}, James~M.}, \au{{Rafikov}, Roman~R.} \&
  \au{{Bai}, Xue-ning}} \yr{2014}  \at{{Particle Concentration at
  Planet-induced Gap Edges and Vortices. I. Inviscid Three-dimensional Hydro
  Disks}}.  \jt{\apj}  \bvol{785}~(2),  \pg{122},  \arxiv{arXiv: 1308.0648}.

\end{thebibliography}


\end{document}